\documentclass[aps,prd,groupedaddress,preprintnumbers,nofootinbib]{revtex4-1}
\usepackage{amsmath, amssymb}
\usepackage{mathtools}
\usepackage[pdftex]{graphicx}
\usepackage{setspace, enumitem}
\usepackage[colorlinks]{hyperref}
\usepackage{color}
\usepackage[normalem]{ulem}
\usepackage{enumitem}
\usepackage{physics}
\usepackage[dvipsnames]{xcolor}
\usepackage{simpler-wick}
\usepackage{float}
\usepackage{amsmath}
\usepackage[capitalise]{cleveref}
\usepackage{calrsfs}
\DeclareMathAlphabet{\pazocal}{OMS}{zplm}{m}{n}
\usepackage{BOONDOX-cal}
\usepackage{comment}
\usepackage{mathrsfs}
\usepackage{lipsum}
\usepackage{mwe}

\usetikzlibrary{decorations.pathreplacing}

\definecolor{AWG}{RGB}{212,175,55} % Gold
\definecolor{JMLB}{RGB}{4, 55, 242} %Ultramarine
\definecolor{MPP}{RGB}{231, 84, 128} % Dark Pink
\definecolor{EDG}{rgb}{0.172549, 0.627451, 0.172549} % Green
\definecolor{MFR}{RGB}{198, 0, 0} % Dark Red 

\definecolor{Gwpink}{rgb}{1,0.5,0.5} %myc8
\definecolor{Gwblue}{rgb}{0,0.666667,0.666667} %myc6
\definecolor{del5}{rgb}{0.4,0.7,1} 
\definecolor{del3}{rgb}{0.96,0.73,1} 
\definecolor{del2}{rgb}{1,0.66,0.61} 

\definecolor{stringblu}{rgb}{0, 0.26, 0.5} % mycolors$3
\definecolor{stringred}{rgb}{0.75, 0, 0} % mycolors$3
\definecolor{stringora}{rgb}{1, 0.498039, 0.054902} % mycolors$3
\definecolor{stringpin}{rgb}{0.99607843137, 0.50980392156862, 0.549019607843} % mycolors$3
\definecolor{verde}{rgb}{0,0.5,0}

\newcommand{\nds}{ N_{{\rm D}7} }
\newcommand{\CYV}{\pazocal{V}}
\newcommand{\CYB}{X_3}
\newcommand{\ori}{\widetilde{X}_3}
\newcommand{\oricycle}{\widetilde{\Pi}}
\newcommand{\CYcycle}{\Pi_4}

\newcommand{\wrap}{\mathcal{w}}
\newcommand{\magn}{\mathcal{m}}

\newcommand{\CYdiv}{\pazocal{D}}
\newcommand{\CYimdiv}{\pazocal{D}^{\prime}}
\newcommand{\oridiv}{\widetilde{\pazocal{D}}}

\newcommand{\fstop}{\, .}

\graphicspath{{Figures/}}

\begin{document}

\preprint{DESY 23-197}

%=============================================================================
% \title{MSCNI\\
% Gravitational Waves from the axiverse}
\title{Gravitational Axiverse Spectroscopy:\texorpdfstring{\\ Seeing the Forest for the Axions}{}}

\author{Ema Dimastrogiovanni}
\affiliation{Van Swinderen Institute for Particle Physics and Gravity, University of Groningen, Nijenborgh
4, 9747 AG Groningen, The Netherlands}

\author{Matteo Fasiello}
\affiliation{Instituto de F\'isica T\'eorica UAM-CSIC, Calle Nicol\'as Cabrera 13-15, 28049, Madrid, Spain}

\author{Margherita Putti}
\affiliation{II. Institut f\"{u}r Theoretische Physik, Universit\"{a}t Hamburg
Luruper Chaussee 149, 22607 Hamburg, Germany}
\affiliation{Deutsches Elektronen-Synchrotron DESY, Notkestr. 85, 22607 Hamburg, Germany}

\author{Jacob M. Leedom}
\affiliation{Deutsches Elektronen-Synchrotron DESY, Notkestr. 85, 22607 Hamburg, Germany}

\author{Alexander Westphal}
\affiliation{Deutsches Elektronen-Synchrotron DESY, Notkestr. 85, 22607 Hamburg, Germany}
%=============================================================================

\begin{abstract}
We consider inflationary models with multiple spectator axions coupled to dark gauge sectors via Chern-Simons (CS) terms. The energy injection into Abelian gauge fields from the axions engenders a multi-peak profile for scalar and tensor spectra. We highlight the constraining power of CMB spectral distortions on the scalar signal and discuss the conditions under which spectator sectors can account for the recently observed stochastic gravitational wave (GW) background in the nHz range.
Given the tantalizing prospect of a multi-peak ``GW forest'' spanning 
several decades in frequency, we elaborate on possible ultraviolet origins of the spectator models from Type IIB orientifolds. String compactifications generically produce a multitude of axions, the ``Axiverse'', from dimensional reduction of p-form gauge fields. The CS coupling of such axions to dark gauge fields in the worldvolume theory of D7-branes can be tuned via multiple brane wrappings and/or quantized gauge field strengths. If string axions coupled to Abelian gauge fields undergo slow-roll during inflation, they produce GW signals with peaked frequency distribution whose magnitude depends on the details of the compactification. We discuss the restrictions on spectator models from consistency and control requirements of the string compactification and thereby motivate models that may live in the string landscape as opposed to the swampland.
\end{abstract}

\maketitle

{
\hypersetup{linkcolor=blue}
\tableofcontents
}

%==============a================
\section{Introduction}
\label{sec:intro}
%==============================
The observable Universe is a strange thing. We see~\cite{COBE:1992syq,Mather:1993ij,Fixsen:1996nj} its incredible isotropy and homogeneity, which appears highly unnatural~\cite{Collins:1972tf}. Inflation~\cite{Guth:1980zm,Linde:1981mu,Albrecht:1982wi} emerged as a compelling explanation to these observations, and remains the premier paradigm in the standard model of cosmology. A period of inflation not only accounts for the observed uniformity, but resolves other cosmological puzzles, such as the non-observation of magnetic monopoles.

While the \textit{mechanism} of inflation itself is rather simple and effective, the \textit{microphysics} of inflation remains largely unknown. Inflation is commonly modeled via the dynamics of a scalar field, the inflaton, or  multi-field generalizations thereof. Successful inflation typically requires that the inflaton satisfies so-called slow-roll conditions, associated to the flatness of its potential, for a sufficient number of e-folds. However, the existence of a flat region in the scalar potential makes it susceptible to perturbations from higher-dimensional operators. Consequently, understanding the microphysics of inflation involves a reckoning with the ultraviolet (UV) physics of our Universe, especially quantum gravity. Indeed, it has been postulated that quantum gravity consistency constraints impose severe restrictions on inflationary models~\cite{Ooguri:2006in}. 

Remarkably, there are several known mechanisms within string theory that successfully realize inflation such as axion monodromy~\cite{Silverstein:2008sg,McAllister:2008hb,Kaloper:2008fb,Kaloper:2011jz}, fibre inflation~\cite{Cicoli:2008gp}, and others~\cite{Conlon:2005jm,Krippendorf:2009zza}. 

A generic prediction of string theory is the existence of axions\footnote{We will refer to axion-like particles (ALPs) as axions.}~\cite{Svrcek:2006yi}. They arise as the zero modes from dimensional reduction of p-form fields~\cite{Witten:1984dg}, degrees of freedom of open strings attached to D-branes, and even as the lowest lying Kaluza-Klein states in extremely warped compactifications~\cite{Hebecker:2015tzo,Hebecker:2018yxs,Carta:2021uwv}. The number of axions in a compactification is tied to the topology of the compactification manifold via the Hodge numbers and can vary from a few to $\pazocal{O}(10^2)$, all the way to $\pazocal{O}(10^5)$ in extreme cases~\cite{Douglas:2006es,Grimm:2012yq}. This potentially huge number of axions is the basis for the so-called string axiverse~\cite{Arvanitaki:2009fg,Acharya:2010zx,Cicoli:2012sz,Demirtas:2018akl,Demirtas:2021gsq}. These axions tend to couple to hidden gauge sectors and gravity via Chern-Simons couplings. The presence of multiple axions can significantly shape the dynamics of the early Universe. At a generic point in the string Landscape viable for cosmology, one could expect the EFT to contain an inflationary sector and a spectator sector consisting of multiple axions and their hidden gauge theories. Such scenarios are multi-spectator generalizations of the rolling axion models of~\cite{Peloso:2016gqs,Namba:2015gja} and of spectator chromonatural inflation (SCNI) models~\cite{Dimastrogiovanni:2016fuu,Obata:2016tmo}, themselves related to natural inflation  \cite{Freese:1990rb,Adams:1992bn,Anber:2009ua,Barnaby:2010vf,Cook:2011hg,Maleknejad:2011jw,Barnaby:2011vw,Adshead:2012kp,Dimastrogiovanni:2012st,Watanabe:2020ctz,Dimastrogiovanni:2023oid}.
We will refer to such models as Multiple Abelian Spectator Axion (MASA) and Multiple non-Abelian Spectator Axion (MnASA) inflationary models, depending on the nature of the hidden gauge groups.

What makes this class of models particularly compelling is the realistic prospect of testing significant portions of their parameter space via upcoming cosmological probes. Spectator axions roll down their potential and, as a result of the CS couplings, dissipate into the gauge sectors. The enhanced gauge quanta can then significantly source gravitational waves and, depending on the specific setup, also curvature fluctuations. Among the most interesting features that ensue are a large chiral GW spectrum, which may also exhibit a blue or  bump-like structure. Multi-spectator models support a rich peak structure for the GW signal, giving rise to what we call a ``GW forest\footnote{This term was used in~\cite{Kitajima:2018zco} for gravitational waves from the axiverse, where a different mechanism was in play w.r.t. the one we shall employ here.}". Crucially, chirality may be put to the test at CMB scales \cite{Gluscevic:2010vv,Thorne:2017jft} as well as at interferometers \cite{Seto:2007tn,Smith:2016jqs,Thorne:2017jft,Domcke:2019zls}. Let us underscore here that this specific feature, which can be traced back to the parity breaking CS term (including the gravitational CS \cite{Bartolo:2017szm,Bartolo:2018elp,Mirzagholi:2020irt}), is a very distinctive signature of this class of models\footnote{One should add, vis-\`{a}-vis birefringence, that at CMB scales low multiples are more effective at constraining primordial chirality \cite{Gluscevic:2010vv}.}.
Large GW non-Gaussianities \cite{Agrawal:2017awz,Agrawal:2018mrg,Dimastrogiovanni:2018xnn,Ozsoy:2021onx} are yet another testable feature of such scenarios. In the case of Abelian gauge sectors the sourcing mechanism for scalar curvature and tensor perturbations, the former being mediated by the axion field, are analogous. This makes for peaked scalar and GW spectra, a possibility that has been investigated also in the context of primordial black holes as a dark matter candidate \cite{Garcia-Bellido:2017aan,Ozsoy:2020kat}.

A potentially generic signature of inflationary models descending from string theory is a \textit{gravitational wave forest} from a plethora of spectator axions during inflation.  Observations of the corresponding peak-like structure in the GW and possibly also scalar sectors would shed light on the number of axions in the EFT as well as their properties -- thus constituting a form of  gravitational spectroscopy of the axiverse. As there is no guarantee in string constructions that additional axions beyond the QCD axion will couple to the Standard Model\footnote{ 
Although such a coupling can be induced if the axiverse solves the QCD axion quality problem via the mechanism in~\cite{Burgess:2023ifd,Choi:2023gin}.}, gravitational wave signals may represent one of the very few observables to test the axiverse. For other discussions on observing the axiverse, see~\cite{Cicoli:2012sz,Marsh:2013taa,Tashiro:2013yea,Yoshino:2014wwa,Kamionkowski:2014zda,Acharya:2015zfk,Daido:2015bva,Yoshino:2015nsa,Emami:2016mrt,Karwal:2016vyq,Gorbunov:2017ayg,Visinelli:2018utg}. Note that an axion-driven GW forest in general has two sources. One is comprised of the spectators we are discussing here. The second arises from the inflationary sector itself if it occurs in several shortly interrupted epochs~\cite{DAmico:2020euu} of slow-roll axion inflation~\cite{DAmico:2021vka,DAmico:2021fhz}.

 In general the salient features of the spectator models GW signal depend heavily on the specifics of the gauge group, the axion initial conditions and mass, and the strength of the axion-gauge coupling. The latter is particularly relevant. As axions enjoy a (perturbative) continuous shift symmetry, they couple to gauge fields via the usual Chern-Simons (CS) terms with a coupling we denote by $\lambda$ (see~\cref{eq:fullL1} below). To get a sizable GW signal, one requires that $\lambda \simeq \pazocal{O}(10)$, or larger still, in the non-Abelian case. Superficially this appears to be an innocuous demand, but in truth it is non-trivial and challenging to realize from a UV perspective~\cite{Agrawal:2018mkd,Bagherian:2022mau}. The fundamental challenge is that $\lambda\propto m \alpha$, where $m$ is some integer and $\alpha$ is the fine-structure constant of the hidden gauge group coupled to the spectator axion. For non-Abelian spectators, the models in the literature require small values of $\alpha$, and so
attaining $\lambda\gtrsim 10$ requires a large integer m. The primary difficulty of UV embeddings of SCNI models lies in realizing a sufficiently large $m$. In contrast, spectator axions coupled to Abelian gauge fields do not self-interact, one can take larger values of $\alpha$ and thereby reduce the demand on the integer $m$. That is not to say that Abelian spectators are without constraints - attempts to boost the Chern-Simons coupling can result in issues such as the descent of Landau poles. We will revisit constraints on both Abelian and non-Abelian spectator models in~\cref{sec:stringy}.\\

\indent If one wishes to make deeper connections between spectator models and the string axiverse, one must explore how to realize spectator sectors within string compactifications. For SCNI models, this task was considered in~\cite{McDonough:2018xzh,Holland:2020jdh}. The axionic portion of the spectator sector can arise from dimensional reduction of p-forms in the $10d$ string theories. The gauge sector of spectator models depends greatly on which corner of the string landscape one works in. We will largely focus on type IIB string theory compactified on orientifolded Calabi-Yau (CY) manifolds with quantized 3-form fluxes, D7-branes and O7-orientifold planes. In this setting, $4d$ axions arise as KK zero modes of the 4-form $C_4$ and 2-form $C_2$ gauge fields. As mentioned above, the number of axions is governed by the number of compact $n$-dimensional sub-manifolds (n-cycles) of the $6d$ CY manifold chosen, as well as the structure of the orientifold projection: some number of 4-form axions are always present, while 2-form axions arise from a non-trivial `projection-odd' sector of the orientifold action. Gauge sectors are realized by the worldvolume theory of D7-branes wrapping 4-cycle submanifolds of the CY and permeating our $4d$ spacetime. The two types of closed string axions differ in the way they couple to the D7-brane worldvolume gauge fields via Chern-Simons terms: the 4-form axions intrinsincally couple to the worldvolume theory, while 2-form axions only acquire such a coupling in the presence of a particular type of quantized magnetic flux on the D7-brane.\\
\indent The ``intrinsic'' size of these CS couplings turns out to be too small to generate GW signals detectable with current or planned experiments. However, both CS couplings increase linearly with the number of times the D7-brane ``wraps'' a 4-cycle, and the 2-form axion CS coupling in addition increases with the amount of magnetic flux used on the D7-brane to generate it. A non-trivial obstacle to realizing Abelian spectator sections in this construction is the presence of St\"{u}ckelberg couplings between axions and $U(1)$ gauge bosons, which cause the gauge boson to eat the axion and combine in a massive spin-1 degree of freedom. This is a generic issue for 2-form axions, whose St\"uckelberg couplings arise geometrically. In contrast, 4-form axions acquire St\"{u}ckelberg couplings only in the presence of certain magnetic fluxes. Under certain assumptions on the topology of the CY compactification manifold, both St\"{u}ckelberg couplings can be avoided.\\
\indent The goals of this work are manifold. First, we explore the cosmology of MASA and MnASA inflationary models to establish  gravitational waves observations as one of the ultimate testing grounds of the string axiverse. We then consider in detail how to realize these models in string compactifications, the inherent constraints placed on such constructions, and their implications for signal detection.\\
\indent The paper is organized as follows. In~\cref{sec:cosmo} we review some of the main models comprising a spectator sector during inflation and generalize them towards the MASA and MnASA scenarios. We focus mainly on the former, deriving the scalar curvature and gravitational wave spectra in both curvature and tensor perturbations. 
The overall signal stems from superimposing the contributions from multiple spectator sectors and gives rise to  peak-like structures. For each sector contribution the amplitude of the signal will be determined by the number of e-folds of the axion rolling and the Chern-Simons coupling between the axion and the gauge field. The positions of the peak(s) depends also on the initial conditions of the axion field. What emerges is a plethora of GWs at different scales populating the GW forest. The ever-growing number of GW experiments will soon cover about 20 decades in frequencies, providing ample opportunities look for such a forest.  The overall spectrum of curvature perturbations shares a similar profile. \\
\indent In ~\cref{sec:signal} we show how constraints from spectral distortions can reduce the allowed parameter space of MASA models. We also highlight the constraints stemming from primordial black holes abundance limits. We detail on the primordial GW signal and its dependence on initial conditions, axions stacking and, most importantly, CS couplings. We also investigate the effect of scalar induced gravitational waves.
In pointing out the ability of MASA models to potentially explain the recent PTA observations~\cite{NANOGrav:2023gor,NANOGrav:2023hvm,Antoniadis:2023ott,Reardon:2023gzh,Xu:2023wog,Unal:2023srk}, we identify the parameter space supporting a GW signal detectable at intermediate and/or small scales.  \\
\indent In~\cref{sec:stringy} we discuss the potential stringy origin of spectator models in the context of type IIB orientifold compactifications. We expound the requirements on the manifold and D7-brane content to furnish a viable spectator sector and examine the viability of 4-form and 2-form axions as spectator axions. We also place constraints on such constructions. First, we demand that any spectator sector not produce anomalously large induced D3-charge tadpoles. We also apply bounds on sub-manifold volumes needed to guarantee the existence of the controlled 4D EFT description such that higher-order quantum corrections (both perturbative and non-perturbative) remain under control. The result of these string theory constraints is the relevance of the orientifold-odd 2-form axions: the 4-form axions can enhance their GW signal only by increasing the wrapping number of the D7-brane generating the $U(1)$ gauge field, which is forbidden by the EFT control constraints. Hence, detectable GW signals will be a sign for the presence of orientifold-odd 2-form axion spectators for which we provide the correlated constraints in their main examples and associated string theory parameter space leading to detectable GW signals. \\
\indent We offer our final remarks and conclusions in~\cref{sec:conclusion}. Additional details relevant to the computation of the primordial power spectra and to the results of \cref{sec:stringy} are provided respectively in Appendices~\ref{app:calculations} and \ref{app:stringconv}, while the calculations of backreaction and perturbativity constraints can be found in Appendix~\ref{app:backreaction}.

%==============================
\section{Gravitational Waves from Multiple Spectator Axions}
\label{sec:cosmo}
%==============================

%%%%%%%%%%%%%%%%%%%%%%%%%%%%%%%
\subsection{Review of Single Axion Spectator Models}
%%%%%%%%%%%%%%%%%%%%%%%%%%%%%%%

The study of axion-like particles in inflationary physics has a rich history, with the natural inflation model \cite{Freese:1990rb} being perhaps the most well-known example. Coupling the axion with a gauge sector is the next logical step given that the symmetries of the theory are preserved and the great interest and motivation in exploring the inflationary particle content. The simplest realization is that of coupling the axion-inflaton to U(1) vector fields \cite{Anber:2009ua} via a Chern-Simons (CS) term, thereby effectively flattening the potential without resorting to a trans-Planckian axion decay constant\footnote{The latter is hard to implement in string theory  constructions \cite{Banks:2003sx,Svrcek:2006yi}.}. The multi-field nature of these models and the parity violation originating from the CS term makes for an interesting inflationary phenomenology with very distinct, testable, signatures:~from the non-trivial spectral shape of scalar and tensor degrees of freedom to chiral gravitational waves, from large non-Gaussianities to primordial black holes.

The interest in exploring the broader class of axion - gauge fields models   together with the formidable power of cosmological probes to constrain our models has led to a flurry of (on-going) research activity. Relaxing the requirement that the axion be the inflaton opens up new intriguing directions.  Spectator axions make these models more malleable in terms, for example, of the scales at which their key signatures are most pronounced. For specific models, the spectator nature of the axion is dictated by the need to overcome the possible tensions with CMB observations. The Lagrangian encompassing the case of spectator axions reads 
%------------------
\begin{equation}
\begin{aligned}
    \pazocal{L}&\supset  \pazocal{L}_{\text{inf}} + \pazocal{L}_{S} \,,\\
    \pazocal{L}_{\text{inf}} &= -\frac{1}{2}(\partial\varphi)^2- V_{\text{inf}}(\varphi)\,,\\
    \pazocal{L}_S &= -\frac{1}{4}F_{a \mu\nu } F^{a \mu\nu} -\frac{1}{2}(\partial\chi)^2 - V_{S}(\chi) - \frac{\lambda}{4f}\chi\; F_{a \mu\nu }\tilde{F}^{a \mu\nu }\,.
  %  \label{fullL2}
\end{aligned}
\label{eq:fullL1}
\end{equation}
%------------------
In the above expressions we have defined $V_{\text{inf}}(\varphi)$ and $V_{S}(\chi)$ as the inflaton $\varphi$ potential and the spectator axion $\chi$ one,  respectively. The spectator gauge boson is coupled to the axion via a Chern-Simons coupling term with $\widetilde{F}^{a\mu\nu} = \frac{1}{2\sqrt{-g}} \epsilon^{\mu\nu\rho\sigma}F_{\rho\sigma}^a$ and $f$ the axion decay constant. The Chern-Simons coupling typically takes the form of 
\begin{equation}
    \lambda=  \mathfrak{q} \,\frac{\alpha}{\pi} \,,
    \label{qq}
\end{equation}
where $\alpha =\frac{g^2}{4\pi}$ is the fine structure constant with $g$ the gauge coupling and $\mathfrak{q}$ is a constant that varies from model to model.
Both Abelian~\cite{Barnaby:2012xt,Mukohyama:2014gba,Namba:2015gja,Peloso:2016gqs,Garcia-Bellido:2017aan} and non-Abelian~\cite{Dimastrogiovanni:2016fuu,Agrawal:2017awz,Thorne:2017jft,Agrawal:2018mrg,Lozanov:2018kpk,Maleknejad:2018nxz,Dimastrogiovanni:2018xnn,Fujita:2018vmv} gauge fields have been considered in the literature.

A key compelling feature of the above spectator models lies in their gravitational wave signature. In single-field slow-roll inflation, the gravitational waves are vacuum modes of the metric that are amplified through the inflationary expansion. A slightly red-tilted GW spectrum ensues, with the planned BBO experiment \cite{Crowder:2005nr} as the only probe which may be able to detect such a signal. Multi-field scenarios can support a much richer GW spectrum but it remains non-trivial to realize mechanisms for which the tensor modes can be significant: any source of GWs must produce curvature perturbations through an unavoidable gravitational coupling to the inflaton field. The sourced perturbations generally obey a non-Gaussian statistics so that the existing strong limits on non-Gaussianity of the primordial perturbations force the sourced scalar modes to be subdominant at large scales. This necessarily
 limits the strength of the sourcing mechanism and its signal. As we shall see,
 axion - gauge field models and in particular spectator models are able to evade these constraints while providing detectable signals. 

 The dynamics common to all\footnote{By that we mean here models where the the inflaton is (not) an axion and models whose gauge sector is (non)-Abelian. These features apply to all the four possible combinations.} models goes as follows. The axion rolls down its relatively steep potential. Its coupling to the gauge sector acts as friction and excites the gauge modes. Due to the parity breaking nature of the CS term, the solution for the left and right-handed polarization of the (vector or tensor) gauge quanta have a different equation of motion, one of them being temporarily amplified. This amplification is transmitted on to scalar curvature fluctuation and gravitational waves, with the details depending on the specific model. The sourced gravitational wave spectrum is chiral and is typically blue in the case of an axion-inflaton whilst bump-like features\footnote{The latter are of particular interests vis-\`{a}-vis primordial black holes.} are possible for a spectator axion. These features make such models attractive from both the experimental and theoretical standpoint, as they may evade potential constraints from quantum gravity~\cite{Ooguri:2006in}.

Models where the axion acts as the inflaton and is driven by the standard cosine potential are naturally much more constrained. This is true for the Abelian case, whilst observational constraints directly rule out \cite{Dimastrogiovanni:2012ew,Adshead:2013nka} the well known chromo-natural (CNI) model \cite{Adshead:2012kp}, at least its simplest realization. There are manifold ways to render axion-gauge field models viable while preserving their tantalizing GW signatures. One may consider a different potential or remain completely agnostic as to the nature of the potential altogether. Another possibility is spontaneous symmetry breaking of the gauge symmetry \cite{Adshead:2016omu}. Yet another natural step is to ask the axion(s) to be a spectator field and let another field be the inflaton \cite{Barnaby:2012xt, Dimastrogiovanni:2016fuu,Obata:2016tmo}. This, as may be expected, relaxes the constraints on the axion dynamics: it is true that GW signatures remain tied to the axion rolling-down its potential, but it is now the slow-roll of another field along another potential that is directly related to observables such as the scalar spectral index $n_s$. Advocating a spectator axion pays off in the case of an SU(2) gauge sector: the spectator model of \cite{Dimastrogiovanni:2016fuu}, SCNI, has a viable cosmology. This comes at a not insignificant cost from the top-down perspective: the required values of the Chern-Simons coupling $\lambda$ are $\lambda\sim\pazocal{O}(10^2)$ when $\alpha\sim \pazocal{O}(10^{-12})$, and this gives an unnaturally high integer $\mathfrak{q}$ in \cref{qq}.  This has been argued to be problematic for embedding the models in an ultraviolet completion, at least for some model-building techniques~\cite{Agrawal:2018mkd,Bagherian:2022mau}. The same issues are not in place for Abelian models.

As we shall see below, in both Abelian and non-Abelian models with spectator axions, the latter are typically taken to be heavy, such that their energy density washes away after they reach their minimum, and the curvature perturbations can be identified with that of the inflaton. However, in general, inflationary constructions can accommodate both heavy and light axions, fully in accordance with the expectations of the axiverse.

In the following, we will first study these models considering one spectator axion, and then generalize  to  multiple spectators. 
The non-Abelian spectator model of \cite{Dimastrogiovanni:2016fuu} will not generally, in the weak backreaction regime (see \cite{Iarygina:2023mtj} for strong backreaction\footnote{There also exists a rich literature on strong backreaction in Abelian models, see e.g. \cite{DallAgata:2019yrr,Domcke:2020zez,Caravano:2022epk,Peloso:2022ovc,vonEckardstein:2023gwk,Garcia-Bellido:2023ser,Figueroa:2023oxc}.}), exhibit distinctive peaks in the GW spectrum but a rather broad profile. This is due to the constraining power of stability and  consistency conditions~\cite{MP:2023} the model ought to satisfy. When generalizing to multiple spectator sectors, we find it convenient to focus on the Abelian case (MASA) given that this configuration shows distinctive peak-like structures whose detection is less taxing on the values
of the Chern-Simons couplings.

%%%%%%%%%%%%%%%%%%%%%%%
\subsubsection{Inflation with Abelian Spectators}
\label{sectiontwo}
%%%%%%%%%%%%%%%%%%%%%%%
We first consider the Abelian models of~\cite{Barnaby:2012xt,Namba:2015gja}: these are described by Eq.~(\ref{eq:fullL1}), the trace over the gauge sector being trivial in this case. The inflaton potential $V_{\text{inf}}(\varphi)$ is assumed to be sufficiently flat to grant a nearly constant Hubble rate $H$.
A simple cosine potential characterizes the spectator axion,
\begin{equation}
    V_{S}(\chi)=\frac{\Lambda^4}{2}\left[\cos\left(\frac{\chi}{f}\right)+1\right]\,,
    \label{axion-cos-potential}
\end{equation}
with $f$ the axion decay constant, and the mass of the axion being $m_\chi^2=
\frac{\Lambda^4}{2 f^2}$\footnote{In Sec.~\ref{sec:stringy} we will be using $\mu^4\equiv\frac{\Lambda^4}{2}$ for consistency with conventional notation in string theory.}, where in
 the slow roll regime $\eta_{V}\equiv M_P^2 |V^{\prime\prime}/V|\propto m_\chi^2\simeq%\frac{\mu^4}{f^2}=
\frac{\Lambda^4}{2 f^2} \cos(\chi/f)$. %\ED{$\mu$ versus $\Lambda$ fix throughout}.
Such a periodic potential is relatively steep, so much so that it has been necessary to postulate a trans-Planckian decay constant to avoid the tension between natural inflation \cite{Freese:1990rb,Adams:1992bn} and CMB data, until recent observations \cite{Planck:2018jri,BICEP:2021xfz} ruled out the model altogether. An efficient mechanism to effectively flatten the potential is to couple the axion to a gauge sector via a Chern-Simons term: the rolling axion ``dissipates'' energy into the gauge sector. Even if the axion plays merely a spectator role, the following dynamics is still in play: the rolling field excites gauge quanta that are non-linearly coupled to tensor modes thus engendering an intriguing GW phenomenology. By virtue of its spectator nature\footnote{Strictly speaking one ought to also  require that its energy density becomes negligible by the end of inflation or, at least, that it stays sub-leading. This is in contradistinction, for example, to the well-know curvaton scenario \cite{Lyth:2001nq}.}, the axion does not directly provide a significant contribution to scalar curvature perturbations. It is nevertheless gravitationally coupled to the inflaton field: the axion will then mediate an interaction between gauge fields and the inflaton, thus sourcing the curvature fluctuation $\zeta$.   
\bigskip

\noindent If the curvature of its potential is tuned to be $\pazocal{O}(H)$ during inflation, the axion rolls from (nearly) the maximum at $\chi=0$ down to (nearly) the minimum of the potential at $\chi=\pi\,f$  in a few e-folds, of the order of $\pazocal{O}\left(H^2/m^2\right)$. The slow roll solution to the $\chi$ equation of motion derived from~\cref{eq:fullL1} is \cite{Namba:2015gja}
\begin{equation}
\begin{aligned}
    \chi&=2 f \arctan(e^{\delta H(t-t_*)})\,, \\
    \dot{\chi}&=\frac{f H\delta}{\cosh(\delta H(t-t_*))}\,,
\end{aligned}
\end{equation}
with $\delta\equiv\frac{\Lambda^4}{6H^2f^2}$ and $t_*$ the moment when $\frac{\chi}{f}=\frac{\pi}{2}$, corresponding to the time when the axion is at its maximal velocity.~The resulting power spectra will show a peak in the vicinity\footnote{A more accurate estimate would amount to $k_{\rm peak}= \pazocal{O}(\rm{a\,few}) \times k_*$ .} of $k=k_*$\cite{Namba:2015gja}, which can be deduced from the previous equation in terms of the initial conditions
\begin{equation}
    k_*=k_{in}\tan\left(\frac{\chi_{in}}{2f}\right)^{-1/\delta} \; .
    \label{k_initial}
\end{equation}
In order for slow roll to hold, we require
\begin{equation}
    \frac{\ddot{\chi}}{3H\dot{\chi}}=-\frac{\delta}{3}\tanh(\delta H(t-t_*)
    )\ll 1 \, \rightarrow \delta\ll 3\,.
\end{equation}
It turns out to be very useful to define, for both Abelian and non Abelian models, the parameter 
\begin{equation}
\xi\equiv \frac{\lambda\dot{\chi}}{2Hf}\,.%= \frac{\lambda\delta}{\cosh(\delta H(t-t_*))}%=\frac{\xi_*}{\cosh(\delta H(t-t_*))}\; ,=\frac{2\xi_*}{\left(\frac{a}{a_*}\right)^{\delta}+\left(\frac{a_*}{a}\right)^{\delta}}\,,
\label{xi-parameter}
\end{equation}
In the Abelian case one finds $\xi_*=%\frac{\lambda\dot{\chi}_*}{2H f}=
\lambda\frac{\delta}{2}$, which is independent of the axion decay constant $f$.\\
\begin{paragraph}*
\noindent Having defined some of the key background quantities, we can briefly discuss the axion mass. Let us take as the typical mass for an ultralight axion  $m_{\chi}\sim 10^{-12}\,$eV. In the present context we ask that the axion rolling last (i) more than a few e-folds and (ii) indicatively less than 60 e-folds. The lower bound is on account of stringent CMB bounds on scalar and tensor spectra: a more prolonged rolling ensures the peak of the signal is at smaller, less constrained, scales. Conversely, the upper bound guarantees that the most interesting phenomenology takes place during the last sixty efolds of inflation. Upon recalling that the axion rolls for about $\Delta N\sim 6 H^2/m_\chi^2$ e-folds, we ought to require that
\begin{equation}
    m_\chi\gtrsim \frac{H}{\sqrt{10}}\; ,
\end{equation}
thus identifying a minimum value for the axion mass in terms of the proxy inflationary scale $H$ and highlighting the difficulty in fitting a light axion within these setups.
\end{paragraph}

%%%%%%%%%%%%%%%%%%%%%%%
\subsubsection{Inflation with non-Abelian Spectators}
%%%%%%%%%%%%%%%%%%%%%%%

The Lagrangian in~\cref{eq:fullL1} includes non-Abelian scenarios such as the model in \cite{Dimastrogiovanni:2016fuu}, with the proviso that this time the trace goes over  gauge indices. We focus on the SU(2) case given the rich literature on this model (see e.g. \cite{Pajer:2013fsa}) and the fact that the key features of its dynamics and observables are shared by a much larger class of theories \cite{Fujita:2022fff}. The axion potential is the same as in Eq.~(\ref{axion-cos-potential}), but without the overall $1/2$ factor, removed in order to conform to the existing literature. One may choose the vector field vacuum expectation value (vev) components as $ \left\langle A_0^a(t)\right\rangle=0$ and $\left\langle A_i^a(t)\right\rangle=\delta_i^a a(t) Q(t)$. Contrary to the Abelian case, the SU(2) setup can accommodate an isotropic background solution upon identifying the gauge and rotation indices \cite{Maleknejad:2011sq}. It turns out the isotropic one is an attractor solution \cite{Wolfson:2020fqz}, further motivating our starting out already in FLRW. The background quantity $\xi$ is defined according to Eq.~(\ref{xi-parameter}) and it is convenient to introduce also

\begin{equation}
m_Q\equiv g\frac{Q(t)}{H}\,, 
\end{equation}
directly associated with the gauge field vev. The two parameters are identical in the large $m_Q$ limit\footnote{For completeness we should add that in the strong backrection regime there exists a different attractor solution \cite{Iarygina:2023mtj} that does not satisfy this identity.}. Examination of the perturbations at the linear level shows that this model is unstable for $m_Q <\sqrt{2}$ \cite{Dimastrogiovanni:2012ew}.\\ One of the polarizations of the gauge tensor perturbations is amplified and sources GWs. 
By solving the background equations of motion in slow roll dynamics for $Q(t)$ and $\chi(t)$, one can find a good approximation for $m_Q$ that reads
\begin{equation}
m_Q\propto \frac{V_{S,\chi}(\chi)^{1/3} }{H^{4/3}}= \frac{m_\chi^{2/3} f^{1/3}}{H^{4/3}}\sin^{1/3}\left(\frac{\chi}{f}\right)\,,
\label{mchi}
\end{equation}
where $m_\chi^2\equiv\Lambda^4/f^2$ is the axion mass. Given the instability for $m_Q<\sqrt{2}$, we can lower the axion mass only as long as the opposite of this inequality holds.
One simple possibility is to lower the value of $H$, which has been the subject of a thorough investigation in \cite{Fujita:2017jwq}. It turns out one may indeed lower the Hubble rate by many orders of magnitude without trespassing into the strong backreaction regime. This is done by simultaneously acting on $g$. On the other hand, if we also require that the GW signal be detectable by upcoming experiments we are forced to tie the value of $H$ to that of $m_Q$ (for a fixed $r$, the smaller is $H$ the larger becomes $m_Q$). Given how the power spectra scale with the two quantities, when exploring larger and larger $m_Q$ values one soon hits e.g. PBH bounds \cite{Byrnes:2018txb}, which implies that $m_Q$ is limited from above (and, correspondingly, $H$ from below). It follows that plausible detectability confines us to a large $H\gtrsim 10^{10}\, {\rm GeV}$  and therefore (via Eq.~(\ref{mchi})) to a relatively heavy  spectator mass.

% colo2

%%%%%%%%%%%%%%%%%%%%%%%
\subsection{A Multitude of Spectators \& Their Power Spectra}
\label{sectionthree}
%%%%%%%%%%%%%%%%%%%%%%%
The previous two subsections featured only a single spectator axion coupled to a gauge sector. As argued in~\cref{sec:intro}, a natural expectation from string compactifications is the existence of a whole axiverse, with the surfeit of axions that it entails. This serves as a strong motivation to extend the above axion gauge field constructions to allow for multiple axionic (and gauge) spectators\footnote{See~\cite{Obata:2014loa} for a related discussion regarding CNI models.}.  
Of course, there are plentiful possibilities, including those where the spectator gauge group is a semi-simple sum of both non-Abelian and Abelian groups. The class of theories we will explore stems from generalizing~\cref{eq:fullL1} as follows
%------------------
\begin{equation}\label{eq:MASAlag}
    \begin{aligned}
    \pazocal{L}&\supset  \pazocal{L}_{\text{inf}} + \sum_i\pazocal{L}_{S_i}\,,\\
    \pazocal{L}_{inf} &= -\frac{1}{2}(\partial\varphi)^2- V_{inf}(\varphi)\,,\\
    \pazocal{L}_{S_i} &= -\frac{1}{4}F_{i a \mu\nu } F^{a \mu\nu}_i -\frac{1}{2}(\partial\chi_i)^2 - V_{S_i}(\chi_i) - \frac{\lambda_i}{4f_i}\chi_i\; F_{ i a \mu\nu }\tilde{F}_i^{a \mu\nu }\; ,
    \end{aligned}
\end{equation}
%------------------

 where $\chi_i$ and $F_i^a$ are the spectator axions and gauge field strengths, respectively, with $i=1,...,N_S$ .
We have also defined the Chern-Simons couplings $\lambda_i$, the axion decay constants $f_i$, and the scalar potentials $V_{S_i}(\chi_i)$, the explicit form of the latter being that of Eq.~(\ref{axion-cos-potential}). For the sake of simplicity, we will focus here on the Abelian case. Note that we are not pursuing the possibility of a direct coupling between the inflaton and the spectators. We are also neglecting mixing between axions and/or gauge fields from different spectator sectors. Such possibilities are of course permitted and can be included to further enrich the landscape of MASA models.

We shall now study the gravitational wave spectrum of MASA models. First, let us revisit the gravitational waves generated in the single spectator case.

\begin{figure}[ht]
    %\centering
     \includegraphics[scale=.6]{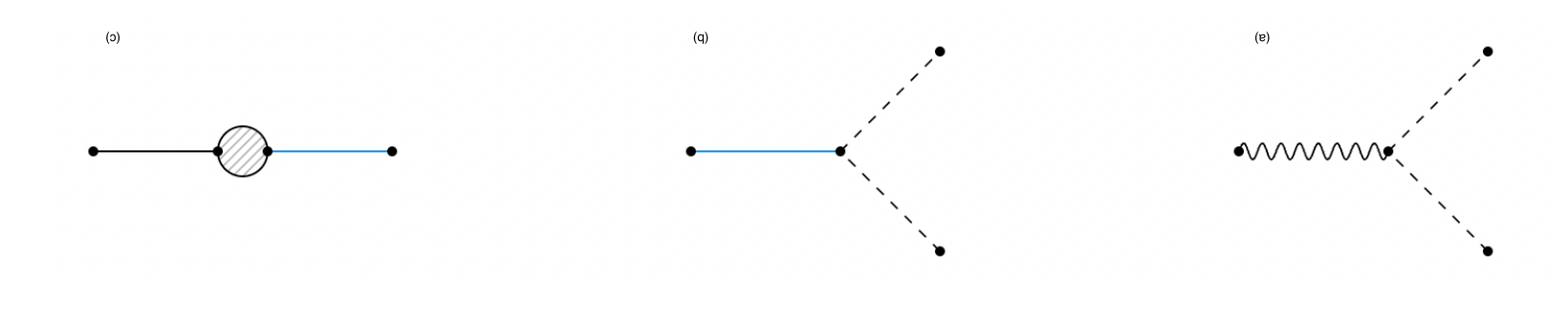}
    \caption{ 
    \textit{Left}: the pictorial representation of the (quadratic) interaction between $\delta\phi$ (black propagator) and $\delta \chi$ (blue). \textit{Center}: the cubic interaction between $\delta\chi$ and the gauge field quanta $A_{\mu}$ (dashed line)  \textit{Right}: the cubic interaction between tensor modes (wiggly propagator) and the gauge field quanta.     }
    \label{fig:pic6}
\end{figure}

\begin{figure}[ht]
  %  \centering
     \includegraphics[scale=.55]{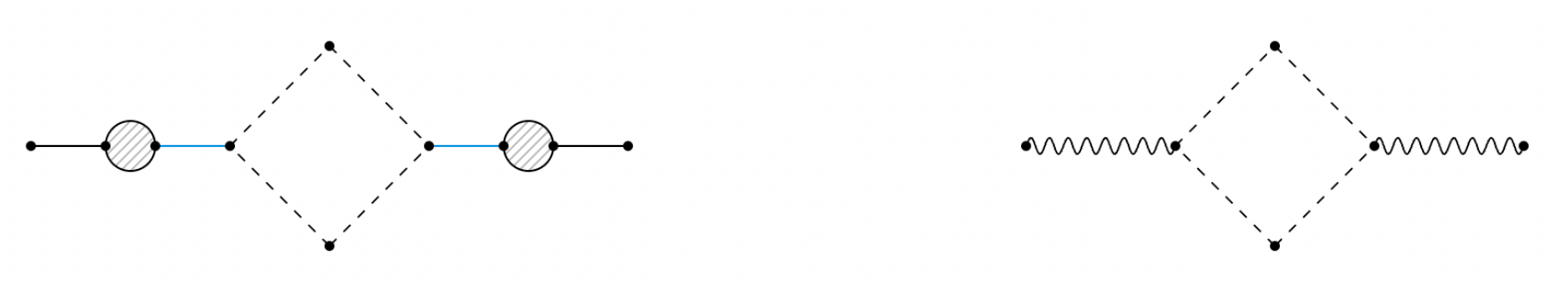}
    \caption{\textit{Left}: the sourced scalar power spectrum mediated by $\delta \chi$. The fluctuation of the inflaton $\phi$ are linearly related to $\zeta$.  \textit{Right}: the sourced tensor  power spectrum. One may use the \textit{in-in} formalism to calculate these observables. However, in what follows we employ the equally valid Green's function method.}
    \label{fig:pic7}
\end{figure}

When $\chi$ is rolling, $\dot{\chi}\neq 0$ and gauge fields are produced via
\begin{equation}
  \delta\chi  \to \delta A+\delta A \,,
\end{equation}
then, in turn, gauge perturbations source scalar and tensor perturbations via
\begin{equation}
\begin{aligned}
&\delta A+\delta A\to\delta \phi\; ({\rm via \, \delta\chi})\, \\
&\delta A+\delta A\to\delta h_{\lambda}\,.
\end{aligned}
\end{equation}
The interaction between the gauge field and the inflaton is purely gravitational so that the direct channel  $\delta A+\delta A\to\delta \phi$ is negligible. By assumption, $\chi$ in this model has an energy density much smaller than that of the inflaton and thus we can identify the scalar curvature perturbation $\zeta$ with the perturbation of the inflaton.  A linear coupling of gravitational nature however still remains between the perturbations of $\chi$ and of $\phi$ as long as $\dot{\chi}\neq 0$, which leads to the partial conversion $\delta\chi \to \delta \phi$. Feynman diagrams in Figs.\ref{fig:pic6} \& \ref{fig:pic7}  help in identifying the interactions involved.

The power spectrum of curvature perturbations $\zeta$ is defined as 
\begin{equation}
    P_\zeta(k)\delta^{(3)}(\vec{k}+\vec{k'}):=\frac{k^3}{2\pi^2}\langle \zeta(\vec{k})\zeta(\vec{k'})\rangle\,,
\end{equation}
and analogously for $h_{\pm}$ (where $+$ and $-$ denote the two gravitational wave polarizations%in the helicity basis
).
The power spectra receive contributions from  vacuum fluctuations (superscript ``(0)'') and the sources (superscript ``(1)''), i.e.:
\begin{equation}
    P_i(k)=P_i^{(0)}(k) +P_i^{(1)}(k) \quad, \, i=\zeta,\,h_+, \, h_-\,.
\end{equation}
The vacuum power spectrum is parametrized as
\begin{equation}
    P_\zeta^{(0)}=A_s\left(\frac{k}{k_0}\right)^{ns-1}\,,
\end{equation}
and the tensor to scalar ration $r$ is defined as 
\begin{equation}
r\equiv {P_T^{\rm total}/ P_{\zeta}^{\rm total}} \,.
\end{equation}

\noindent Let us specialize our discussion of~\cref{eq:MASAlag} to the case with $N_S=2$.  For both axions we take $\dot{\chi}_i\neq0$ and the slow roll solutions will be analogous to the case with one spectator sector
\begin{equation}
\begin{aligned}
    \chi_i &= 2f_i \arctan(e^{\delta_i H(t-t_*)})\\
    \dot{\chi}_{i}&=\frac{f_i H\delta_i}{\cosh{(\delta_i H(t-t_*^i)}}\; ,\label{chistar}
\end{aligned}
\end{equation}
with $\delta_i=\frac{\Lambda_i^4}{6H^2f_i^2}$ and $t_*^i$ is again the moment when $\chi_i=\frac{\pi f}{2}$. 
The validity of the slow-roll approximation requires 
\begin{equation}
    \frac{\ddot{\chi}_i}{3H\dot{\chi}_i}=-\frac{\delta_i}{3}\tanh(\delta_i H(t-t_*^i)
    )\ll 1 \, \rightarrow \delta_i\ll 3\,.
\end{equation}
As before, we define
\begin{equation}
\xi_i=\frac{\lambda_i\dot{\chi}_i}{2Hf_i}=\frac{\xi_*^i}{\cosh(\delta_i H(t-t_*^i))}=\frac{2\xi_*^i}{\left(\frac{a}{a_*^i}\right)^{\delta_i}+\left(\frac{a_*^i}{a}\right)^{\delta_i}}\,,
\end{equation}
with $\xi_*^i=\frac{\lambda_i\dot{\chi}_*^i}{2H f_i}=\frac{\lambda_i\delta_i}{2}$.
\subsubsection{Amplification of tensor perturbations}
Let us move on to tensor perturbations. The amplification w.r.t. the vacuum comes from the standard kinetic term of the gauge fields in~\cref{eq:MASAlag}. Indeed, the Chern-Simons term does not contain the metric tensor directly as the indices of $\tilde{F}$ are raised with the antisymmetric tensor $\epsilon^{\mu\nu\rho\sigma}$. Naturally, the CS coupling is still crucial in light of its effect on  gauge fields at the level of their wavefunction. Following e.g. \cite{Namba:2015gja}, we employ the Coulomb gauge and perform the mode expansion 
\begin{equation}
    \hat{A}_i^{(l)}(\tau, \vec{x})=\int \frac{d^3 k}{(2 \pi)^{3 / 2}} \mathrm{e}^{i \vec{k} \cdot \vec{x}} \hat{A}_i^{(l)}(\tau, \vec{k})=\sum_{\lambda= \pm} \int \frac{d^3 k}{(2 \pi)^{3 / 2}}\left[\epsilon_i^{(\lambda)}(\vec{k}) A^{(l)}_\lambda(\tau, \vec{k}) \hat{a}^{(l)}_\lambda(\vec{k}) \mathrm{e}^{i \vec{k} \cdot \vec{x}}+\text { h.c. }\right],
\label{eq:gaugemode1}
\end{equation}
where $l=1,2$. In a spatially flat, inflating, Universe with Hubble parameter $H$ and scale factor $a(\tau ) = -1/(H \tau)$, the mode functions satisfy (for both $l=1,2$)
\begin{equation}
     A_{ \pm}^{(l) \prime \prime }+\left(k^2 \pm \frac{4k\xi_*^{(l)}}{\tau\left[\left(\tau/\tau_*^{(l)}\right)^{\delta_l}+\left(\tau_*^{(l)}/\tau\right)^{\delta_l}\right]}\right)A_{\pm}^{(l)}=0\; ,
\label{eq:gaugemodeEQ1}
\end{equation}
where the prime indicates a derivative with respect to the conformal time $\tau$. Taking $\xi_*\geq0$, only the positive helicity mode is amplified. Dropping the index $l$ for the moment, we can solve~\cref{eq:gaugemodeEQ1} using the WKB approximation:
\begin{equation}\begin{aligned}
A_{+}(\tau, k) &\simeq\left[\frac{-\tau}{8 k \xi(\tau)}\right]^{1 / 4} \tilde{A}(\tau, k), \\ A_{+}^{\prime}(\tau, k) &\simeq\left[\frac{k \xi(\tau)}{-2 \tau}\right]^{1 / 4} \tilde{A}(\tau, k)\,,
\label{eq:WKBsol}
\end{aligned}
\end{equation}
where
\begin{equation}\label{eqgauge1}
\tilde{A}(\tau, k) \equiv N\left[\xi_*, x_*, \delta\right] \exp \left[-\frac{4 \xi_*^{1 / 2}}{1+\delta}\left(\frac{\tau}{\tau_*}\right)^{\delta / 2}(-k \tau)^{1 / 2}\right]\,,
\end{equation}
with $N$ a time-independent normalization factor. The electric and magnetic field can be defined as 
\begin{equation}
    \hat{E}_i\equiv -\frac{1}{a^2} \hat{A}_i^{\prime}, \quad \hat{B}_i\equiv\frac{1}{a^2} \epsilon_{i j k} \partial_j \hat{A}_k\,.
\label{eq:EMfields}
\end{equation}
The mode expansion for the transverse, traceless tensor perturbations of the metric $\hat{h}_{ij}$ defined in~\cref{app:calculations} reads 
\begin{equation}
    \hat{h}_{i j}(\tau, \vec{k})=\frac{2}{M_p\, a(\tau)} \int \frac{d^3 k}{(2 \pi)^{3 / 2}} \mathrm{e}^{i \vec{k} \cdot \vec{x}} \sum_{\lambda=+,-} \Pi_{i j, \lambda}^*(\hat{k}) \hat{Q}_\lambda(\tau, \vec{k})\,,
\label{eq:hexpand}
\end{equation}
with $\Pi_{i j, \lambda}^*(\hat{k})=\epsilon_i^{(\pm)}(\hat{k})\,\epsilon_j^{(\pm)}(\hat{k}) $ the polarization operators. The tensor modes equation of motion reads
\begin{equation}
    \left(\frac{\partial^2}{\partial \tau^2}+k^2-\frac{2}{\tau^2}\right) \hat{Q}_\lambda(\vec{k}, \tau)= \hat{\pazocal{S}}_\lambda^{(1)}(\tau, \vec{k})+\hat{\pazocal{S}}_\lambda^{(2)}(\tau, \vec{k})\,,
\end{equation}
where 
\begin{equation}
   \hat{\pazocal{S}}_\lambda^{(l)}(\tau, \vec{k}) \equiv-\frac{a^3}{M_p} \Pi_{i j, \lambda}(\hat{k}) \int \frac{d^3 x}{(2 \pi)^{3 / 2}} \mathrm{e}^{-i \vec{k} \cdot \vec{x}}\left[\left(\hat{E}_i \hat{E}_j+\hat{B}_i \hat{B}_j\right)_l\right] \,,
   \label{sourcedh}
\end{equation}
and where we have re-introduced the spectator label $l$. The sourcing explicit in Eq.~(\ref{sourcedh}) is the one represented\footnote{There is one such diagram for each value run by the spectator label ``$l$''.} in the right panel of Fig.~\ref{fig:pic7}.  Decomposing the solution into homogeneous and sourced parts, the vacuum mode is given by 
\begin{equation}
\begin{aligned}
\hat{Q}_\lambda^{(0)}(\vec{k}) & =h_\lambda(\tau, k) \hat{a}_\lambda(\vec{k})+h_\lambda^*(\tau, k) \hat{a}_\lambda^{\dagger}(-\vec{k}), \\
h_\lambda(\tau, k) & =\frac{\mathrm{e}^{-i k \tau}}{\sqrt{2 k}}\left(1-\frac{i}{k \tau}\right)\,.
\end{aligned}    
\end{equation}
The inhomogeneous solution is given by 
\begin{equation}
    \hat{Q}_\lambda^{(1)}(\tau, \vec{k})=\int^\tau d \tau^{\prime} G_k\left(\tau, \tau^{\prime}\right) \left[\hat{\pazocal{S}}_\lambda^{(1)}(\tau, \vec{k})+\hat{\pazocal{S}}_\lambda^{(2)}(\tau, \vec{k})\right]\,,
\end{equation}
where the Green's function is defined in \cref{app:calculations}.
The power spectrum $\pazocal{P}$ is defined as 
\begin{equation}
    \pazocal{P}_\lambda(k) \delta_{\lambda \lambda^{\prime}} \delta^{(3)}\left(\vec{k}+\vec{k}^{\prime}\right) = \frac{k^3}{2 \pi^2}\left\langle\hat{h}_\lambda(\vec{k}) \hat{h}_{\lambda^{\prime}}(\vec{k}^{\prime})\right\rangle\,.
\end{equation}
Given that in our case the two source terms are to a good approximation uncorrelated, one may compute the two corresponding inhomogeneous solutions separately and sum them to obtain the resulting spectrum. The calculation of the sourced scalar spectrum is a bit more involved in that the inflaton is only gravitationally coupled to the axion. As a result, the effect of the gauge quanta on the scalar curvature $\zeta$ is mediated by the field $\chi$.

\subsubsection{Curvature power spectrum}
We take the spatially flat gauge, such that the scalar sector of the metric can be written solving for the non dynamical variables $\varphi, B$ as: 
\begin{equation}
    ds^2=a^2(\tau)[-(1+2\varphi)d\tau^2+2\partial_i Bdx^id\tau+\delta_{ij}dx^idx^k]\,.
\end{equation}
The remaining physical modes can be decomposed as 
\begin{equation}
\begin{aligned}  
\hat{\phi}(x,\tau)&=\phi(\tau)+\int\frac{d^3 k}{(2\pi)^{3/2}}e^{-\vec{k}\cdot \vec{x}}\frac{\hat{Q}_\phi(\vec{k})}{a(\tau)}\\
\hat{\chi}_l(x,\tau)&=\chi_l(\tau)+\int\frac{d^3 k}{(2\pi)^{3/2}}e^{-\vec{k}\cdot \vec{x}}\frac{\hat{Q}_{\chi_l}(\vec{k})}{a(\tau)}\,.
\end{aligned}
\end{equation}
We can then rewrite\footnote{Note that we are omitting the kinetic term of gauge fields as these do not play a key role in sourcing scalars.} the action of ~\cref{eq:MASAlag} paired with the Einstein-Hilbert action as
\begin{align}
       S &= S_{\text{free}}^{}+S_{\text{int}}^{(1)}+S_{\text{int}}^{(2)}\nonumber \\
     S_{\text{free}}^{}\left[\hat{Q}_i\right] &= \frac{1}{2} \int d \tau d^3 k\left[\hat{Q}_l^{\prime \dagger} \hat{Q}_l^{\prime}-\hat{Q}_l^{\dagger}\left(k^2 \delta_{lk}+\tilde{M}_{lk}^2\right) \hat{Q}_k\right] \label{eq:curvmode}\\
      S_{\mathrm{int}}^{(l)}&=-\int d^4 x \sqrt{-g}\; \lambda_l \frac{\chi_l}{4 f_l} F_{\mu \nu}^{l} \tilde{F}^{l\mu \nu }=\int d^4 x a^4 \lambda_l \frac{\chi_l}{f_l} \hat{\mathbf{E}}_l \cdot \hat{\mathbf{B}}_l\; , \nonumber
\end{align}
where we have defined $(\hat{Q}_1,\hat{Q}_2,\hat{Q}_3)\equiv(\hat{Q}_\phi,\hat{Q}_{\chi_1},\hat{Q}_{\chi_2})$ and $(\phi_1,\phi_2,\phi_3)\equiv(\phi,\chi_1,\chi_2)$. The electric $\hat{\mathbf{E}}_l$ and magnetic $\hat{\mathbf{B}}_l$ field vectors are those of~\cref{eq:EMfields} and one finds 
\begin{equation}
    \tilde{M}_{lk}^2\equiv-\frac{a^{\prime \prime}}{a} \delta_{lk}+a^2 V_{, lk}+\left(3-\frac{\phi_i^{\prime} \phi_i^{\prime}}{2 M_p^2} \frac{a^2}{a^{\prime 2}}\right) \frac{\phi_l^{\prime} \phi_k^{\prime}}{M_p^2}+\frac{a^3}{M_p^2 a^{\prime}}\left(\phi_l^{\prime} V_{, k}+\phi_k^{\prime} V_{, l}\right)\,,
\label{eq:massmatrix}
\end{equation}
with $V_{, l} \equiv \partial V / \partial \phi_l$. Introducing slow-roll parameters for each of the fields $\phi_l$ as in~\cref{app:calculations}, we obtain, to leading order, the following equations of motion 
\begin{equation}
\begin{aligned}
    \left(\frac{\partial}{\partial\tau^2}+k^2-\frac{2}{\tau^2}\right)\hat{Q}_\phi &\simeq \frac{6}{\tau^2}\sqrt{\epsilon_\phi \epsilon_{\chi_1}} \hat{Q}_{\chi_1}+\frac{6}{\tau^2}\sqrt{\epsilon_\phi \epsilon_{\chi_2}} \hat{Q}_{\chi_2}\,,\\
    \left(\frac{\partial}{\partial\tau^2}+k^2-\frac{2}{\tau^2}\right)\hat{Q}_{\chi_l}&\simeq \lambda_l\frac{a^3}{f_l}\int \frac{d^3 x}{(2\pi)^{3/2}}e^{-i\vec{k}\cdot\vec{x}}(\vec{\hat{E}}\cdot\vec{\hat{B}})_l\equiv \hat{S}_{\chi_l}(\tau,\vec{k}) \,.
\label{eq:slowEoM}
\end{aligned}
\end{equation}
Considering the dominant terms in the mass matrix and in the equations leads us to the two independent equations of motion for the axions. Given that we assume that the two axions' energy densities will vanish after CMB, in the spatially flat gauge the scalar curvature perturbation are given just by the perturbations of the inflaton, 
\begin{equation}
    \hat{\zeta}(\tau,\vec{k})\simeq-\frac{H}{\dot{\phi}}\hat{\delta\phi}(\tau,\vec{k})=\frac{H\tau}{\sqrt{2\epsilon_\phi}M_{pl}}\hat{Q}_\phi(\tau,\vec{k})\,.
\end{equation}
It turns out to be convenient to write the solution as
\begin{equation}
    \hat{Q}_\phi=\hat{Q}_\phi^{(0)} +\hat{Q}_\phi^{(1)}\,,
\end{equation}
where $ \hat{Q}_\phi^{(0)}$ is the homogeneous solution to the inflaton equation of motion and $ \hat{Q}_\phi^{(1)}$ is the particular solution. Expanding the homogeneous solution operator as 
\begin{equation}   
\hat{Q}_\phi^{(0)}(\tau,\vec{k})=Q_\phi^{(0)} (\tau,k) a(\vec{k})+Q_\phi^{(0)*} (\tau,k) a^\dagger(-\vec{k})
\,,
\end{equation}
and imposing Bunch-Davis initial conditions, we find
\begin{equation}
    Q_i^{(0)}(\tau,k)=\frac{e^{-ik\tau}}{\sqrt{2k}}\left(1-\frac{i}{k\tau}\right)\,.
\end{equation}
The particular solution is found, using the retarded Green's function from~\cref{app:calculations}, to be given by
\begin{equation}
  \hat{Q}_\phi^{(1)} =6\sqrt{\epsilon_\phi}\int d\tau' G_k(\tau,\tau')\left(\frac{\sqrt{\epsilon_{\chi_1}}}{\tau'^2}\int d\tau''G_k(\tau',\tau'')\hat{S}_{\chi_1}(\tau'',\vec{k})+\frac{\sqrt{\epsilon_{\chi_2}}}{\tau'^2}\int d\tau''G_k(\tau',\tau'')\hat{S}_{\chi_2}(\tau'',\vec{k})\right)\,.
\label{eq:QphiPart}
\end{equation}
The corresponding solution for the curvature perturbation is
\begin{equation}\label{eqscalar1}
     \hat{\zeta}^{(1)} =\frac{3\sqrt{2}H\tau}{M_{pl}}\int d\tau' G_k
     \left(\frac{\sqrt{\epsilon_{\chi_1}}}{\tau'^2}\int d\tau''G_k\hat{S}_{\chi_1}+\frac{\sqrt{\epsilon_{\chi_2}}}{\tau'^2}\int d\tau''G_k\hat{S}_{\chi_2}\right)\,,
\end{equation}
where the source functions $\hat{S}_{\chi_l}(\tau,\vec{k})$ are defined in~\cref{eq:slowEoM}. The  standard definition of the power spectrum is
\begin{equation}
    P_\zeta(k)\delta^{(3)}(\vec{k}+\vec{k}')=\frac{k^3}{2\pi^2}\langle\zeta(k)\zeta(k')\rangle\,.
\end{equation}
\noindent Much like for the tensor power spectrum, the sourcing is such that the scalar power spectrum is also additive, so that we can write
\begin{equation}
\begin{aligned}  P_\zeta=P_\zeta^{(0)}+\sum_{l=i}^N P_\zeta^{(l)}\,,\\
P_{GW}=P_{GW}^{(0)}+\sum_{l=1}^N P_{GW}^{(l)} \; .
\end{aligned}
\end{equation}

\subsubsection{Axion and Gauge Field Mixing}
In the next section we will rely on the notion that both (i) the coupling between different axions and (ii)  the coupling between different  gauge sectors are only gravitational and can therefore be disregarded. The coupling comes with a slow-roll suppression in the scalar sector  (see e.g. Eq.~(\ref{eq:slowEoM})) and turns out to be negligible also in the case of gauge fields mixing. 

One may intuitively arrive at the latter conclusion as follows. First of all, in the Abelian case any coupling between gauge sectors starts out at the level of the cubic Lagrangian, so we are already in the realm of loop corrections. Secondly, the CS terms is not directly involved given that the field strengths are not contracted through the metric. In solving the constraint equation no one gauge sector plays a leading role with respect to the other one(s). It follows that neglecting mixing across sectors is tantamount to neglecting the sub-leading gauge self-interactions as routinely done in models with only one gauge sector.  

In neglecting the mixing between the different axions extra care needs to be exerted. The leading contribution to the $\delta\chi_i$ power spectra comes from the non-linear interaction with the gauge fields $A^{(i)}_{\mu}$, where ``i'' is the index that runs through the various axion sectors, and is schematically proportional to $\lambda_i^2$. If we want to probe the contribution of another $A^{(j\not=i)}_{\mu}$ gauge sector to the power spectrum of a given field $\delta\chi_i$, we need the mediation of the field $\delta\chi_j$. The price to pay is, as mentioned, a suppression\footnote{Note that here $\epsilon^2$ is a placeholder for a suppression of the order of slow-roll parameters and is to be understood, conservatively, as satisfying $\epsilon <10^{-2}$.} of order $\epsilon^2$. The $\delta\chi_j$-mediated contribution to the $\delta\chi_i$ power spectrum will have to then be compared to the contribution due directly to the $A^i_{\mu}$ sector. In other words we have $ \lambda_i^2 \leftrightarrow \epsilon^2 \lambda_j^2$. So long as there is not a significant hierarchy in place for the values of the coupling constants $\lambda_k$, one can safely neglect the contribution coming from mixing the axion and gauge sectors. 
\bigskip

%%%%%%%%%%%%%%%%
%\subsection{Mixed Axion Calculations?}
%%%%%%%%%%%%%%%%
%==============================
\section{Detecting the Spectator axiverse}
\label{sec:signal}
%==============================

In~\cref{sec:cosmo}, we introduced MASA models and explored their gravitational wave spectrum. A key point is that the lack of direct mixing between axions implies that the total gravitational wave spectrum is, to a good approximation, a  superposition of the spectrum from each spectator axion. This raises interesting possibilities for signals arising from MASA models, namely multiple peaks and signal boosting. 

Specific initial conditions and parameter choices lead to different features in the power spectra.
For example, if we examine axions with comparable initial conditions and parameter values, the resulting power spectra will lead to an amplified signal strength within a specific range of scales. In principle, the number of spectator sectors $N_S$ can then serve as another handle for signal enhancement, independent of values of the Chern-Simons couplings.
However, the crucial question is whether such a scenario is  plausible or if it is too finely tuned to be regarded as compelling and relevant for cosmological model building. In fact, when considering random initial conditions and parameters, we can expect to observe a mix of signals at distinct wavelengths, some of which may exhibit greater enhancements than others. These enhanced signals could arise from scenarios involving stronger Chern-Simons couplings or the stacking described above. 

Once the number of e-folds for which the axion ``i'' rolls is fixed, the amplitude of the peaks in the power spectra will primarily depend on the Chern-Simons coupling parameter $\lambda_i$, which is encoded in $\xi^i_*$. The signal strength depends exponentially on $\lambda_i$ whilst in contrast the stacking of axions, although capable of boosting the signal, is approximately only linear. Given a sufficient number of axions the outcome for the signal becomes essentially stochastic thus motivating the use of statistical methods in this context, akin to \cite{McAllister:2012am} for multi-field inflation.

\subsection{Primordial power spectra}
\label{Primordial power spectra}

\begin{figure}[ht]   
    \includegraphics[scale=0.5]{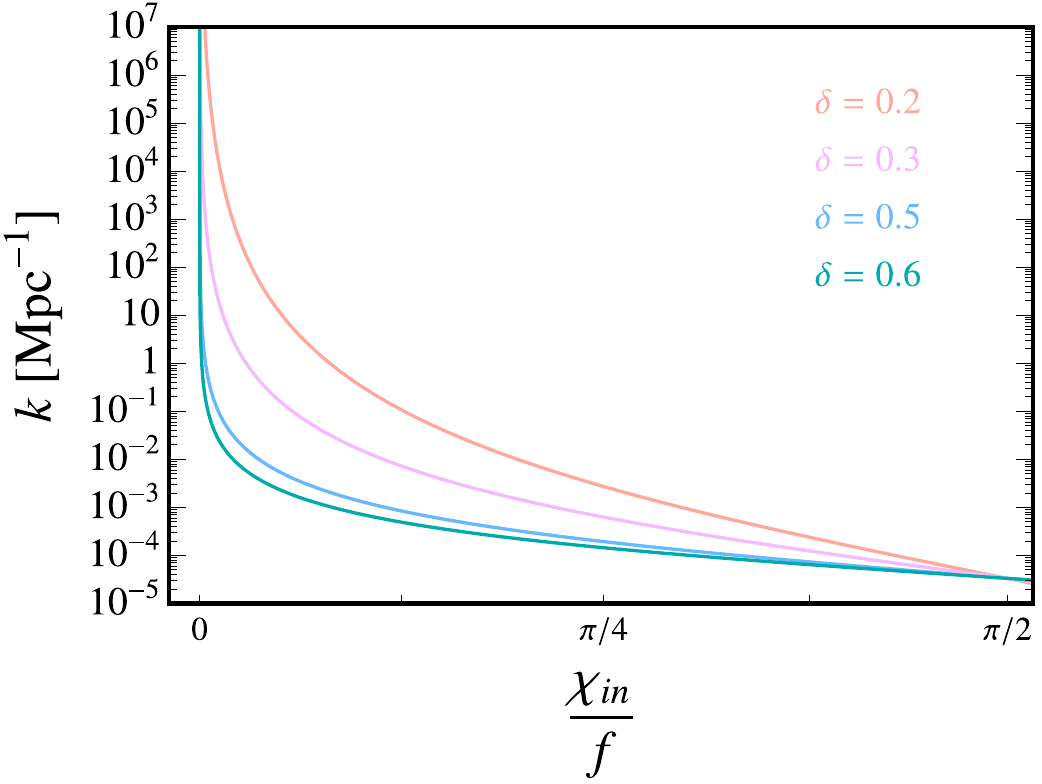}
    \caption{$k_*$ as a function of the initial axion phase, for each choice of $\delta$ considered: red for $\delta=0.2$, pink for $\delta=0.3$, blue for $\delta=0.5$ and turquoise for $\delta=0.6$. $k_*$ peaks for small values of $\chi_{in}/f$. The domain with the lowest absolute value of the slope is the most favorable to stacking.}
    \label{fig:ks_chiin}
\end{figure}

As we showed above in \cref{k_initial}, the position of the peak of the GW signal ($k_*$) is set by the initial phase of the axion, $\chi_{in}/f$, and the number of e-folds during which it rolls, $1/\delta$. Further analysis (see Fig.~\ref{fig:ks_chiin}) shows the existence of a preferred range in $k_*$ corresponding to a higher degeneracy in the initial conditions. This is the most favorable domain for axion stacking. %colo4

The power spectra can be parametrised through~\cite{Namba:2015gja,ozsoy2021synthetic}

\begin{align}
P_\zeta^{(1)} \left( k \right) &=\left[\epsilon \, P_\zeta^{(0)} \left( k \right) \right]^2 f_{2,\zeta} \;\;\;,\;\;\; 
P_T^{(1)} \left( k \right) =\left[\epsilon \, P_\zeta^{(0)} \left( k \right) \right]^2 \left( f_{2,+}+f_{2,-} \right)\,, 
\end{align}
where the dimensionless functions $f_{2,j}$ are well fitted by
\begin{equation}
    f_{2, j}\simeq f_{2, j}^{c}\left[\xi_{*}, \delta\right] \exp \left[-\frac{1}{2 \sigma_{2, j}^{2}\left[\xi_{*}, \delta\right]} \ln ^{2}\left(\frac{k}{k_{*} x_{2, j}^{c}\left[\xi_{*}, \delta\right]}\right)\right]   \;\;,\;\; j = \zeta ,\, h_+ ,\, h_- \,.
\label{f2j}
\end{equation}
The three functions $f_{2, j}^{c}$, $\sigma_{2, j}^{2}$ and $x_{2, j}^{c}$ control the height, the width, and the location of the bump. An analytic fit for the dependence of these functions on $\xi_*$ was described in~\cite{Namba:2015gja}, for the two given values $\delta = 0.2 ,\, 0.5$. We compute these functions for $\delta=0.3$ and $\delta=0.6$ (\cref{table:cosmotable1,table:cosmotable3} ). 
%%%%%%%%%%%%%%%%%%%%%%%%%%%%%%%%
% \MP{Added caption to the table}
\begin{table}[ht!]
\centering
\begin{tabular}{|c|c|c|c|}
\hline$\{2, i\}$ & $\ln \left|f_{2, i}^c\right| \simeq$ &$ x_{2,ij}^c \simeq$ & $\sigma_{2,i} \simeq$ \\ \hline 
 $\{2,\zeta\}$& $-4.926+9.339\xi_*+0.0839\xi_*^2$& $-28.747+12.37\xi_*-1.0989\xi_*^2$&$-3.584+1.633\xi_*-0.155\xi_*^2$\\
\hline$\{2,+\}$ &$ -4.516+8.882 \xi_*+0.108 \xi_*^2$& $-1.324+1.619 \xi_*-0.041 \xi_*^2$& $1.242-0.227 \xi_*+0.017 \xi_*^2$\\
\hline$\{2,-\} $&$ -9.515+8.781 \xi_*+0.115 \xi_*^2$&$ -0.279+0.526 \xi_*-0.009 \xi_*^2 $&$ 1.081-0.135 \xi_*+0.010 \xi_*^2 $\\ \hline\end{tabular}
\caption{$\xi_*$ dependence of the functions appearing in \cref{f2j}, for $\delta=0.3$.}
\label{table:cosmotable1}
\end{table}

\begin{table}[ht!]
\centering
\begin{tabular}{|c|c|c|c|}
\hline$\{2, i\}$ & $\ln \left|f_{2, i}^c\right| \simeq$ &$ x_{2,ij}^c \simeq$ & $\sigma_{2,i} \simeq$ \\ \hline 
 $\{2,\zeta\}$& $-8.819+9.662\xi_*-0.076\xi_*^2$& $0.434+0.962\xi_*-0.020\xi_*^2$&$0.969-0.180\xi_*+0.015\xi_*^2$\\
\hline$\{2,+\}$ &$ -0.759+7.005 \xi_*+0.118 \xi_*^2$& $-0.216+1.096 \xi_*+0.006 \xi_*^2$& $0.670-0.085 \xi_*+0.005 \xi_*^2$\\
\hline$\{2,-\} $&$ -6.010+6.743 \xi_*+0.136 \xi_*^2$&$ 0.536+0.214 \xi_*+0.025 \xi_*^2 $&$ 0.414+0.058 \xi_*-0.006 \xi_*^2 $\\ \hline\end{tabular}
\caption{$\xi_*$ dependence of the functions appearing in \cref{f2j}, for $\delta=0.6$.}
\label{table:cosmotable3}
\end{table}
%%%%%%%%%%%%%%%%%%%%%%%%%%%%%%%%
%
\noindent We also compute the second order scalar induced gravitational waves, given by:
\begin{equation}
    P_T^{II}(k)=1.4 \int_0^{\infty} d v \int_{|1-v|}^{1+v} d u \frac{\pazocal{T}(u, v)}{u^2 v^2} P_{\zeta}(v k) P_{\zeta}(u k)
    \label{secondGW}
\end{equation}
where the  transfer function $\pazocal{T}(u, v) $ is standard (see e.g.\cite{Pi:2020otn}). 
The total GW power spectrum then amounts to the sum of the following contributions:
\begin{equation}
    P_T(k)=P_T^{(0)}(k)+P_T^{(1)}(k)+ P_T^{II}(k)\,.
\end{equation}

\begin{figure}[ht]
    %\centering
    \includegraphics[scale=0.33]{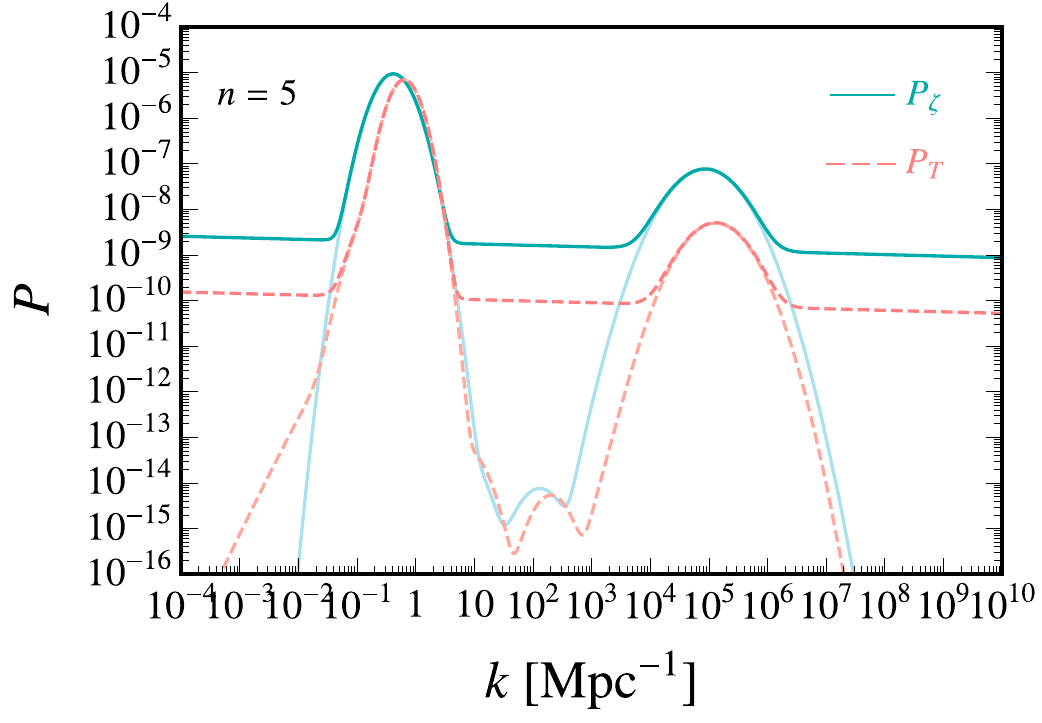}
    \includegraphics[scale=0.33]{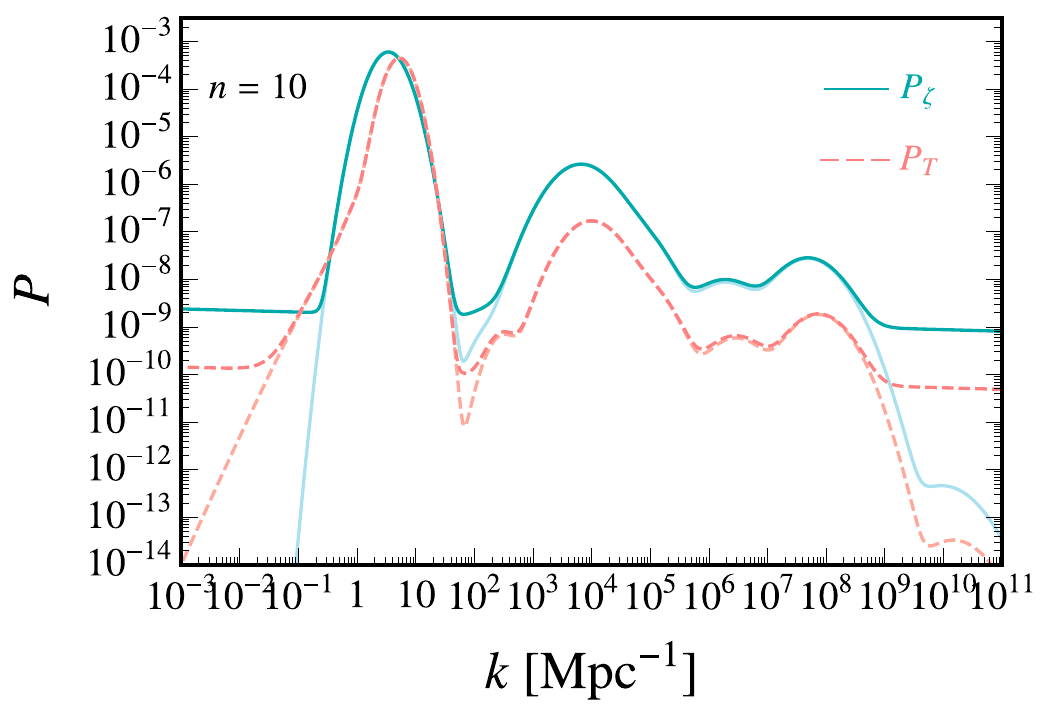}
    \includegraphics[scale=0.33]{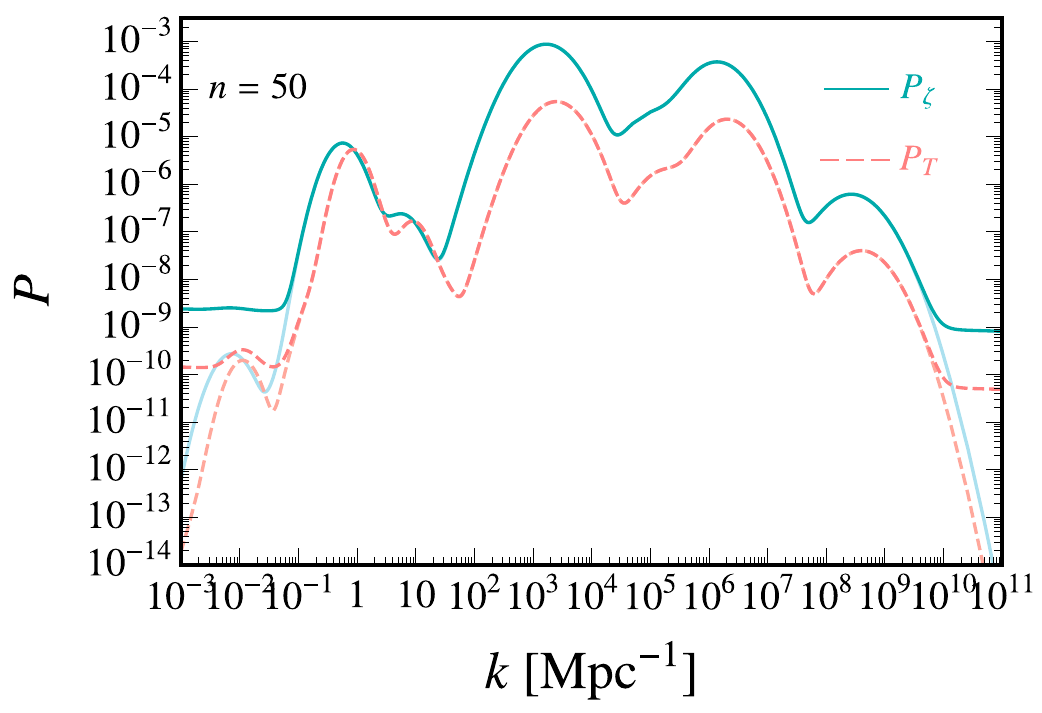}
    \caption{Typical power spectra for a number of spectator sectors $n=5,\,10,\,50$.
    Continuous turquoise lines indicate the total curvature power spectrum, while 
    dashed pink lines show the total $P_T$. Light pink and light turquoise lines mark the sourced contributions alone, without the vacuum term. The values for $\delta$, $\xi_*$ and $\chi_{in}/f$ have been drawn randomly as described in the text. One can notice that not all the spectator sectors contribute in a significant way: some will indeed have a Chern-Simons coupling such that the sourced signal is weaker than its vacuum counterpart.}

    \label{fig:naxion}
\end{figure}

By performing random draws of the parameters, $\xi_*\in[2.5,5]$, $\delta\in \{0.2,0.3,0.5,0.6\}$, and $x_{in}\equiv \chi_{in}/f\in [0,\,\pi/4]$, one can gather what the overall signal could look like for various numbers of spectator sectors in \cref{fig:naxion}. For simplicity, we randomly draw in a flat distribution between those intervals. In \cref{fig:naxion} one can notice how, as the number of axions increases, non-regular patterns on and between the peaks become more visible as a result of the sum and stacking of the signals from multiple spectators.

We note that the contribution from scalar induced GW for these spectra are somewhat visible for the first peak in the $n=5$ case, they are fully visible in the first peak for $n=10$, while they are not visible in the $n=50$ case.

\subsection{Spectral Distortions}
In the discussion above we we delved into the detection of gravitational wave (GW) signals. However, a peculiarity of these models is that generation of GWs is intrinsically linked with the generation of curvature perturbations, as the axion ultimately responsible for the production of GWs is gravitationally coupled with the inflaton.
In the context of our MASA model, this implies that alongside the anticipated GW forest, one also expects a \textit{scalar} forest. This duality arises because each bump in the tensor power spectrum corresponds to a corresponding bump in the curvature power spectrum. Consequently, such a  correlation can be used as a diagnostic tool, allowing one to discern whether the observed GW signal emanates from a MASA model.

In this subsection, we discuss the signatures and constraints that can be imposed on the model by considering cosmic microwave background (CMB) spectral distortions (SDs). SDs manifest as deviations in the CMB frequency spectrum from that of a perfect black body, and they result from processes involving energy exchange between matter and radiation \cite{Zeldovich:1969ff,Sunyaev:1970eu,Illarionov2,Illarionov1,Danese1,Burigana:1991eub,Hu:1992dc} or changes in the number of photons \cite{Hu:1995em,Chluba:2015hma}. Broadly, SDs can be classified into two types, $\mu$ distortions and $y$ distortions, depending on the epoch of their production. The $\mu$ distortion, characterized by a frequency-dependent chemical potential, occurs between the decoupling of double Compton scattering and Bremsstrahlung ($z\sim 2\times 10^6$) and the thermalization decoupling by Compton scattering ($z\sim 5\times 10^4$). On the other hand, the $y$ distortion emerges at lower redshifts, specifically $z\lesssim 5\times 10^4$. 
The COBE/FIRAS mission has constrained distortions to $\mu\leq 9\times 10^{-5}$ and $y\leq1.5\times10^{-5}$ at a $95\%$ confidence level \cite{Fixsen:1996nj}. Presently, discussions revolve around potential future missions such as PIXIE~\cite{Kogut:2010xfw}, Super-PIXIE \cite{Kogut:2019vqh}, PRISM~\cite{PRISM:2013fvg}, COSMO~\cite{Masi:2021azs}, and BISOU~\cite{Maffei:2021xur}. These missions have the capability to detect distortions down to $\mu\sim 10^{-8}$ and $y\sim 10^{-9}$, significantly enhancing our ability to investigate various processes related to the origin and evolution of the universe \cite{Chluba:2019kpb,Chluba:2019nxa,Delabrouille:2019thj}. It's noteworthy that SDs also arise due to the Silk damping effect \cite{Silk:1967kq} of small-scale primordial perturbations upon horizon re-entry after inflation \cite{Sunyaev:1970plh,Daly:1991uob,barrow1991primordial,Hu:1994bz,Chluba:2012gq,Khatri:2012rt}. For a given primordial curvature power spectrum $P_\zeta$, the resulting average distortions are equal to: 
\begin{equation} 
\mu_{\zeta} = \int_{k_{\rm min}}^\infty d \ln k \, P_\zeta \left( k \right) \,  W_\zeta^\mu \left( k \right) \;\;,\;\; 
y_{\zeta} = \int_{k_{\rm min}}^\infty d \ln k \, P_\zeta \left( k \right) \,  W_\zeta^y \left( k \right) \;, 
\label{mu-y-zeta}
\end{equation}
where $k_{\rm min} \simeq 1 \, {\rm Mpc}^{-1}$ and $W_{\zeta}^{\mu,\,y}$ are the scalar window functions, for which we use the analytic approximations provided in ref.~\cite{Chluba:2013dna} 
\begin{equation}
\begin{aligned}
W_\zeta^\mu \left( k \right) &\simeq 2.28  \left[ {\rm exp} \left( - \frac{\left( \frac{\hat k}{1360} \right)^2}{1+\left( \frac{\hat k}{260} \right)^{0.3}+  \frac{\hat k}{340}} \right) - {\rm exp} \left( - \left( \frac{\hat k}{32} \right)^2 \right) \right] \;, \\
W_\zeta^y \left( k \right) &\simeq 0.406 \, {\rm exp} \left( - \left( \frac{\hat k}{32} \right)^2 \right) \;.
\end{aligned} 
\end{equation}
Here ${\hat k}$ is the comoving wavenumber in units of ${\rm Mpc}^{-1}$. 
Primordial tensor modes also contribute to CMB distortions~\cite{Ota:2014hha}. Unlike scalar modes, tensors induce a local quadrupolar anisotropy without requiring photon diffusion. Interactions between photons and electrons, influenced by this anisotropy, lead to dissipation~\cite{Ota:2014hha,Chluba:2014qia}, resulting in the emergence of distortions. 
 \indent The scalar window function experiences a sharp fall at around the diffusion scale, $k_D \simeq 4.0 \times 10^{-6} \left( 1 + z \right)^{3/2} {\rm Mpc}^{-1}$, so the damping of scalar modes contributes to the distortions up to $k\sim 10^4$Mpc$^{-1}$. The dissipation of tensor modes, instead, continues to source SDs up to $k\sim 10^6$Mpc$^{-1}$. 
However, the contribution from GWs turns out to be well below future observational limits for this model \cite{MP:2023}.\\ 
\indent We report here the predictions for the average distortions arising from the dissipation of curvature perturbations in the cases illustrated in \cref{fig:naxion}. One obtains the following values:
\begin{equation}
\begin{aligned}
 \mu&=2.7 \times 10^{-7}\,, \quad y=2.3 \times 10^{-9} \quad \text{for } \, N_s=5 \text{  spectators}\,,\\
 \mu&=1.5 \times 10^{-5}\,, \quad y=1.2 \times 10^{-4} \quad \text{for } \, N_s=10 \text{  spectators}\,,\\
 \mu&= 1.2 \times 10^{-3} \,,\quad y=3.2 \times 10^{-9} \,\quad \text{for }\, N_s=50 \text{ spectators}\,.
\end{aligned}
\end{equation}
The $N_s=10$ and $N_s=50$ scenarios are at odds with existing constraints.
This shows how spectral distortion constraints have the ability to limit the allowed range for the Chern-Simons coupling, in conjunction with the number of axions in the model, whenever the signal reaches its peak within the spectral distortions window. An exhaustive analysis of these constraints would involve accounting for the interplay of several elements of the theory, including the number of axions, their initial conditions, and parameters in the potential. Undertaking such an exhaustive analysis goes beyond the scope of the present paper and is deferred to future work. Our goal here is rather to underscore the necessity of taking spectral distortion into account when investigating predictions of the axiverse, both as a constraining factor and as a signature itself.

\subsection{Gravitational wave signals and future detections}

In this section, we will investigate the conditions on the model parameters that ensue from requiring a GW signal be at the level of the recent PTA observations of a stochastic background \cite{EPTA:2023fyk,NANOGrav:2023gor,Reardon:2023gzh}, or above the sensitivity limits of planned interferometers. 

Let us begin at PTA scales and consider for simplicity the scenario with one spectator sector. We present in \cref{table:cosmotable2} a representative sample parameters set that delivers a GW signal with a strength comparable to the one observed, corresponding to a primordial power spectrum of order $P_T\sim10^{-3}$ \footnote{We verified that the chosen parameters fall within the constraints posed by working in a regime of \textsl{(i) weak backreaction} for the spectator fields, and of (ii) \textsl{perturbative control} for the theory. See Appendix~\ref{app:backreaction} for details.}.\\
%%%%%%%%%%%%%%%%%%%%%%%%%%%%%%%%
%
\begin{table}[ht!]
\centering
\begin{tabular}{|c|l|c|c|c|l|}
\hline$\delta$ && $\chi_{in}\simeq$&$ \xi_*\simeq$& $\lambda\simeq$&$\mathfrak{q}\simeq$\\ \hline 
 $0.2$ && $1.9\times 10^{-2}$& $5$&$50$ & $1570$\\
\hline$0.3$ &&$ 1.9 \times 10^{-3}$& $5.2$& $35$ & $1089$\\
\hline$0.5 $ &&$ 1.8 \times 10^{-5}$&$ 5.5 $&$ 22$ & $691$\\ \hline
 $0.6$ && $ 1.8 \times 10^{-5}$& $5.9$&$20$ & $617$\\\hline\end{tabular}
\caption{Set of parameters used to reproduce, for each given $\delta$, the PTA signal in single-spectator MASA models. }
\label{table:cosmotable2}
\end{table}
%%%%%%%%%%%%%%%%%%%%%%%%%%%%%%%%

\begin{figure}[ht]
    \centering   \includegraphics[scale=0.43]{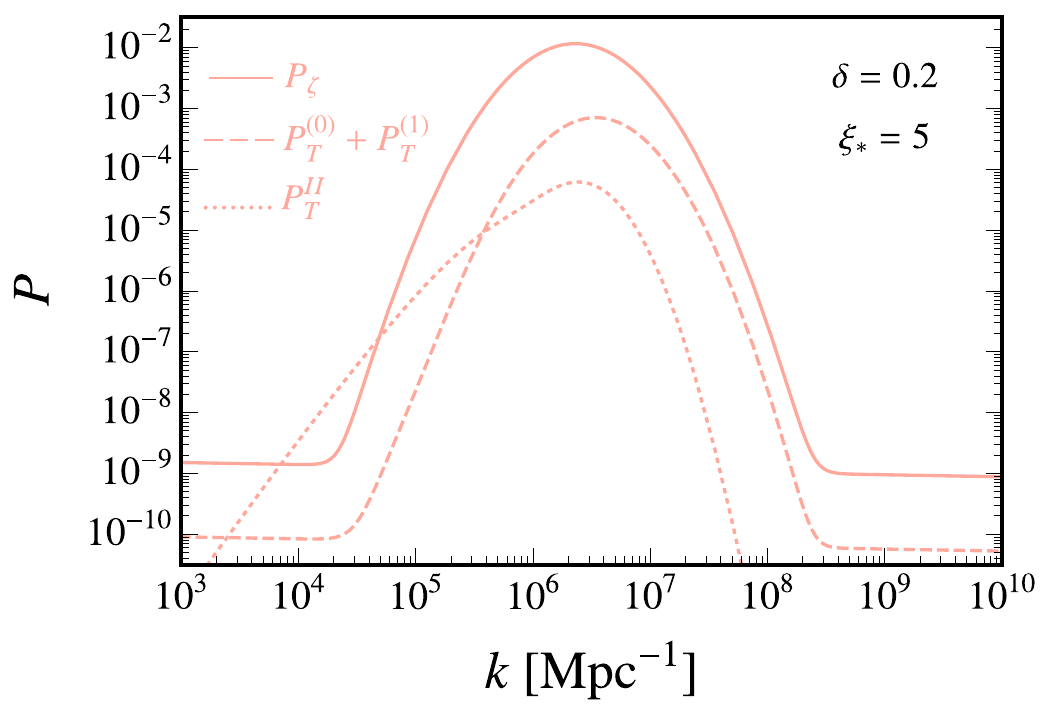}
    \includegraphics[scale=0.43]{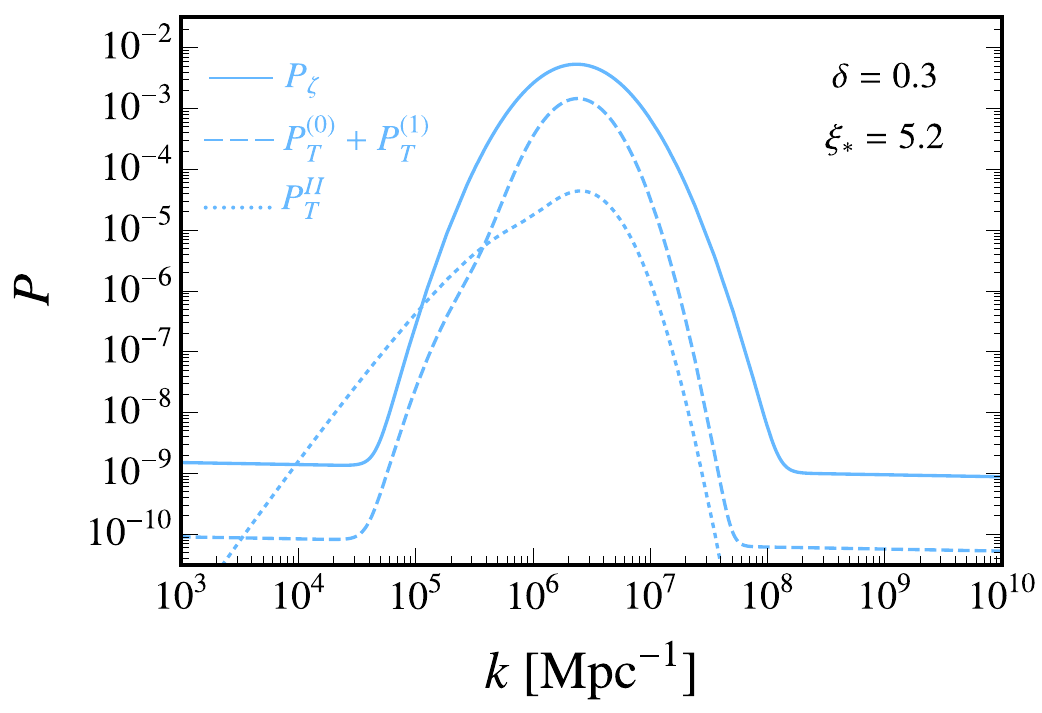}
    \includegraphics[scale=0.43]{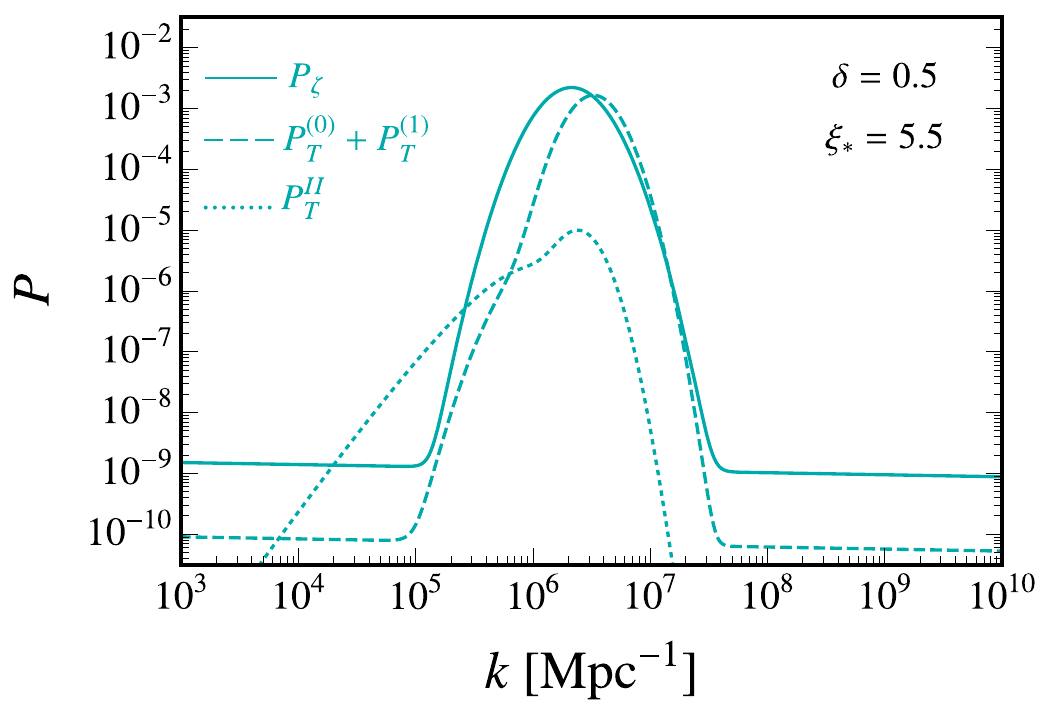}
       \includegraphics[scale=0.43]{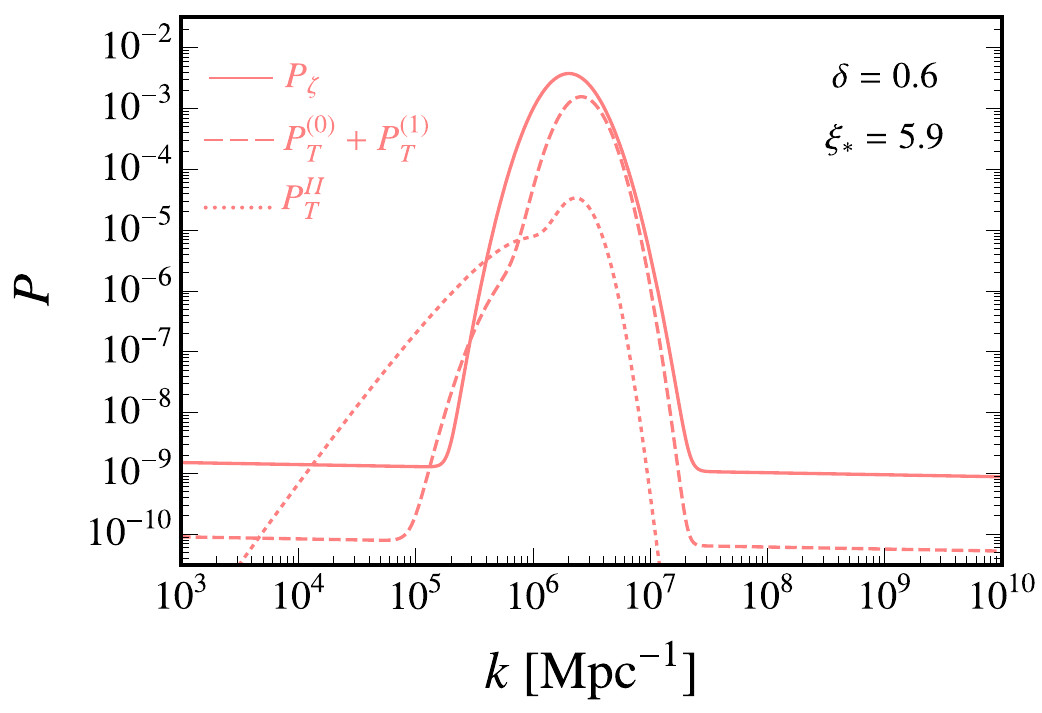}
    \caption{Curvature perturbation (solid line) and GW (dashed line) power spectra for $\delta=0.2$ (top left), $\delta=0.3$ (top right), $\delta=0.5$ (bottom left) and $\delta=0.6$ (bottom right). The dotted line corresponds to the induced gravitational waves computed in \cref{secondGW} }
    \label{fig:peaks_PTA}
\end{figure}

\noindent Here $\mathfrak{q}$ is related to the Chern-Simons coupling via $\lambda= \mathfrak{q} \alpha /\pi $, with the fine structure constant taken to be  $\alpha=0.1$. For every chosen value of $\delta$ and for a given target value of $P_{T}$, a value of $\xi_{*}$ will follow (and, from there, a value for $\lambda$). Given the relation $\xi_*=\lambda \delta /2$, higher values of $\delta$ require a lower $\lambda$ to reach a given amplitude of the signal. 
One can construct, keeping the power spectrum amplitude fixed and interpolating the parameters found above, an approximate relation between the duration of the rolling for the axion, encoded in $\delta$, and the parameter $\lambda$, finding:
\begin{equation}
    \lambda=90.2-246.3 \delta+216.7\delta^2\; ,
    \label{eq:lambdafit}
\end{equation}
a description valid for $\delta$ values ranging between 0.2 and 0.6. \\
\indent The scalar and tensor power spectra corresponding to the benchmark values in \cref{table:cosmotable2} are plotted in \cref{fig:peaks_PTA}. The plots show that smaller values of $\delta$ correspond to increasingly high values of $P_{\zeta}$ and to a lower value of the tensor-to-scalar ratio ($r=P_T/P_\zeta$). We note  that a change in $\delta$ affects the scalar fluctuations more strongly than the tensor fluctuations. This can be easily understood as follows: by increasing $\delta$ one is decreasing the time during which $\chi$ rolls, thus lowering the amplitude of the produced signal as well its width (the smaller $\Delta N\sim1/\delta$, the fewer the modes that exited the horizon while $\dot{\chi}$ was non negligible). Inflaton perturbations are sourced by the $\delta\chi$ modes only while $\dot{\chi}\neq 0$, therefore a decrease of $\Delta N$ affects the sourced scalar modes more than the tensor modes because it decreases both the number of modes that are sourced, and the time interval during which the $\delta\chi$ modes can be converted into inflaton perturbations. \\
\indent We find that, for the set value of $P_{T}$, $P_{\zeta}$ is slightly above the PBH bound in the $\delta=0.2$ case, while larger values of $\delta$ are viable. The left tail of the peak falls within SDs scales, producing a $\mu$ distortion of order $10^{-8}$, possibly at reach for future probes.\\ 
\indent The range considered for $\delta$ (and, as a result, for $\xi_{*}$, $\lambda$ and $\mathfrak{q}$) corresponds to the extent of parameter space supporting a signal compatible with PTA observations: we are working under the condition $\delta\ll 3$ (hence the chosen maximum, $\delta\lesssim 0.6$), while values of $\delta\lesssim 0.2$ are ruled out by constraints on $P_{\zeta}$. A value of $\mathfrak{q}$ of order $10^{2}-10^{3}$ (see Table \ref{table:cosmotable2}) is therefore a prerequisite for a signal compatible with PTA observations. The authors of \cite{Agrawal:2018mkd} discussed the UV embedding of spectator-axion models pointing out that the previously defined $\mathfrak{q}$ (as opposed to $\lambda$) should be the parameter of choice in that context. We will address the feasibility of a $\mathfrak{q}$ of the order of $10^{2}-10^{3}$ in the next sections, where we compute the restrictions on $\lambda$ and $\mathfrak{q}$ from UV embeddings. The scalar-induced second order GWs encoded in $P_T^{II}$ are visible on the left tail of the spectrum in Fig.~\ref{fig:peaks_PTA}. \\

\noindent Let us now discuss what would be the requirements on the model parameters for the signal to be visible by GW detectors at smaller scales. 

\begin{figure}[ht]
    \centering   
    \includegraphics[scale=0.6]{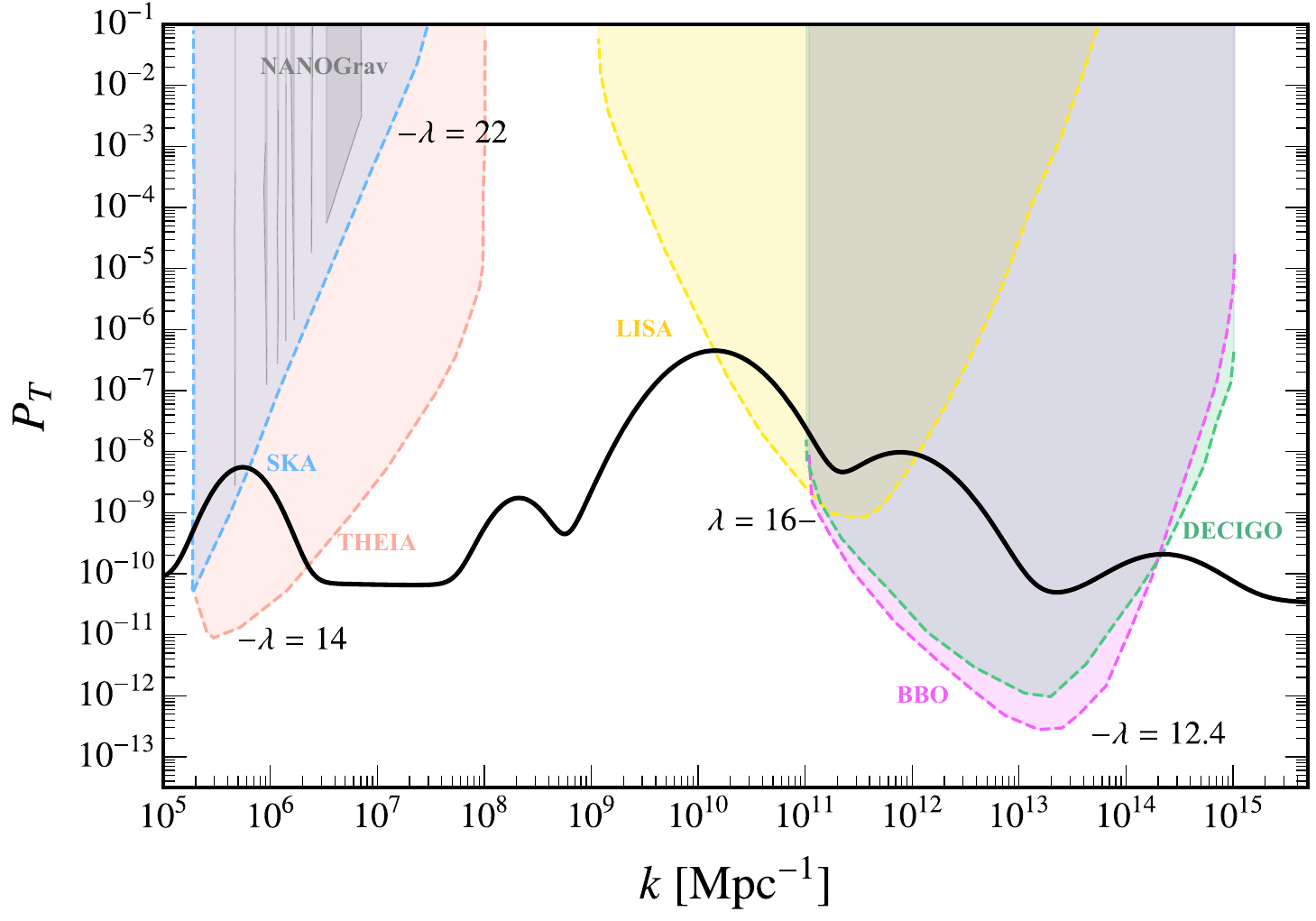}
   \caption{Sensitivity and observations of  stochastic GW backgrounds of current and future detectors, such as LISA~ \cite{amaroseoane2017laser}, SKA~\cite{Janssen:2014dka}, THEIA~\cite{Theia:2017xtk}, BBO~\cite{Corbin:2005ny} and DECIGO~\cite{Seto:2001qf}. Next to the sensitivity curves we show the values of $\lambda$ necessary for the signal in our model to reach those amplitudes when $\delta=0.5$. 
   The black curve shows what a typical signal may look like for $n=10$ spectator sectors with $\xi_*\in[3.5,5]$.}
    \label{fig:GW}
\end{figure}

We provide in \cref{fig:GW} the values of $\lambda$ that are needed, for fixed  $\delta=0.5$, in order for the signal to reach the sensitivity limits of a number of experiments. These values have been derived under the simplifying assumption of one spectator sector (or, which is effectively equivalent, of one spectator sector whose signal dominates over those from other sectors). One can notice how strongly sensitive the signal is to $\lambda$: a small change in this parameter can produce a dramatic change in the spectrum amplitude. As an example, for $\delta=0.5$, at LISA scales ($ 10^{10}- 10^{13}$Mpc$^{-1}$) one requires $\xi_*\gtrsim 4$, so $\lambda\gtrsim 16$ and $\mathfrak{q}\sim 6\times 10^{2}$, to reach the minimum detectable amplitude of $P_T\sim 10^{-9}$. The BBO sensitivity limit (matched up by $P_T\sim  10^{-13}$ around scales of $k\sim 10^{13} $Mpc$^{-1}$) would require $\xi_*\sim3.1$, corresponding to $\lambda\sim12.4$ and $\mathfrak{q}\sim 4\times 10^{2}$. It is also worth stressing that from the argument offered in Sec.~\ref{Primordial power spectra}, a $k_*$ at interferometer scales may not be favoured if parameters are drawn randomly and assuming, as we did, a flat distribution.

As we will elaborate further in the following sections, when embedding these models in string theory, one would like to lower the required CS coupling in order to broaden, as much as possible, the parameter space. To this aim, one can exploit the stacking of the signals: assigning multiple axions initial conditions and parameters in such a way as to have similar $k_*$ leads to peaks at similar scales which then add up. The simplest possibility is to have spectators with similar parameters and initial conditions.
As an example, for a detectable signal by BBO with $\delta=0.5$ and one axion, the required CS coupling is $\lambda\sim 12.4$. On the other hand, incorporating ten spectator axions with comparable initial conditions reduces the necessary CS coupling to $\lambda\sim 11.4$. 
While stacking undeniably helps reduce the required CS coupling for signal observation, it introduces the need for rather specific initial conditions. In the absence of a precise distribution of initial conditions, one may question the naturalness of requiring ten different axions with identical initial field values. Moreover, requiring ten distinct axions with a CS coupling significantly greater than one might prove to be more challenging than having a single axions with a slightly higher CS coupling. 
\\

The general feature of any "stringy" inflationary model is the presence of axions, or more precisely ALPs, that can couple to dark $U(1)$ sectors through CS coupling. The number of these axions and their precise properties depend on the model and the compactification considered. 
However, regardless of the specifics of the UV properties, one can arrive at a general feature: the presence of a gravitational wave forest. Indeed, the natural end point of the above discussion is to expect GWs signals and curvature perturbation peaks throughout the whole frequency spectrum. 
We refrain from giving a specific shape of the signal as this depends on the number of axions, the CS coupling, and the amount of e-folds for which the axions roll. Instead, we want to emphasize that we expect to find a gravitational wave forest by searching for GWs and $\zeta$-peaks at different scales. However, it is essential to acknowledge a significant challenge posed by our analysis. To detect this signal today, a sufficiently large Chern-Simons coupling value is required. This observation raises valid concerns from the perspective of UV physics, as it complicates the integration of these models into a coherent UV framework.
We will address these questions in the following sections.

%==============================
\section{Cultivating Forests in the String Landscape}
\label{sec:stringy}
%==============================
In the preceding sections, we have described MASA inflationary models and their experimental signatures. To tie these observables to the string axiverse, we must discuss embeddings of spectator models within string theory. In this section we consider this task, as well as the constraints on such embeddings and their implications for observable signals.

Since MASA models are Abelian variations of the SCNI model, we first review the attempts to embed this model in a UV framework. Inclusion at the level of $\pazocal{N}=1$ supergravity was discussed in~\cite{DallAgata:2018ybl}. Constructions within type IIB orientifold compactifications were pursued in~\cite{McDonough:2018xzh,Holland:2020jdh}. In particular,~\cite{Holland:2020jdh} sought to embed SCNI models in a Large Volume Scenario (LVS)~\cite{Balasubramanian:2005zx} with the inflaton itself as either a blow-up mode that realizes K\"{a}hler inflation or the fibre modulus in fibre inflation~\cite{Cicoli:2008gp}. The non-Abelian spectator sector is realized by a stack of magnetized D7-branes with an axion arising from dimensional reduction of the 2-form gauge potential $C_2$.

In the following we will mimic and extend this construction. We will largely focus on realizing Abelian spectator sectors due to their distinctive signals discussed above. However, we shall start broadly and attempt a categorization and analysis of axionic spectator candidates and constructions to realize the gauge theory of the spectator sector.

Before proceeding into the details of string models, it is worth considering to what extent one should expect to be able to embed models  with large Chern-Simons couplings into string theory. As briefly discussed in~\cref{sec:intro}, there have been several arguments restricting possibilities to realize large Chern-Simons couplings~\cite{Agrawal:2018mkd,Bagherian:2022mau}. To motivate the difficulty, recall that in the typical quantum field theory perspective of the QCD axion, the Chern-Simons coupling to electromagnetism has a coefficient
$\lambda\propto \text{Tr}(Q_{PQ}Q^2_{EM})\alpha_{EM}$, where the trace runs over the charged fermions in the EFT. Naively, a large CS coupling can then be realized by one of the following strategies: a). large Peccei-Quinn charges $Q_{PQ}$, b). large electromagnetic charges $Q_{EM}$, or c). a large number of fermions. These latter two options are problematic since these charged fermions also enter the self-energy of the photon and a large number of fermions or large charges will bring down the scale of the QED Landau pole~\cite{Agrawal:2017cmd}. The issue with large PQ charges is more subtle, but involves the exponential suppression of fermion masses -- see the appendix of~\cite{Agrawal:2017cmd} for further discussion.

However, this is not to say that it is impossible to increase CS couplings beyond naive expectations. With the strategies above, one may obtain an $\pazocal{O}(10-10^2)$ enhancement. From the QFT perspective, several other model-building strategies have been utilized to boost CS couplings, mostly in the context of the electromagnetic coupling of the QCD axion~\cite{Farina:2016tgd,Agrawal:2017cmd,DiLuzio:2021pxd,DiLuzio:2021gos} and ultralight axion dark matter~\cite{Dror:2020zru}. Such strategies involve models with (i) kinetic mixing of axions~\cite{Babu:1994id}, (ii) KNP alignment~\cite{Kim:2004rp} or clockworking~\cite{Choi:2015fiu,Kaplan:2015fuy}, and (iii) introduction of a discrete symmetry~\cite{Hook:2018jle}. 

While these mechanisms can indeed enhance the CS coupling of an axion to gauge fields, they typically induce unpleasant side effects into the EFT, akin to the descent of the Landau pole. Discrete symmetries introduce many new degrees of freedom, leading to a non-trivial cosmological history. Kinetic mixing appears unrestricted apart from the necessary inclusion of an additional light axion in the EFT. Yet it remains unclear if sufficient mixing can be realized in a string construction~\cite{Agrawal:2017cmd}. However, we will address this mechanism below. In the context of spectator models,~\cite{Agrawal:2018mkd,Bagherian:2022mau} argue that EFT restrictions and cosmological observations completely rule out the possibility of utilizing clockwork to realize SCNI models .  

Given the above arguments restricting the possibilities of realizing large CS couplings in QFT, and therefore spectator models, we may wonder if string theory somehow evades these arguments. Such a statement would be surprising given that the ongoing swampland program posits broad restrictions on EFTs coupled to gravity\footnote{For reviews, see~\cite{Brennan:2017rbf,Palti:2019pca,vanBeest:2021lhn,Agmon:2022thq}.}. Indeed, in alignment with the swampland paradigm, we will find that methods to realize MASA/MnASA models in string theory are extremely restricted. 
%==============================
\subsection{Spectator Sectors from D7-branes}
\label{subsec:stringSCNI}
%==============================

%---------------------------
\subsubsection{States \& Couplings}
%---------------------------
We will consider $4d$ effective field theories obtained by compactifying $10d$ type IIB string theory on a $6d$ space $\ori$. To maintain $\pazocal{N}=1$ supersymmetry, we take $\ori$ to be an orientifold defined via the action of a holomorphic involution on a Calabi-Yau 3-fold $\CYB$. Below the string scale, the theory is described by a supergravity effective field theory. This EFT contains moduli fields and, in the presence of Dp-branes, gauge sectors. While we will not be concerned with realizing the particle physics of our Universe, we assume that there exists some stack of branes that contain the Standard Model or some suitable unified gauge theory extension (for review, see~\cite{Marchesano:2022qbx}).

We are primarily interested in the axionic content of the EFT. From dimensional reduction of the higher-form gauge potentials $C_4$, $C_2$, and $B_2$, the EFT inherits the ``even axions" $\rho_\alpha$ as well as the ``odd-axions" $c^a$, and $b^a$. The span of the indices is determined by the topology of the orientifold via the Hodge numbers as $\alpha = 1,..,h^{1,1}_+$ and $a = 1,..,h^{1,1}_-$. For more details, see~\cref{app:stringconv}.  There is also the $C_0$-axion that descends from the $10d$ 0-form. These axions are organized in $\pazocal{N}=1$ chiral supermultiplets with scalar components~\cite{Grimm:2004uq,Jockers:2004yj}
\begin{equation}
    \begin{aligned}
          S &= C_0 +i e^{-\phi}\,,\\
          G^a &= c^a - Sb^a\,,\\
          T_\alpha &= \tau_\alpha +  i\bigg(\rho_\alpha -\frac{1}{2} \kappa_{\alpha bc}c^b b^c\bigg) + \frac{i}{2(S-\bar{S})} \kappa_{\alpha bc}G^b(G^c - \bar{G}^c)\\
                        &= \tau_\alpha + i(\rho_\alpha-\kappa_{\alpha b c} c^b b^c) + \frac{i}{2}S\kappa_{\alpha b c} b^b b^c\,.
    \end{aligned}
\label{eq:chiralmults}
\end{equation}
Here we have introduced the dilaton $\phi$ as well as the real scalars $\tau_\alpha$. The dilaton vev determines the string coupling $g_s = \langle e^\phi\rangle$ while $\langle\tau_\alpha\rangle$ are the volumes of 4-dimensional submanifolds (4-cycles) $\oricycle_\alpha$ in $\ori$. In addition to the above axion content, there are also open string axions arising from the worlvolume theory of branes, but we will not be concerned with these. 

The $\{\rho_\alpha, c^a, b^a\}$ fields are candidates for the axionic content of the spectator sector, but they must couple to gauge fields via a Chern-Simons coupling to furnish examples of the MASA models described in the previous sections of this paper. As in~\cite{McDonough:2018xzh,Holland:2020jdh}, we 
can realize the gauge theory portion of the spectator sector by wrapping $\nds$ D7-branes on a divisor $\oridiv$ of $\ori$\footnote{A divisor $\oridiv$ is a formal sum of 4-cycles $\oricycle_\alpha$ in $\ori$ whose coefficients are the wrapping numbers of the branes. For more formal details, see~\cref{app:stringconv}.}. The worldvolume theory of such a stack contains a gauge theory with a unitary, orthogonal, or symplectic gauge group. The particular group realized by the brane stack depends on further details of the compactification.

The action of the $4d$ EFT descends from dimensional reduction (and orientifold projection) of the Dirac-Born-Infeld (DBI) and Chern-Simons (CS) actions of the D7-branes as well as the $10d$ bulk type IIB action. It has the schematic form
\begin{equation}
    \begin{aligned}
    \pazocal{S}_{EFT} &\supset \pazocal{S}_{moduli}+ \pazocal{S}_{axions}+\pazocal{S}_{gauge} + \pazocal{S}_{potential}\,.\\
    \end{aligned}
\end{equation}
Here $\pazocal{S}_{moduli}$, $\pazocal{S}_{moduli}$, and $\pazocal{S}_{gauge}$ correspond to contributions to the EFT action from moduli, axions, and the D7-brane stack worldvolume gauge theory, respectively. In particular, $\pazocal{S}_{moduli}$ contains the kinetic terms of the $\tau_\alpha$ whereas $\pazocal{S}_{axions}$ contains the kinetic terms for the axions $\rho_\alpha$, $c^a$, and $b^a$. We will discuss the contributions to the scalar potential of the EFT, encoded in $\pazocal{S}_{potential}$, in a subsequent section. For the worldvolume gauge theory, we have
%---------------------
\begin{equation}
    \begin{aligned}
    \pazocal{S}_{gauge} &\supset \int{\rm d}^4x\sqrt{-g}\left[-\frac14 {\rm Re }[f_{\oridiv}] F^A_{\mu\nu}F^{A\mu\nu}-\frac14 {\rm Im}[f_{\oridiv}]F^A_{\mu\nu}\widetilde{F}^{A\mu\nu}\right]\,,
    \end{aligned}
    \label{eq:D7bosons}
\end{equation}
%---------------------
where $F_{\mu\nu}^A$ are the gauge fields arising from the D7-brane stack. The holomorphic gauge kinetic function $f_{\oridiv}$ depends on the chiral superfields, and so the worldvolume gauge theory coupling and fine structure constant are determined by moduli vevs as 
%---------------------
\begin{equation}
     g_{\oridiv}^{-2} = \langle\text{Re}[f_{\oridiv}]\rangle \qquad \alpha_{\oridiv} = \frac{g_{\oridiv}^2}{4\pi}\,.
\label{eq:D7fine}
\end{equation}
%---------------------
In the absence of magnetic flux, the tree-level gauge kinetic function is determined by the fields in~\cref{eq:chiralmults} as  
\begin{equation}
    f^{(1)}_{\oridiv} = \frac{\wrap^\alpha}{2\pi} T_{\alpha}\,,
\label{eq:gaugkin1}
\end{equation}
where $\wrap^\alpha$ corresponds to the wrapping number of the D7-brane stack on the divisor $\oridiv$ in $\ori$ (see~\cref{app:stringconv}). In particular, we note that ${\rm Im}[f_{\oridiv}^{(1)}] \supset (2\pi)^{-1}\wrap^\alpha\rho_\alpha$, so we could already identify the $C_4$-axions $\rho_\alpha$ as candidate spectator axions whose couplings to gauge fields can be potentially be increased by wrapping the D7-brane stack multiple times. 

A distinct candidate spectator axion can be identified in compactifications with $h^{1,1}_- >0$ by allowing magnetic flux in the D7-branes. This is done by turning on quantized field strength units of $F_2^A$ on a 2-cycle of the extra dimensions~\cite{Jockers:2004yj,Jockers:2005zy,Grimm:2011dj,Long:2014dta}. This 2-cycle must be non-trivial in homology either with respect to the full CY (and thus exist as a pullback from $\ori$ to $\oridiv$) or at least in relative homology with respect to the divisor wrapped by the D7-brane itself. Suppressing the gauge index, the quantization condition of $F_2^A$-flux reads
\begin{equation}
    \int_{\Pi_2^{\mathscr{A}}} \frac{\ell_s^2}{2\pi}F_2 +\frac12 \int_{\Pi_2^{\mathscr{A}}} c_1(\Pi^{\mathscr{A}}) = \magn^{\mathscr{A}}\in \mathbb{Z}\,.
\end{equation}
Here $\ell_s = 2\pi \sqrt{\alpha^\prime}$ is the string length and $\mathscr{A}$ labels the set of 2-cycles on which gauge flux can be put, such that $\mathscr{A}=\alpha, a, r_\vee$ splits into the orientifold-even pullback 2-cycles $\alpha=1...h^{1,1}_+$, orientifold-odd pullback 2-cycles $a=1...h^{1,1}_-$ and the 4-cycle-local 2-cycles $r_\vee$ which are trivial in the CY.\footnote{The half-integer shift of the 2nd term of the LHS is controlled by the 1st Chern class $c_1$ of the 4-cycle in question, and it is zero unless the 4-cycle is not spin, in which case it accounts for the Freed-Witten quantization condition, see~\cite{Blumenhagen:2008zz}.} We can then expand such a $F_2$-flux in terms of the relevant 2-forms 
%---------------------
\begin{equation}
\frac{\ell_s^2}{2\pi}F_2=\magn^\alpha\omega_\alpha+\magn^a\omega_a+
\magn^{r_\vee}\omega_{r_\vee}\,.
\end{equation}
%---------------------
Such gauge flux induces couplings between the 2-form axions $c^a$ and $b^a$ and the $4d$ gauge field strength $F_{\mu\nu}^A$.\\ These additional couplings are captured by extending the gauge kinetic function beyond~\cref{eq:gaugkin1} to 
%------------------
\begin{equation}
\begin{aligned}
    f^{(2)}_{\oridiv} &=\frac{\wrap^\alpha}{2\pi}\left\{T_{\alpha} + i\kappa_{\alpha bc} \bigg(G^b \magn^c 
+ \frac{S}{2} \magn^b\magn^c\bigg) \right\}\\
            &= \frac{\wrap^\alpha}{2\pi}\bigg\{  \tau_{\alpha} + e^{-\phi}\kappa_{\alpha bc}\bigg( b^b\magn^c - \frac{1}{2} b^b b^c - \frac{1}{2}\magn^b\magn^c\bigg)      \bigg\}\\
                &\qquad+ i\frac{\wrap^\alpha}{2\pi}\bigg(\rho_{\alpha} +\kappa_{\alpha bc} c^b (\magn^c-b^c) + C_0 \kappa_{\alpha bc}\bigg( \frac{1}{2} b^b b^c + \frac{1}{2}\magn^b \magn^c - b^b \magn^c\bigg)\bigg)\,,
\end{aligned}
\label{eq:gaugkin2}
\end{equation}
where $w^\alpha\in\mathbb{Z}$ are once again the wrapping numbers of the D7-brane stack. Here we have assumed that quantized flux is placed only on odd 2-cycles. Plugging this into eq.~\eqref{eq:D7bosons} we acquire additional CS couplings
\begin{equation}
    S_{EFT}\supset \int_{M_4}{\rm d}^4x\sqrt{-g_4}\left[-\frac14\bigg(\frac{\wrap^\alpha}{2\pi} \kappa_{\alpha bc}\magn^c\bigg)c^bF_{\mu\nu}\widetilde{F}^{\mu\nu}\right]\,.
    \label{eq:D7bosons2}
\end{equation}
%---------------------
Thus we can increase the effective CS coupling of a $C_2$-axion to $4d$ gauge fields by turning on internal quantized $F_2$ gauge flux on the D7-brane stack, as well as increasing the wrapping number of the stack.

From the discussion above, we see that type IIB orientifolds with D7-branes do indeed provide the particle content for spectator models. However, there is a restriction on these constructions that must be taken into account - the axions and massless gauge bosons may gain large masses via the St\"{u}ckelberg mechanism~\cite{Plauschinn:2008yd}. This can occur with both the $C_2$- and $C_4$-axions, but the nature of the mechanism is distinct between the two types of axions. To be concrete, let us imagine a stack of $\nds$ D7-branes on $\oridiv$ giving rise to a $U(\nds) = SU(\nds)\times U(1)$ gauge theory in the EFT. From~\cref{app:stringconv}, the $C_4$- and $C_2$-axions associated with $\oridiv$ will have kinetic terms determined by the covariant derivatives 
\begin{equation}
    \begin{aligned} \nabla c^a&= dc^a -q^a A\,, \hspace{2.56cm} q^a=\frac{\nds}{2\pi}\wrap^a\,,\\
    \nabla \rho_\alpha&=d \rho_\alpha- i q_{\alpha}A\,, \hspace{2cm} q_{\alpha}= -\frac{\nds}{2\pi}\kappa_{\alpha bc}\magn^b\wrap^c\,.
    \end{aligned}
\label{eq:mainStuck}
\end{equation}
Where $A$ is the 1-form gauge potential of the diagonal $U(1)$ gauge interaction and we have again omitted some terms by forbidding flux on even 2-cycles. Appearing here is the magnetization on odd 2-cycles, $\magn^b$, as well as odd wrapping numbers $\wrap^a$ defined in~\cref{eq:itsawrap}. From these expressions, we see that the $C_4$-axions have a \textit{flux-induced} St\"{u}ckelberg mechanism in that the gauging of the axion shift symmetry only occurs in the presence of non-zero magnetic flux. On the other hand, $C_2$-axions have a \textit{geometric} St\"{u}ckelberg mechanism that arises purely from the details of the orientifold $\ori$ in the form of the odd wrapping number $\wrap^a$. In either case, the $U(1)$ factor of the gauge group becomes massive by eating an axion.

For stringy spectator models with non-Abelian gauge fields, the St\"{u}ckelberg mechanism is only dangerous insofar as a candidate spectator axion may be lost. For Abelian spectator sectors, the St\"{u}ckelberg couplings above present a major obstacle that must evaded to ensure that both axions and massless $U(1)$'s exist in the spectrum of the EFT. This requirement furnishes our first restriction on spectator models in string theory. Thus we will focus on Abelian spectators here and classify potential models via different methods to realize $U(1)$ gauge theories in the EFT. We also describe how to ensure a viable candidate spectator axion appears in the EFT. We do not claim that this list is exhaustive, but we merely outline schematic requirements. 

From the form of~\cref{eq:mainStuck}, we see that the general strategies to fulfill this task are to i). ensure that the odd wrapping numbers $\wrap^a$ vanish ii). require that certain intersection numbers $\kappa_{\alpha bc}$ vanish and/or iii). assume a structure that leads to cancellation of St\"{u}ckelberg couplings. To implement these strategies, we must refine our discussion of the divisor wrapped by the D7-branes. The divisor $\oridiv$ of $\ori$ descends from a divisor $\CYdiv$ of $\CYB$ and its image divisor $\CYimdiv$ under the orientifold involution. Adopting the organization scheme of~\cite{Grimm:2011tb}, the classes of spectator models depends on the precise relation between $\CYdiv$ and $\CYimdiv$ as we now describe.
\begin{itemize}
    \item \textbf{Class I Spectators:}\\
        The most naive approach to realizing an Abelian spectator model is to place a single D7-brane on a divisor $\oridiv$ that does not lie on top of an orientifold plane. This yields a $U(1)$ gauge theory in the $4d$ EFT and one can consider $C_4$- or $C_2$-axions as spectators. However, there is a danger that the St\"{u}ckelberg mechanisms described above remove the photon from the massless spectrum.

        To avoid this, $\oridiv$ must be chosen such that the divisors  $\CYdiv$ and $\CYimdiv$ in $\CYB$ are in the same homology class of $\CYB$, i.e. $[\CYdiv]=[\CYimdiv]$. If this condition is fulfilled, then the odd wrapping numbers in~\cref{eq:mainStuck} vanish and the St\"{u}ckelberg couplings are eliminated.   
        
        A generalization of this setup can be achieved by considering $N_{U(1)}$ D7-branes, all wrapping distinct 4-cycles $\Pi^\alpha$ of $\CYB$, where $\alpha = 1,..,N_{U(1)}$. Naively this gives $N_{U(1)}$ Abelian gauge sectors, labelled as $U(1)_\alpha$, but some will become massive due to the St\"{u}ckelberg mechanism. The linear combinations that do not become massive are those that mimic the mechanism just described: any linear combination $\sum_\alpha p_\alpha U(1)_\alpha$ such that the corresponding homology element maps to itself under the orientifold involution~\cite{Grimm:2010ez}. Mathematically, the criterion is $ \sum_\alpha p_\alpha ([\Pi^\alpha]-[\Pi^{\alpha^\prime}])=0$. In this class, we avoid both the geometric and flux St\"{u}ckelberg mechanisms and both $C_4$- and $C_2$- are candidates  spectator axions. 
    \item \textbf{Class II Spectators:}\\
        In Class I,  the  St\"{u}ckelberg couplings are circumvented through a judicious choice of $\oridiv$ such that $[\CYdiv]=[\CYimdiv]$. An alternative scenario is a divisor $\oridiv$ such that $[\CYdiv]\neq[\CYimdiv]$.  Branes on such divisors can interact with both $C_4$- and $C_2$-axions in the EFT, but the odd wrapping number is necessarily non-zero and the geometric St\"{u}ckelberg mechanism is active. Thus in the absence of flux, a $C_2$-axion is eaten and the $U(\nds)$ gauge theory realized by the $\nds$ D7-branes wrapping $\oridiv$ loses its $U(1)$ factor and becomes $SU(\nds)$. In the case of $\nds=1$, the gauge theory is replaced by a massive photon.
        
        To realize a $U(1)$ gauge factor in the EFT for such scenarios, we consider two options. The first is by breaking the $SU(\nds)$ via flux to a subgroup containing a $U(1)$ factor. However, care must be taken since flux may induce St\"{u}ckelberg couplings that remove the new photons from the massless field content. Thus one may have to assume the vanishing of certain intersection numbers in~\cref{eq:mainStuck} or simply obtain a sufficiently large number of $U(1)$ factors from the breaking of $SU(\nds)$.
        
        The second possibility depends on the topology of the orientifold space itself. We can consider a scenario where $\oridiv$ corresponds to a solitary 4-cycle  $\oricycle$ in $\ori$ such that the homology class of $\oricycle$ has additional volume-minimizing representatives. In other words, there exists two or more distinct $\oricycle^{(i)}\in [\oricycle]$.  D7-branes can be wrapped on the various $\oricycle^{(i)}$, and each stack  produces a $U(1)$ factor in the EFT gauge group. One linear combination of the $U(1)$'s eats the $C_2$-axion of $[\oricycle]$  and becomes massive. If magnetic flux is turned on, another linear combination of $U(1)$'s may gain a mass by eating the $C_4$-axion of $[\oricycle]$ . This can be circumvented if certain intersection numbers vanish. Alternatively, one could consider scenarios with at least 3 homologous cycles, as was used to realize aligned natural inflation in~\cite{Long:2014dta}.
        
        We label the above possibilities as \textbf{Class IIa} and \textbf{Class IIb}, respectively. 

        As for the axionic state of the spectator sector, in the absence of flux, one could consider the $C_4$-axions as spectators. To have a $C_2$-axion as a spectator candidate requires a bit of care as the geometric St\"{u}ckelberg coupling is present. For both sub-classes, we can consider a $C_2$-axion that is associated to some other divisor, $\oridiv^{alt}$, in $\ori$. There are two obvious possibilities.  

        The first is that there exists a divisor and image-divisor pair in $\CYB$ such that the even combination $\oridiv^{alt}$ is rigid and is stabilized by an ED3 instanton. If we assume that such a $\oridiv^{alt}$ has certain trivial intersection numbers with divisors that support D7-branes, then the geometric St\"{u}ckelberg mechanism can be avoided. For this $C_2$-axion to couple to the worldvolume theory of $\oricycle$, we must also assume that appropriate intersection numbers between $\oridiv$  and $\oridiv^{alt}$ exist.

        The second possibility is that $\oridiv^{alt}$ arises from a cycle and image-cycle pair that are homologous in $\CYB$, as in the Class I spectator theories. Then the $C_2$-axion could couple to the worldvolume theory of $\oricycle$ assuming appropriate intersection numbers are non-trivial. 
        
    \item \textbf{Class III Spectators:}\\
        The final class corresponds to the case where $\CYdiv$ is invariant under the orientifold involution so that $\CYdiv = \CYimdiv$ pointwise. This class shares certain features with both of the previous classes. First, the odd wrapping numbers $\wrap^a$ of $\oridiv$ vanish, so both the geometric and flux St\"{u}ckelberg mechanism are absent. However, $\oridiv$ sits on top of an orientifold plane, and local D3-tadpole cancellation gives rise to orthogonal or symplectic gauge groups~\cite{Gimon:1996rq} and there is no diagonal $U(1)$ factor. To obtain a $U(1)$ factor in the EFT, we must implement flux to break the gauge group.
        
        Once again $C_4$-axions are candidate spectators. To obtain $C_2$-axion spectators, one must assume the existence of a separate 4-cycle $\oridiv^{alt}$ with the properties outlined in Class II above. 
\end{itemize}

%========================================================
\begin{figure}
\centering
\begin{tikzpicture}
 %   \draw[help lines] (-7,-7) grid (7,7);
%==============
%Case 1
%==============
    \node[inner sep=0pt] (case1) at (-5.5,0)
    {\includegraphics[width=.26\textwidth]{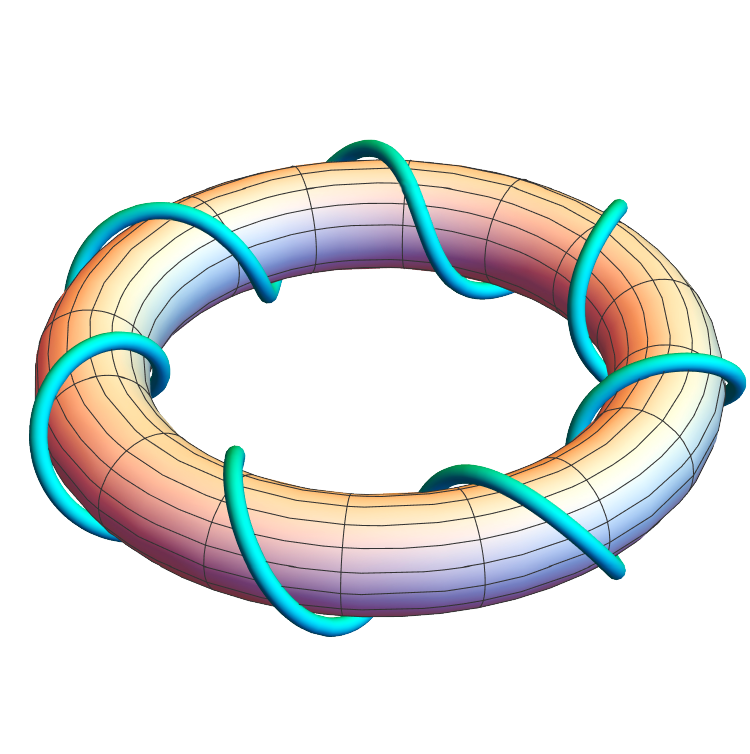}};
    
\draw [decorate,thick,decoration={brace,amplitude=4pt,mirror,raise=8pt},xshift=4pt,yshift=-2.5pt]
(-6.05,-1.3) -- (-5.8,-1.23) node [black,midway,yshift=-0.6cm,xshift=0.2cm] {\footnotesize
$U(1)$};
\node[] (Y0) at (-5.5,-2.5) {\scalebox{1}{\textbf{Class I}}};
%==============
%Case 2
%==============
\node[inner sep=0pt] (case2) at (0,0) {\includegraphics[width=.25\textwidth]{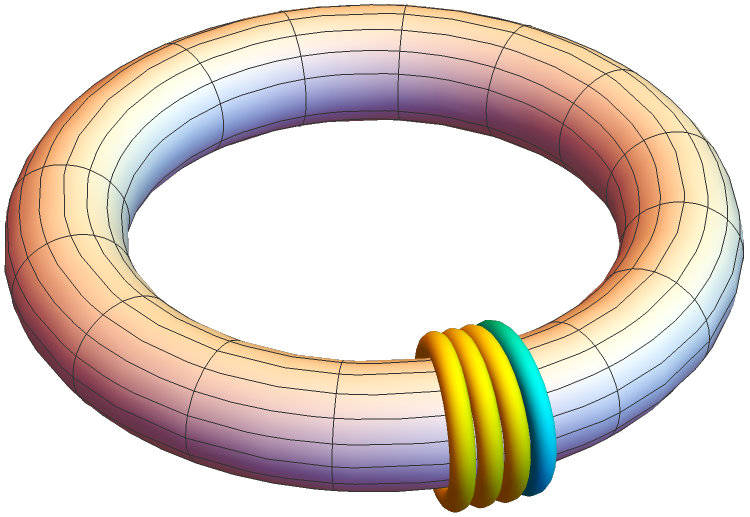}};
\draw [decorate,thick,decoration={brace,amplitude=3pt,mirror,raise=8pt},yshift=-0pt]
(-4.7+5.5,-1.2) -- (-4.5+5.5,-1.1) node [black,midway,yshift=-0.5cm,xshift=0.5cm] {\footnotesize
$U(1)$};
    \draw [decorate,thick,decoration={brace,amplitude=4pt,mirror,raise=8pt},yshift=0pt]
(-5.1+5.5,-1.27) -- (-4.65+5.5,-1.2) node [black,midway,yshift=-0.7cm,xshift=0cm] {\footnotesize
$SU(\nds)$};
\node[] (Y1) at (0,-2.5) {\scalebox{1}{\textbf{Class IIa}}};
%\node[] (Y1) at (0,-3) {\scalebox{1}{\textbf{Class III}}};
%==============
%Case 3
%==============
\node[inner sep=0pt] (case3) at (5.5,0)
    {\includegraphics[width=.25\textwidth]{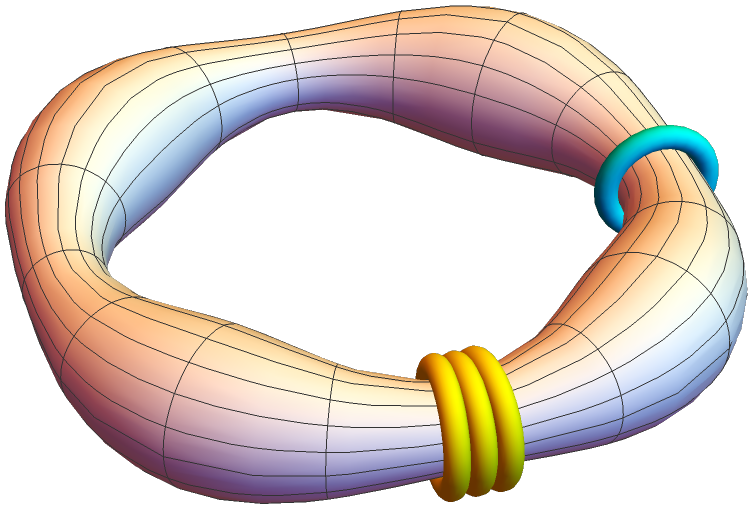}};
\draw [decorate,thick,decoration={brace,amplitude=4pt,mirror,raise=8pt},yshift=0pt]
(-5.13+11,-1.27) -- (-4.68+11,-1.15) node [black,midway,yshift=-0.7cm,xshift=0cm] {\footnotesize
$U(\nds)$};
\draw [decorate,thick,decoration={brace,amplitude=4pt,raise=8pt},yshift=0pt]
(7.3,0.65) -- (7.35,0.3) node [black,midway,yshift=-0.cm,xshift=0.85cm] {\footnotesize
$U(1)$};
\node[] (Y2) at (5.7,-2.5) {\scalebox{1}{\textbf{Class IIb}}};
\end{tikzpicture}
\caption{Pictorial representation of the D7-brane(s) configurations giving rise to spectator sector classes discussed in the text. Class III has a representation similar to Class IIa but with a different gauge group.}
\label{fig:enter-label}
\end{figure}
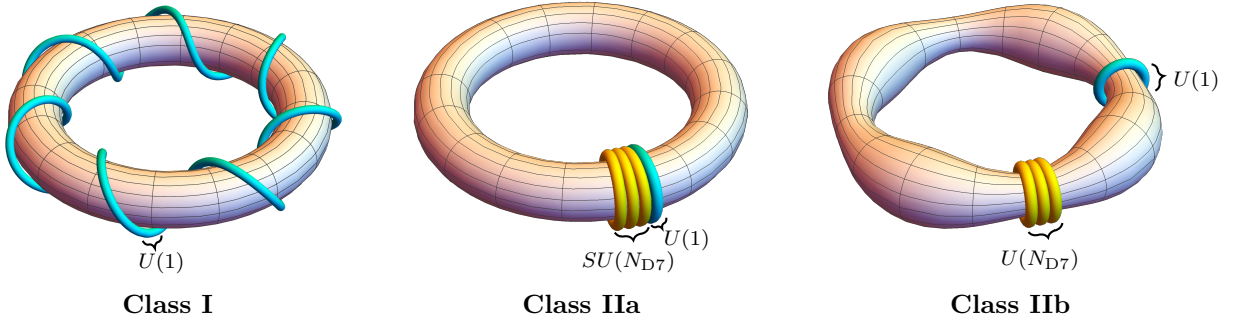
%========================================================
There is an important caveat to the above discussion - namely, the vevs of the $b^a$ axions. Magnetization on the brane induces a D-term~\cite{Jockers:2005pn,Grimm:2011dj}
\begin{equation}
    D_{\oridiv} \propto \kappa_{\alpha bc} v^\alpha (b^c - \magn^c)\wrap^b\,, 
\end{equation}
where the $v^\alpha$ determine 2-cycle volumes and the $\wrap^b$ are wrapping numbers on odd cycles, as explained in~\cref{app:stringconv}\footnote{Here we have excluded contributions from charged matter fields.}. If these D-terms are non-zero, one can set $\langle b^a\rangle = \magn^a$ to cancel them. In doing so, we see from~\cref{eq:gaugkin2} that the CS coupling of the $C_2$-axion is set to zero. This is not an issue for Class I or Class III models as the odd wrapping numbers $\wrap^a$ automatically vanish. For Class II models, demanding a coupling of the $C_2$-axions to the gauge field may require vacua with some $\langle b^a\rangle \neq \magn^a$ or manifolds that have certain intersection numbers set to zero. As a side note, typically, K\"{a}hler modulus stabilization enforces $\langle b^a\rangle = 0$. Cancellation of D-terms then becomes a construction-dependent question on the charged matter in the brane.

From this section, we see that even demanding the existence of spectator sectors is non-trivial from the string theory viewpoint and many restrictions are already enforced. Our first restrictions arose from demanding the necessary field content and led to our discussion on the classes above. The second restriction here arises from demanding that a CS coupling exist between axions and gauge fields. There are also constraints from D3-tadpoles and control of the compactification, which we outline below. 

%---------------------
\subsubsection{Axion Decay Constants}
%---------------------
In this and the subsequent sub-section, we outline the origin of spectator model parameters in string constructions. The decay constants for the axion states described in the previous section are determined by their kinetic terms
\begin{equation}
    S_{axions}\supset M_p^2\int d^4x \sqrt{-g}\bigg(- K^{\alpha\beta}\; \partial_\mu\rho_\alpha \partial^\mu \rho_\beta - K_{ab}\; \partial_\mu c^a \partial^\mu c^b \bigg)\,,
\label{eq:axkinbas}
\end{equation}
where again $\alpha,\beta = 1,..,h^{1,1}_+$ and $a,b = 1,..,h^{1,1}_-$. For more explicit formulae for the K\"{a}hler metrics, see~\cref{app:stringconv}. 

For simplicity, we consider an orientifold with $h^{1,1}_-=1$ and/or $h^{1,1}_+=1$, then the kinetic terms of the even $C_4$-axion, $-K^{11}(\partial\rho)^2$ , and only odd $C_2$-axion ,$-K_{11} (\partial c)^2$, allows us to identify the axion decay constants 
\begin{equation}
    F_{o} = \langle 2K_{11}\rangle^{1/2} M_p \,,\qquad F_{e} = \langle 2K^{11}\rangle^{1/2} M_p \,,
\label{eq:axdecays}
\end{equation}
then the re-scaled axions
\begin{equation}
    \chi_{o} = F_o c  \,,\qquad  \chi_{e} = F_e\rho   \,,  
\label{eq:canonax}
\end{equation}
have canonical kinetic terms.

The above steps can be generalized for orientifolds with $h^{1,1}_->1$ and/or $h^{1,1}_+>1$. The only complication is that one must implement orthogonal transformations to diagonalize the metrics $K^{\alpha\beta}$ and $K_{ab}$ before re-scaling the axions. The orthogonal transformation in principle affects the coupling of the axion to the spectator gauge fields, but without a specific compactification manifold we cannot determine their precise impact.  

%---------------------
\subsubsection{Axion Masses}
%---------------------
The shift symmetries of the $C_4$- and $C_2$-axions have their origins in the gauge symmetries of the corresponding $10d$ gauge potentials and are thereby protected at all orders in string perturbation theory. Thus, to give a mass to these potential spectator axions, which break the shift symmetries, we must introduce non-perturbative effects via either Euclidean or Lorenztian objects.

From Euclidean objects, axion potentials can be generated by Euclidean D3-branes (ED3s), Euclidean D1-branes (ED1s), or bound states thereof. An ED3 brane wrapping a 4-cycle of $\ori$ gives a contribution to the superpotential of the form
\begin{equation}
    W_{ED3} \sim \pazocal{A} e^{-2\pi T}\fstop
\label{eq:ED3supes}
\end{equation}
Here we neglect the typical dependence of the 1-loop Pfaffian factor $\pazocal{A}$ on other fields of the theory, such as the complex structure moduli. Such a superpotential yields a periodic potential for the $C_4$-axion $\rho$ embedded in the $T$ supermultiplet of the 4-cycle\footnote{Here and below we assume the superpotential has the standard type IIB flux compactification structure of $W=W_0+W_{np}$, where $W_0$ is a constant and $W_{np}$ is the non-perturbative contribution.}:
\begin{equation}
    \begin{aligned}
        \delta V_{ED3} \sim \mu^4 e^{-2\pi\tau}\cos(2\pi \rho) = \mu^4 e^{-2\pi\tau}\cos\bigg(2\pi \frac{\chi_e}{F_e}\bigg)\,.
    \end{aligned}
\label{eq:ED3pot}
\end{equation}

To obtain a mass for a $C_2$ spectator axion, we turn to ED1-branes wrapping the orientifold-even  2-cycles. As an illustration, consider a scenario with intersection numbers such that the volume of the wrapped 2-cycle $v$ and the volume of its dual 4-cycle $\tau$ are related as $\tau = \frac{1}{2}\kappa_{+++}v^2$.  We also let the 2-cycle have a non-zero intersection number $\kappa_{+--}$ with an orientifold-odd combination 2-cycle supporting the 2-form axion chiral multiplet $G$. Such an object contributes to the K\"ahler potential and schematically might take the form~\cite{Camara:2008zk,McAllister:2008hb}
\begin{equation}
    K=-3\ln \left[T+\bar T-\frac{3}{2(S-\bar S)}\kappa_{+--}(G+\bar G)^2+e^{-2\pi \sqrt{\frac{T+\bar T}{\kappa_{+++}}}}\cos\left(2\pi\frac{G+\bar G}{2}\right)\right]\,.
\label{eq:GKahED1}
\end{equation}
If this ED1-brane instanton contributes to the scalar potential of $G$, its contribution from the path integral has to scale as
\begin{equation}
    \begin{aligned}
    \delta V_{{\rm ED}1}\sim \mu^4\;{\rm Re}\,e^{-S_{{\rm ED}1}}&= \mu^4\; {\rm Re}\,e^{-2\pi v -2\pi i c}\\%=\mu^4e^{-2\pi v}\cos\left(2\pi \frac{c}{M_p}\right)\\
    &= \mu^4e^{-2\pi v}\cos\left(2\pi\frac{\chi_o}{F_o}\right)\,,
    \end{aligned}
    \label{eq:c2axionpot}
\end{equation}
by direct evaluation of $S_{{\rm ED}1}$ (eq.~\eqref{eq:Dp} with $\ell_s=1$). In the second line, we have rewritten the argument in terms of the axion with canonically normalized kinetic term.
%Note that~\cref{eq:c2axionpot} implies that the axion $\chi$ has a discrete shift symmetry set by $\chi_0\rightarrow \chi_0+F_0$. 

It is possible that the a single instanton effect can give masses to both $C_4$- and $C_2$-axions. The full contribution of an ED3 instanton includes summing over all possible magnetic fluxes of the divisor wrapped by the ED3, which can be thought of as a sum over ED3-ED1 bound states. This requires a modification of~\cref{eq:ED3supes} to $W_{ED3}\rightarrow W_{ED3/ED1}\simeq \Theta(G)e^{-2\pi T}$, where $\Theta(G)$ is a holomorphic theta function from the theory of modular forms~\cite{Grimm:2007hs,Grimm:2007xm}. In the presence of D7-branes such that axion shift symmetries are gauged due to the presence of a St\"{u}ckelberg mechanism, the sum over fluxes defining $\Theta(G)$ must be suitably restricted to include only gauge-invariant instantons~\cite{Grimm:2011dj}. Setting aside this subtlety, the ED3-ED1 bound states gives rise to terms with the schematic form $W_{ED3/ED1}\sim e^{2\pi (G+T)}$. Such a term contributes to mass-mixing between the even and odd axions. This mixing forces one to rotate into the mass eigenbasis, which introduces some coefficient in the Chern-Simons coupling. In the following, we will neglect this complication and consider $C_2$-axion masses arising only from ED1 contributions to the K\"{a}hler potential.

Alternatively, spectator axions could obtain potentials from Lorentzian branes via gaugino condensation. This requires that the spectator axion couples to both the spectator $U(1)$ as well as a condensing hidden non-Abelian gauge sector. We will take this to be the case and assume that there is a stack of D7-branes wrapping a 4-cycle $\oricycle^{\pazocal{G}}$ of $\ori$ realizing a non-Abelian gauge group $\pazocal{G}$. The gaugino condensate then contributes to the non-perturbative part of the superpotential as 
\begin{equation}
    W_{np} \sim \pazocal{A} e^{-\frac{(2\pi)^2}{c(\pazocal{G})}f_{\pazocal{G}}}\,,
\label{eq:condsupes}
\end{equation}
where $f_{\pazocal{G}}$ and $c(\pazocal{G})$ are the gauge kinetic function and the dual Coxeter number of $\pazocal{G}$, respectively. For $\pazocal{G} = SU(N_{\pazocal{G}})$, $c(\pazocal{G}) = N_{\pazocal{G}}$. If we are considering a scenario without worldvolume flux, then $f_{\pazocal{G}} \propto T_{\oricycle^{\pazocal{G}}} = \tau_{\pazocal{G}}+i \rho_{\pazocal{G}}+\cdots$ and the $C_4$-axion in $T_{\oricycle^{\pazocal{G}}}$ obtains a potential similar to that of~\cref{eq:ED3pot} but with an extra factor of $c(\pazocal{G})^{-1}$ in the argument of the cosine. 

If instead the D7-brane stack is magnetized and has a non-zero intersection number $\kappa_{\pazocal{G}}\equiv \kappa_{\pazocal{G}--}$ with the odd cycle supporting an odd chiral supermultiplet $G$, then $f_{\pazocal{G}}$ can take the form of $f_{D7}^{(2)}$ in~\cref{eq:gaugkin2}. Then the scalar potential will have a contribution
\begin{equation}
    \begin{aligned}
        \delta V_{cond} 
        &\simeq \mu^4 e^{-\frac{2\pi \wrap_{\pazocal{G}}}{c(\pazocal{G})} \tau_{\pazocal{G}}}\cos\bigg(2\pi \frac{\wrap_{\pazocal{G}}}{c(\pazocal{G})}\bigg[ \rho_{\pazocal{G}} + \kappa_{\pazocal{G}} \magn_{\pazocal{G}} c \bigg]\bigg)\\ 
        &= \mu^4e^{-\frac{2\pi \wrap_\pazocal{G}}{c(\pazocal{G})} \tau_{\pazocal{G}}}\cos\bigg(2\pi \frac{\wrap_{\pazocal{G}}}{c(\pazocal{G})}\bigg[\frac{\chi_e}{F_e} +\kappa_{\pazocal{G}}\magn_\pazocal{G} \frac{\chi_o}{F_o}\bigg] \bigg) \fstop
    \end{aligned}
\label{eq:condaxpot}
\end{equation}
where $\wrap_{\pazocal{G}}$ and $\magn_{\pazocal{G}}$ are the wrapping number and magnetization on $\oricycle_{\pazocal{G}}$. Such a potential introduces mass mixing between the two axions. The masses and eigenstates are
\begin{equation}
    \begin{aligned}
         \chi_1 &= \frac{1}{\sqrt{1+\epsilon^2}}\bigg(\chi_0 - \epsilon \chi_e\bigg) \qquad m_1^2 = 0 \,,\\
        \chi_2 &= \frac{1}{\sqrt{1+\epsilon^{2}}}\bigg(\chi_e + \epsilon \chi_o\bigg) \qquad m_2^2 = \left(\frac{2\pi\wrap_{\pazocal{G}}}{c(\pazocal{G})F_e}\right)^2(1+\epsilon^2)\,.
    \end{aligned} 
\label{eq:condstates}
\end{equation}
We have also defined the parameter $\epsilon = \kappa_{\pazocal{G}} \magn_{\pazocal{G}} F_e/F_o$, which tracks the mixing between the axions. Below we will take $\kappa_{\pazocal{G}}\magn_{\pazocal{G}}\simeq 1$ so that $\epsilon\simeq F_e/F_o$, which is generally expected to be less than unity due to the natural hierarchy between the decay constants of even and odd axions -- see~\cref{app:stringconv} for more details. If we assume that both $\chi_e$ and $\chi_o$ couple to a $U(1)$ gauge sector, then the mass eigenstates couple as 
\begin{equation}
    \begin{aligned}
        \pazocal{S} &\supset -\frac{1}{4}\int \bigg(\frac{\lambda_e}{F_e}\chi_e + \frac{\lambda_o}{F_o}\chi_o\bigg) F_{\mu\nu}\widetilde{F}^{\mu\nu}\\ 
                        &\hspace{0.5cm}= -\frac{1}{4}\int  \bigg( \frac{F_e\lambda_o - \epsilon F_o\lambda_e}{F_oF_e\sqrt{1+\epsilon^2}} \bigg)\; \chi_1 F_{\mu\nu}\widetilde{F}^{\mu\nu}  -\frac{1}{4}\int  \bigg(\frac{F_o \lambda_e+\epsilon F_e\lambda_o}{F_eF_o\sqrt{1+\epsilon^2}}\bigg)\; \chi_2 F_{\mu\nu}\widetilde{F}^{\mu\nu}\,.
    \end{aligned}
\end{equation}

To classify the spectator axion(s) and their mass generation mechanism, we introduce the cases displayed in~\cref{table:cases}.

\begin{table}[t!]
\centering
    \begin{tabular}{ |p{1cm}||p{1.3cm}|p{4cm}| }
 \hline
 Case & Axion(s) & Mass Mechanism \\
 \hline
 1   &  $C_4$   & ED3 Instantons\\
 2 & $C_2$ & ED1 Instantons\\
 3 & $C_4$ & Gaugino Condensation\\
 4 & $C_2$\&$C_4$ & Gaugino Condensation\\
 \hline
\end{tabular}
\caption{Spectator axions organized by mass generation mechanism.}
\label{table:cases}
\end{table}

Finally, we note that we have made several assumptions on the structure of $\ori$ in order to permit the above non-perturbative effects. For example, a 4-cycle must be rigid to support ED3 instantons~\cite{Witten:1996bn}, which places requirements on the sheaf cohomology of the divisor. More precisely, a sufficient (but not necessary) condition for generation of a superpotential by an ED3 instanton is that the 4-cycle $D$ is rigid, and in addition $D \neq D’$ pointwise but $[D] = [D’]$. The Hodge numbers of the divisor determine the number of chiral fields charged under the worldvolume theory of a non-Abelian D7-branes stack, which in turn could affect the existence of a gaugino condensate. In the following we will assume the required structures whenever necessary and leave a detailed study of their abundance for our scenarios to future work. 

%---------------------
\subsubsection{From Strings to Spectators}
%---------------------
We now combine all of the above to make contact with the spectator Lagrangian~\cref{eq:fullL1}.
First, let us consider a single $C_4$- or $C_2$-axion coupled to a hidden $U(1)$ gauge theory. From the discussion of the previous sections, we have contributions to the EFT Lagrangian of the form
\begin{equation}
    \pazocal{L}_{EFT}\supset \Lambda^4\cos\bigg(2\pi\pazocal{J}\frac{\chi}{F}\bigg) - \frac{g_{\oridiv}^2}{8\pi}\pazocal{M}\frac{\chi}{F}F_{\mu\nu}\widetilde{F}^{\mu\nu} \fstop
\label{eq:stringspeclag}
\end{equation}
Note that we have re-scaled the worldvolume gauge field from~\cref{eq:D7fine} as $A_\mu\to g_{\oridiv} A_\mu$ to canonically normalize the gauge field kinetic term. We have also re-scaled the axion as per~\cref{eq:canonax} to obtain a canonically-normalized axion $\chi$. Here $\pazocal{M}$ depending on the $10d$ origin of $\chi$ and encodes either just the wrapping number $\wrap$ of the brane or combinations from~\cref{eq:D7bosons2} such as $\wrap\kappa_{+--}\magn_1$. The factor $\pazocal{J}$ specifies extra information on the non-perturbative effect giving the axion a mass:
\begin{equation}
    \pazocal{J}= \begin{cases}
            1 \\
            \frac{\wrap}{c(\pazocal{G})}
    \end{cases}
     \quad 
    \begin{aligned}
        &\text{ED1 or ED3 Instantons}\\
        &\text{Gaugino Condensation (only $C_4$)}\,.
    \end{aligned}
\end{equation}
For the second line above, we assume that any $C_2$-axions in the EFT do not couple to the condensing gauge group $\pazocal{G}$. If $C_2$-axions do couple via magnetic flux, then $\pazocal{J}$ becomes matrix-valued. We address this general case below. 

For now, we define a re-scaled decay constant
\begin{equation}
    f = \frac{F}{2\pi\pazocal{J}}\,,
\end{equation}
such that the axion has the conventional periodicity $\chi \rightarrow \chi + 2\pi f$. The spectator parameters in~\cref{eq:MASAlag} and then given by
\begin{equation}
    \begin{aligned}
        m &= \frac{2\pi\pazocal{J}\Lambda^2}{F} =\frac{\Lambda^2}{f}\,, \\ 
        \lambda &= \frac{g^2_{\oridiv}}{4\pi^2\pazocal{J}}\pazocal{M} = \frac{\alpha_{\oridiv}}{\pi\pazocal{J}}\pazocal{M}\,.
    \end{aligned} 
\label{eq:stringspecCS}
\end{equation}

The above can be generalized to cases with multiple spectators and more complicated gaugino condensation scenarios. We can consider the following subset of terms in the EFT Lagrangian
\begin{equation}
    \pazocal{L}_{EFT}\supset -\sum_{a=1}^{N_{np}}\Lambda^4_a\cos\bigg[ 2\pi \sum_{b=1}^{N_{ax}}\pazocal{J}_{ab}  \frac{\chi_b}{F_b}\bigg] - \sum_{c=1}^{N_{U(1)}}\sum_{b=1}^{N_{ax}} \frac{\alpha_c}{\pi}\pazocal{M}_{cb}\bigg(\frac{\chi_b}{F_b}\bigg) F_{(c)}\wedge F_{(c)}\fstop
\label{eq:genax}
\end{equation}
where we have included $N_{ax}$ axions that couple to $N_{U(1)}$ Abelian gauge sectors via a coupling matrix $\pazocal{M}_{cb}$, and $N_{np}$ non-perturbative effects giving cosine potentials that include mass-mixing via the matrix $\pazocal{J}_{ab}$. The potential in~\cref{eq:genax} defines a mass matrix for the axions
\begin{equation}
    (M^2)_{ij} = \sum_a \Lambda_a^4 \bigg(\frac{4\pi^2}{F_i}\bigg)\pazocal{J}_{ai}\pazocal{J}_{aj}\,.
\end{equation}
This can be diagonalized via an orthogonal matrix $O_{jb}$ such that $(O^T)_{ai}M_{ij}O_{jb} = m_a^2\delta_{ab}$ and so the axion mass eigenstates are given by 
\begin{equation}
    \chi^{\prime}_a = \sum_i(O^T)_{ai}\chi_i\fstop
\end{equation}
These states have Chern-Simons couplings
\begin{equation}
    \pazocal{L}\supset \sum_{c,b} \frac{\alpha_c}{\pi}\pazocal{M}_{cb}O_{bi}\frac{\chi_i^\prime}{F_b}F_{(c)}\wedge F_{(c)}\,.
\end{equation}

With this, all the parameters defined a spectator model has been identified from our D7-brane constructions. What remains to be shown is that the parameters necessary for an observable signal can be generated. First however, we must consider the inherent limitations of these setups.

%=====================================
\subsection{Constraints on Spectator Models in String Theory}
%=====================================
In the previous subsection, we discussed how to realize spectator model parameters in string theory, in particular how the large Chern-Simons coupling can potentially be realized by magnetizing D7-branes and/or a sufficiently high wrapping number. However, this enhancement comes at a price. 

Focusing on $C_2$ spectators, a boosted Chern-Simons coupling can be achieved by magnetized D7-branes. However, such a flux induces an effective D3-brane charge that contributes to the D3-brane tadpole cancellation condition. This is reviewed in~\cref{app:stringconv}, here we merely list the effective charge from the magnetized branes, which reads as\footnote{For simplicity we still assume work with a single pair of even and odd cycles.}:
%---------------------
\begin{equation}
    Q_{D3,ind} = \wrap\kappa_{+--}(\magn_1)^2 \nds\fstop
\label{eq:QD7}
\end{equation}
%---------------------
Where we have made the simplifying assumptions of considering branes on particular 4-cycle and flux $\magn_1$ on an odd 2-cycle that insects only with this wrapped 4-cycle. Other contributions to the D3-tadpole include the number of D3-branes as well as the spacetime curvature contributions from the D7-branes and O7-planes. The cancellation of the D3-tadpole is non-trivial and involves an interplay between the number of branes, their magnetizations, and the cycles they wrap. In the absence of a detailed compactification, we instead take an approximate approach and argue for an upper bound on the allowed D3-tadpole that a spectator model can induce. This is not a strict requirement, but nonetheless a spectator model that violates the bound would be hard pressed to find a home in the string landscape. 

To get an estimate for the allowed amount of magnetization, we turn to F-theory compactified on an elliptically fibred Calabi-Yau 4-fold $Y_4$. In such scenarios, the D3-tadpole cancellation condition takes the form~\cite{Sethi:1996es} 
%---------------------
\begin{equation}
   N_{{\rm D}3} + \int_{Y_4} G_4 \wedge G_4 = \frac{\chi(Y_4)}{24}\,.
 \label{eq:FTadpole}  
\end{equation}
%---------------------
On the left hand side of the above relation, $N_{{\rm D}3}$ is the total number of D3-branes while the second term is a 4-form flux that contains~\cref{eq:QD7} in the type IIB limit. On the right hand side, $\chi(Y_4)$ is the Euler characteristic of $Y_4$\footnote{In F-theory models with non-Abelian gauge groups, $Y_4$ is singular and one must calculate the Euler characteristic of $\bar{Y}_4$, the resolution of $Y_4$.} and in the type IIB limit this encodes the  curvature D3-charge from the D7-branes and O7-planes~\cite{Collinucci:2008pf}.

~\cref{eq:FTadpole} implies an absolute upper bound on the amount of magnetic flux allowed in a CY 4-fold $Y_4$ as $Q_{D3,ind}< \chi(Y_4)/24$. Thus the size of CS couplings on a given 4-fold are inherently limited by the topology. However, this is too lax a constraint. For a working type IIB string compactification, we must also stabilize the complex structure moduli of the CY 3-fold and the axio-dilaton via the 3-form fluxes $H_3,F_3$. These contribute to the D3-charge as well, taking the form
\begin{equation}
    Q_{{\rm D}3,H_3\&F_3}\sim \int_{X_3}F_3\wedge H_3\,.
\end{equation}
Therefore, to maintain a sufficient amount of tadpole to allow moduli stabilization, we take a conservative approach and estimate that the amount of flux that can be used to generate large CS couplings is $\sim {\cal O}(0.1)\times \frac{\chi(X_4)}{24}$. Thus we will enforce the following effective constraint
%------------------
\begin{equation}
  Q_{{\rm D}3,ind.} < Q_{eff}(Y_4) :=0.1\times \frac{\chi(Y_4)}{24}\,.
  \label{eq:effTadpole}
\end{equation} 
%------------------
A related but distinct question to the one above is if there is a universal upper bound on the Euler characteristics of elliptically fibred CY 4-folds. Such a bound would provide a universal constraint on CS couplings. No such upper bound currently exists, but we can consider the 4-fold with the largest known characteristic. In terms of the Hodge numbers of a CY 4-fold, one has  
%---------------------
\begin{equation}
    \chi(Y_4) = 6(8 + h^{1,1} + h^{3,1} - h^{2,1})\,.
\label{eq:FEuler}  
\end{equation}
%---------------------
There exists a CY 4-fold $\hat{Y}_4$ with Hodge numbers~\cite{Candelas:1997eh} 
\begin{equation}
    (h^{1,1},h^{2,1},h^{3,1}) = (303148,0,252)\,,
\end{equation}
which gives $\chi(\hat Y_4)/24 = 75852\simeq 10^5$. If we take this CY 4-fold as representing an estimate for the largest possible $\chi(Y_4)$, then in analogy with~\cref{eq:effTadpole} we can bound spectator models by the effective usable charge
\begin{equation}
    \hat{Q}_{eff} := 0.1\times \frac{\chi(\hat{Y}_4)}{24} \simeq 10^4\,,
\label{eq:maxtadpole}
\end{equation}
as 
 \begin{equation}
     \textbf{Constraint I:} \quad Q_{{\rm D}3,ind} \le \hat{Q}_{eff}\fstop
\label{eq:const:tadpole}
 \end{equation}
We now consider the implications of the above arguments on attempts to embed non-Abelian spectator models into type IIB compactifications~\cite{Holland:2020jdh}. The authors consider two separate scenarios within the LVS framework -- K\"{a}hler inflation~\cite{Conlon:2005jm} and fibre inflation. In both examples, the spectator sector is given by the EFT of a stack of magnetized, multiply-wound D7-branes, with the spectator axion given by one of the descendents of $C_2$. In particular,~\cite{Holland:2020jdh} give the required model parameters as a triplet $(m,\nds,w)$. 
\begin{equation}
    (m,\nds,w)\sim  \begin{cases}
        (10^4,10^5,25 )\\
        (10^2,10^3,1)
    \end{cases}
    \qquad 
    \begin{aligned}
        &\text{K\"{a}hler Inflation}\\
        &\text{Fibre Inflation}\,.
    \end{aligned}
\end{equation}

With these parameters, one can utilize~\cref{eq:QD7} to get an estimate of the required tadpole for the models in~\cite{Holland:2020jdh}. However, there is a degeneracy in their parameter $m$ and the product of intersection numbers and D7-brane magnetization, $\kappa_{+--}\magn_1$ as described above. 

Therefore, we must make some assumptions to estimate the tadpole. One can imagine two scenarios for the models of~\cite{Holland:2020jdh}. In the first scenario, the intersection numbers are $\pazocal{O}(1)$ and $\magn_1\simeq m$. Then we have
\begin{equation}
\text{Scenario I:}\qquad 
    Q_{{\rm D}3,ind}\simeq 
    \begin{cases}
        25(10^4)^2 10^5 \sim 10^{14} \\
        (10^2)^2 10^3 \sim 10^7 
    \end{cases}
    \qquad 
    \begin{aligned}
        &\text{K\"{a}hler Inflation}\\
        &\text{Fibre Inflation}\,.
    \end{aligned}
\end{equation}
Alternatively, one could consider a cycle with large intersection numbers so that $\kappa_{+--}\simeq m$ and $m_1\simeq \pazocal{O}(1)$,
\begin{equation}
  \text{Scenario II:}\qquad  Q_{{\rm D}3,ind}\simeq 
    \begin{cases}
        25(10^4) 10^5 \sim 10^{10} \\
        (10^2) 10^3 \sim 10^5 
    \end{cases}
    \qquad 
    \begin{aligned}
        &\text{K\"{a}hler Inflation}\\
        &\text{Fibre Inflation}\,.
    \end{aligned}
\end{equation}
In both scenarios, K\"{a}hler inflation requires a tadpole orders of magnitude larger than $\chi(\hat{Y}_4)/24\simeq 10^5$. This is also true for fibre inflation in Scenario I. In Scenario II, fibre inflation is borderline consistent with tadpole cancellation, but if we incorporate moduli stabilization and enforce $Q_{{\rm D}3,ind}< \hat{Q}_{eff}$, then even this case is problematic. 

While the above not does constitute a proof that the SCNI models of~\cite{Holland:2020jdh} are inconsistent, it does provide a strong hint that such models may be impossible to realize given our current understanding of string models. One may therefore be tempted to place them in the string swampland as opposed to the landscape.\\

Turning back to MASA models, we also supplement~\cref{eq:const:tadpole} with conditions to ensure control of the stringy MASA construction. First, we require that the various $U(1)$ gauge factors are under perturbative control below the string scale. Since the gauge field theory loop expansion parameter is $\alpha_{U(1)}/2\pi$, we impose
\begin{equation}
   \textbf{ \textbf{Constraint II:}}\quad \frac{\alpha_{U(1)}}{2\pi} \lesssim 1 
   \label{eq:const:perturb}
\end{equation}
for each $U(1)$ gauge theory factor.  We must also demand that the expansions of non-perturbative effects giving the spectator axions masses remain under control, i.e. the higher-order terms are suppressed relative to the leading order one.

For spectator axion masses obtained via ED1 instantons, the scale of the non-perturbative effect is controlled by the wrapped 2-cycle volume as $e^{-2\pi v}$. To ensure that the next higher order instanton is sufficiently suppressed that it can be ignored, we demand $2\pi v \gtrsim 2$. Assuming a simple intersection structure such that $\tau = \frac{1}{2}\kappa_{+++} v^2$, this translates into a bound
\begin{equation}
   \textbf{ \textbf{Constraint III:}}\quad  \frac{\pi^2}{\kappa_{+++}w_1\alpha_1} \gtrsim 1
\label{eq:const:ED1}
\end{equation}
where we have replaced $\tau$ with the fine-structure constant of the U(1) gauge theory wrapped on the 4-cycle associated with $\tau$. For ED3 instantons, we place a similar constraint to the previous one except that the relevant exponential depends on 4-cycle volumes controlled by the K\"{a}hler moduli $\tau$. Thus we impose $2\pi \tau \gtrsim 2$ for any 4-cycle supporting an ED3 instanton, which translates to a bound
\begin{equation}
    \textbf{ \textbf{Constraint IV:}}\quad \frac{\pi}{2w_1\alpha_1} \gtrsim 1 
\label{eq:const:ED3}
\end{equation}
Where we have again replaced $\tau$ with the Abelian fine-structure constant of the worldvolume theory of the D7-brane that is also wrapping the 4-cycle.

Finally, for cases that include gaugino condensation arising from a non-Abelian gauge group $\pazocal{G}$ we impose\\
\begin{equation}
   \textbf{ \textbf{Constraint V:}}\quad  \frac{\pi}{2} \frac{1}{c(\pazocal{G})\alpha_{\pazocal{G}}} \gtrsim 1
\end{equation}
to suppress high-order gauge instanton effects.

Concerning the axion sector, we should raise a final issue which arises from boosting the CS coupling of these axions by increasing the wrapping number $\wrap$ of the relevant D7-branes. 
This concerns specifically the even axions, as a large $\wrap$ could be the unique way in which the CS coupling could be boosted, but is valid also for the $C_2$ CS couplings. 
The point is that a D7-brane which is wrapping a given 4-cycle $\oricycle$ several times, in a certain sense can be thought of as `partitioning' the cycle volume ${\cal V}_{\oricycle}$ among its wrappings. For an isotropically shaped 4-cycle with a single size $L_\Sigma=R_{\oricycle}/\sqrt{\alpha'}\sim {\cal V}_{\oricycle}^{1/4}$ we can interpret this as an average distance between two adjacent wrapping loops $d_\wrap=({\cal V}_{\oricycle}/\wrap)^{1/4}\sim L_{\oricycle}/\wrap^{1/4}$. As long as SUSY is unbroken, a multi-wrapped D7-brane is still a BPS object and as such has no potential energy change associated with the wrapping number and its associated wrapping distance $d_\wrap$. However, once SUSY is broken at high scales during inflation where our scenario takes place, the SUSY breaking will communicate at some level to the D7-brane in question as well. In that situation, stretching the single D7-brane over multiple wrappings costs energy, and thus we expect a potential energy which increases with $\wrap$ and shows a potential barrier as a function of the adjacent loop distance $d_\wrap$. Hence, in this regime we expect a tunneling instability toward recombination of adjacent loops of the wrapped-up D7-brane. While the strength of this barrier is clearly dependent on the amount of SUSY breaking communicated to the D7-brane, a very first conservative guess may be that suppressing the tunneling instability requires separating the wrappings by more than a string length which implies a constraint $d_\wrap > 1$. Imposing this, by the above scaling argument, implies an upper bound on the wrapping number 
\begin{equation}
    \wrap < {\cal V}_{\oricycle}={\rm Re}\, T = {\cal O}(10\ldots100)\,.
    \label{wrap1}
\end{equation} 
We can now insert this bound into the relation determining the D7-brane $U(1)$ gauge coupling $g^{-2}=\wrap {\rm Re}\, T_{\oricycle}/2\pi=\wrap {\cal V}_{\oricycle}/2\pi=\wrap L_{\oricycle}^4/2\pi$, which gives the following bound once inserted in \cref{wrap1}  
\begin{equation}
    \wrap < 1/\sqrt{2\alpha_1}\,.
\end{equation}
For values of the gauge coupling $\alpha_1={\cal O}(1\dots 5)$ picked out by the constraint plots below, enforcing this constraint rigorously would restrict us to having only singly wrapped D7-branes. Consequently, this would eliminate the even axions from the spectrum of axions detectable by GW emission, presenting a qualitative argument against the viability of the $C_4$ axions as potential spectators.
However, as elaborated further below, even axions face elimination as candidates for observable GW signals already due to the detrimental impact of a strong CS coupling on control of the EFT.  
In the case of odd axions instead, their viability remains more resilient. Since they do not depend solely on multiple winding to enhance their CS coupling and typically only require a wrapping factor $\wrap \sim \pazocal{O}(1)$ to PTA amplitudes, we do not view this constraint as overly stringent.

%-----------------------------
\subsection{Type IIB MASA Models}
%-----------------------------
We now examine possibilities to realize MASA models in compactifications of type IIB string theory. To this end, we will not consider a concrete compactification manifold, but instead examine potential scenarios by assembling the above ingredients. Effectively, this will provide us with a set of selection rules for the existence of viable MASA models in type IIB string theory on CY orientifolds.

The pairing of a specific class with a corresponding case establishes the essential groundwork, delineating the minimum set of physical parameters needed to formulate a viable spectator model. It is imperative, however, to ensure the compatibility of these pairings by addressing additional considerations. In the subsequent discussion we delve into these intricacies while also incorporating the constraints described of the previous subsection.

Intuitively from the QFT-centric arguments of~\cite{Agrawal:2018mkd,Bagherian:2022mau}, one may expect that realizing spectator models with large signals in string theory would be quite challenging. Indeed it is. In fact, it is simple to show that visible spectator models utilizing $C_4$-axions are quite difficult to realize. The crucial point is that for a stack of D7-branes wrapping a 4-cycle $\oricycle$, the Chern-Simons coupling of the $C_4$-axion is entirely determined by the vev of the related K\"{a}hler modulus
\begin{equation}
    \lambda_{C_4}\simeq \frac{1}{\langle\tau \rangle }\,.
\label{eq:C4CS}
\end{equation}
In particular, the CS coupling is independent of both the wrapping and magnetization of the precise D7-brane configuration one is considering. To have a visible GW signal in the near-term future, we would need $\lambda_{C_4} = \pazocal{O}(10)$, which implies $\langle\tau\rangle \simeq 0.1$. If we assume that the $C_4$-axion gets a mass from an ED3 instanton, then control of the EFT is encapsulated by~\cref{eq:const:ED3}. This requires $\langle\tau\rangle \gtrsim 0.3$, rendering visible GWs and EFT control mutually exclusive. An identical argument can be made for $C_4$-axions that obtain masses via gaugino condensation.   

Of course the details of the above can be complicated by kinetic or mass mixing between axions. However, it seems unlikely that sufficient kinetic mixing can exist.
Naively, if one has a two-axion model of the form
\begin{equation}
    \pazocal{L}\supset  \frac{1}{2}\bigg((\partial a_1)^2+(\partial a_2)^2 +2\epsilon \partial_\mu a_1\partial^\mu a_2\bigg) + \Lambda^4 \cos\left(\frac{a_2}{f_2}\right) - \frac{\lambda_1}{4 f_1}a_1 F_{\mu\nu}\widetilde{F}^{\mu\nu}\,,
\label{eq:mixlag}
\end{equation}
up to $\pazocal{O}(\epsilon^2)$ corrections, the kinetic terms can be  diagonalized by a shift $a_1\rightarrow a_1-\epsilon a_2$, which induces a Chern-Simons coupling for the $a_2$ axion as 
\begin{equation}
    \pazocal{L} \supset \frac{\epsilon f_2}{f_1}\frac{\lambda_1}{4f_2}a_2 F_{\mu\nu}\widetilde{F}^{\mu\nu}\,.
\end{equation}
If $f_2/f_1 \gg \epsilon^{-1}$, then the $a_1$ axion would have a CS coupling $\lambda_2 = \epsilon \lambda_1 f_2/f_1$ that would be boosted relative to $\lambda_1$. 

Nevertheless, achieving a boosting effect in type IIB compactifications appears to be a challenging prospect. From the orientifold kinetic terms displayed in~\cref{app:stringconv}, we see that $C_4$- and $C_2$-axions do not mix directly via the tree-level K\"{a}hler metric. Therefore, if our goal is to have a $C_4$- spectator axion realized via the kinetic mixing outlined in
~\cref{eq:mixlag}, we might consider the kinetic mixing of two $C_4$-axions arising from a non-diagonal term in the K\"{a}hler metric. 
Broadly speaking, $C_4$-axions can be classified as either \textit{local} or \textit{non-local}, depending on whether they arise from blow-up cycles or cycles that determine the overall compactification volume, respectively. Typically, local axions have decay constants that scale as $f\sim M_p\pazocal{V}^{-1/2}$ while the decay constants of non-local axions scale as either $f\sim M_p \pazocal{V}^{-2/3}$ (isotropic compactifications) or 
$f\sim M_p \pazocal{V}^{-1}$ (anisotropic compactifications)~\cite{Cicoli:2012sz}, where $\pazocal{V}$ is the compactification volume in string units. 

To get a handle on some possibilities, let us consider a Swiss cheese CY with volume form
\begin{equation}
    \pazocal{V} = \alpha_0 \tau_b^{3/2} - \alpha_1 \tau_{s_1}^{3/2}-\alpha_2 \tau_{s_2}^{3/2} -\beta_1 \tau_{s_1}\tau_{s_2}^{1/2} - \beta_2 \tau_{s_2}\tau_{s_1}^{1/2}\,.
\label{eq:CYvol}
\end{equation}
Here $\tau_b$ controls the size of the CY volume $\pazocal{V}\simeq \pazocal{V}_0 = \alpha_0\langle\tau_b^{3/2}\rangle$ while $\tau_1$ and $\tau_2$ determine the size of blow-up cycles.  This CY gives 3 $C_4$-axions: two local axions \{$\rho_{s_1}$,\,$\rho_{s_2}$\} and a non-local axion $\rho_b$. The K\"{a}hler metric is determined by derivatives of $K = -2\ln(\pazocal{V})$. We have $\alpha_1,\alpha_2>0$ and we need $\langle\tau_{s_1}\rangle\gg \langle\tau_{s_2}\rangle$ for arbitrary $|\beta_1|\lesssim \alpha_1$ and $0\leq \beta_2\ll \alpha_2$ or $\beta_2<0$ to retain positive kinetic terms for both axions from the two small blow-up cycles. There are three mixing scenarios one can consider from~\cref{eq:CYvol}. 

\begin{itemize}
    \item The first is a mixing between the non-local axion and one of the local ones. We set $\beta_1=\beta_2=\alpha_2=0$ and discard the second blow-up cycle and associated axion.  To allow for an enhancement factor,  we  wrap branes on $\oricycle_b$  so that $\rho_b$ has a CS coupling with the worldvolume gauge fields. If we assume that $\rho_{s_1}$ obtains a mass from some non-perturbative effect, then we have an action of the form~\cref{eq:mixlag}  with $\{\rho_b,\, \rho_{s_1}\}$ corresponding to $\{a_1,\,a_2\}$. The mixing parameter of~\cref{eq:mixlag}  is $\epsilon \sim \langle\tau_{s_1}/\tau_b\rangle^{3/4}$ so that after unmixing the axions, $\rho_{s_1}$ couples to the gauge fields with CS coupling $\lambda_{s_1} \propto \langle\tau_{s_1}/\tau_b\rangle^{1/2} \lambda_b$. Since we must have the hierarchy $\langle\tau_b\rangle\gg \langle\tau_{s_1}\rangle$, the CS coupling of $\rho_{s_1}$ is suppressed relative to that of  $\rho_b$, not enhanced.   
    \item We can also consider the mixing between the two local axions $\rho_{s_1}$ and $\rho_{s_2}$. First we consider the ``strong" Swiss-cheese scenario which corresponds to the volume form in~\cref{eq:CYvol} with $\beta_1=\beta_2=0$.  Since the decay constants of the local axions are inversely proportional to the size of the cycle they are supported on, we can wrap branes on $\oricycle_{s_1}$ and assume a hierarchy $\langle\tau_{s_1}\rangle\gg\langle\tau_{s_2}\rangle$ to obtain a setup akin to~\cref{eq:mixlag} with $\rho_{s_1}$ and $\rho_{s_2}$ playing the roles of $a_1$ and $a_2$, respectively. The mixing parameter is then $\epsilon\sim \langle\tau_{s_1}\tau_{s_2}\rangle^{3/4}/\pazocal{V}_0$ and diagonalization gives $\rho_{s_2}$ a CS coupling $\lambda_{s_2}\simeq \langle\tau_{s_1}\tau_{s_2}^{1/2}/\pazocal{V}\rangle \lambda_{s_1}$. Since $\langle\pazocal{V}\rangle\gg \langle\tau_{s_1}\rangle,\langle\tau_{s_2}\rangle$, we again have suppression as opposed to enhancement. 
    \item Finally, we can consider the mixing of the two local axions but including the $\beta_i$ terms in~\cref{eq:CYvol}. We again consider the hierarchy $\langle\tau_{s_1}\rangle\gg\langle\tau_{s_2}\rangle$ and branes wrapping $\oricycle_{s_1}$. If $\beta_1\gtrsim \alpha_1$ we get for the kinetic mixing parameter $\epsilon\sim \langle\tau_{s_2}/\tau_{s_1}\rangle^{1/4}\ll 1$ and a CS coupling for $\rho_{s_2}$ of  strength  $\lambda_{s_2}\simeq \langle\tau_{s_1}/\tau_{s_2}\rangle^{1/2}\;\lambda_{s_1}$. Thus with the assumed hierarchy in the sizes of the blow-up cycles, we do see an enhancement of the $\rho_{s_2}$ CS coupling relative to that of $\rho_{s_1}$. However, there is an important caveat in that $\lambda_{s_1}\propto \langle\tau_{s_1}\rangle^{-1}$ so that $\lambda_{s_2}\propto \langle\tau_{s_1}\tau_{s_2}\rangle^{-1/2}$. Thus attempting to increase the enhancement factor $\epsilon f_{s_2}/f_{s_1}$ by enlarging $\langle\tau_{s_1}\rangle$ will in truth drive the gauge coupling down and spoil any attempts to realize a large CS coupling.
\end{itemize}

Note that this last case was considered by~\cite{Agrawal:2017cmd} in the context of the QCD axion and the authors found that an enhancement of axion coupling to QED may be possible. Their scenario works as discussed because they fix the fine structure constant to that of QED. In our scenario,  the strength of the gauge coupling is determined by the size of the 4-cycles and the enhancement factor is rendered ineffectual. While the above arguments do not entirely forbid enhancement of $C_4$-axion CS couplings in the type IIB orientifold landscape, it suggests that achieving enhancement via kinetic mixing is not simple to realize. 

We end this discussion on $C_4$-axions with some caveats. First, we have discussed very simple scenarios, and one could consider more complicated setups, such as more general divisors. This would involve multiple wrapping numbers and $C_4$-axions. However, it is not clear that this complication helps. The worldvolume gauge theory will couple to some linear combination of $C_4$-axions, and the gauge coupling will also be determined by a linear superposition of  $\tau_\alpha$ also determined by the wrapping numbers. Thus, an attempt to create a large CS coupling by boosting wrapping numbers will simply alter the dominant fields in the superpositions and cancel out when determining the CS coupling, similar to the discussion around~\cref{eq:C4CS} above.  Next, in arguing against $C_4$-axions as viable spectators, we are more specifically referring to the impossibility of observing GWs at near-term detectors as summarized in~\cref{fig:GW}. If $\lambda_{C_4}\simeq 1$, which is more reasonable from the standpoint of control of the EFT, then $P_T=\pazocal{O}(10^{-22})$. This is far below the reach of any proposed detector, but we do not exclude the possibility of extremely advanced detection methods that would be able to reach this level and see peaks arising from $C_4$-axion spectators. We also note that the above argument only rules out visible $C_4$-axions due to demanding an approximation scheme that permits one to neglect higher-order non-perturbative effects. If one could reliably calculate an infinite number of instanton terms, then this constraint could be neglected and the possibility of more visible GWs from $C_4$ spectators opens up. A caveat to these statements is that one still requires small values of $\langle\tau\rangle$, which could result in a tower of states becoming light and ruining the EFT, as predicted by the Swampland Distance and Emergent String conjectures~\cite{Ooguri:2006in,Lee:2019wij}. The precise statement then is that visible GWs from $C_4$ axions requires one to live near the center of K\"{a}hler moduli space and far from asymptotic regions.   

A similar argument can be made to rule out $C_2$-axion spectators in certain compactifications. 
For simplicity, let us assume a construction with $h^{1,1}_-=1$ and where the vev of the sole $B_2$-axion vanishes in a particular vacuum. Then from~\cref{eq:gaugkin2}, we see that the gauge coupling constant is $g^{-2} = \frac{\wrap}{2\pi}\bigg( \langle\tau_{\oricycle}\rangle - \frac{\kappa_{+--}}{2g_s}(\magn_1)^2 \bigg)$, where we have used a shorthand for the relevant intersection number. As is evident from this expression, the sign of $\kappa_{+--}$ plays an essential role in the viability of a controlled spectator model. If $\kappa_{+--}<0$, then increasing magnetization leads to a decrease in the gauge coupling constant. Since $\lambda\propto \alpha \simeq g^2$,  this  increases the difficulty of realizing a large CS coupling. In fact, from~\cref{eq:D7bosons2} the CS coupling has the form 
\begin{equation}
    \lambda_{C_2} = \frac{\kappa_{+--}}{\pi} \frac{g_s\magn_1}{ (2 g_s\langle\tau_{\oricycle}\rangle-\kappa_{+--}\magn_1^2)}\,.
\label{eq:quadchern}
\end{equation}
For $\kappa_{+--}<0$ and fixed $\langle\tau_{\oricycle}\rangle$, $\lvert\lambda_{C_2}\rvert$ increases with the magnetization $\magn$ as long as $\magn< m_{max}$ and then starts to decrease with further increasing $\magn_1$. At the maximum $\magn_{max}\sim \sqrt{-g_s\langle\tau_{\oricycle}\rangle/\kappa_{+--}}$ we find $\lvert\lambda_{C_2}\rvert_{max} \sim 1/m_{max}$. Hence, even by tuning $g_s$ and $\langle\tau_{\oricycle}\rangle$ to arrange for a long regime of $\lvert\lambda_{C_2}\rvert$ growing linearly with $\magn_1$, its magnitude will be driven deeper into the region $\lvert\lambda_{C_2}\rvert\ll1$. In fact, since $\lvert\lambda_{C_2}\rvert_{max}$ is inversely proportional to  $\langle\tau_{\oricycle}\rangle^{1/2}$, satisfying the control constraints of the previous section drives one inexorably to small CS coupling, similar to the above discussion on $C_4$-axions. Thus the associated GW signal is far too small  to be observed in the near future and we see there is a general tension between observable spectator axions and  models with $h^{1,1}_-=1$ and  $\kappa_{+--}<0$. To avoid this issue, one has two obvious paths. The first is to consider compactifications with $h^{1,1}_->1$. In such cases, it may be possible to have intersection numbers such that one $C_2$-axion obtains a large CS coupling without introducing the quadratic flux terms if certain intersection numbers are zero or if opposing fluxes can be placed to cancel the quadratic terms. The second option is to consider compactifications with $\kappa_{+--}>0$.  For $\kappa_{+--}>0$, the magnitude of~\cref{eq:quadchern} can be increased significantly for fixed $g_s$ and $\magn$ assuming some tuning of $\langle\tau_{\oricycle}\rangle$ is permitted. This may even occur in such a way that the control conditions discussed in the previous subsection are easily satisfied. 

Having discussed the severe limitations on realizing spectator models in type IIB models, we now discuss two potentially viable scenarios.

\begin{itemize}
    \item \textbf{Class I Spectators:} \textbf{Case 2}\\
    This scenario contains a single $U(1)$ gauge factor and a $C_2$ axion. We allow the D7-brane to have multiple wrappings parametrized by $\wrap_1$ and magnetic flux $\magn_1$ in its worldvolume\footnote{Here we will neglect the term in the gauge kinetic function that is quadratic in the flux. As described in the preceding paragraphs, such a term will either completely invalidate this model, or greatly relax control restrictions.}. The axion receives a mass via ED1 instantons wrapping a 2-cycle. The Chern-Simons coupling is given by~\cref{eq:stringspecCS} with $\pazocal{J} =1$ and $\pazocal{M} = \wrap_1 \kappa_{+--} \magn_1$ so that  $\lambda = \frac{\alpha_{U(1)}}{\pi}\wrap_1\kappa_{+--} \magn_1$.

     We now want to consider observable GW signals in this setup while keeping in mind the induced D3-charge of the brane stack. To this end, we can fix the CS coupling to a value required to produce a specific GW peak, i.e. $\lambda = \lambda_{\text{GW}}$. Then we can trade one microscopic parameter, say the flux $\magn_1$, for $\lambda_{GW}$. We can use this to replace $\magn_1$ in the expression for induced D3-brane charge, and arrive at a scaling relation for the D3-charge necessary to produce a desired GW peak amplitude. Choosing reference parameters  
                \begin{equation}
                    (\lambda_{\text{GW}},\alpha_{D7}, \kappa_{+--}, \wrap_1) = (22,0.8,1,1)\,,
                \label{eq:refpara} 
                \end{equation}
                we obtain a D3-charge ``Drake Equation" (DDE)\footnote{Named in analogy with the Drake equation~\cite{drake_project_1961}.}:  
                \begin{equation}
                    Q_{{\rm D}3,ind.} \simeq 7.5\times 10^3  \left(\frac{1}{\wrap_1\;\kappa_{+--}}\right)\left(\frac{0.8}{\alpha_{D7}}\right)^2\left(\frac{\lambda_{\text{GW}}}{22}\right)^2 \,.
                \label{eq:QD3Drake-eq1}  
                \end{equation}
    To motivate a string embedding of this scenario, we demand that~\cref{eq:QD3Drake-eq1} satisfy~\cref{eq:const:tadpole}. We also enforce constraints on perturbativity~\cref{eq:const:perturb} and ED1 control~\cref{eq:const:ED1}. An example parameter space including these bounds is displayed in~\cref{fig:stringparaspace}   below.
\item \textbf{Class IIb Spectators:} \textbf{Case 4}\\
We now consider a scenario that involves both $C_4$- and $C_2$-axions. We consider homologous 4-cycles $\oricycle^{(1)}$ and $\oricycle^{(2)}$ in $\ori$ such that $\oricycle^{(1)}$ supports a stack of $\nds$ D7-branes and $\oricycle^{(2)}$ has a single D7-brane wrapped on it. We also consider a third cycle $\oricycle^{(3)}\in\ori$ that is the even combination of cycles $\CYcycle^{(3)}$ and $\Pi_4^{\prime (3)}$ in $\CYB$. We assume that $\oricycle^{(3)}$ supports an ED3 instanton that stabilizes its K\"{a}hler modulus and provides a mass for the associated $C_4$-axion. We will also assume there is a non-trivial even-odd-odd intersection number (called here $\kappa_{-}$ for convenience) such that the $C_2$-axion from the odd cycle associated to $\oricycle^{(3)}$ couples to the gauge theories on $\oricycle^{(1)}$ and $\oricycle^{(2)}$. For ease of discussion, we also assume that certain intersection numbers of $\oricycle^{(1)}$ vanish to avoid a flux-induced St\"{u}ckelberg coupling.

The EFT of this configuration again contains an $SU(\nds)\times U(1)_1\times U(1)_2$ gauge sector. There are also three axions - a $C_4$- and a $C_2$-axion from $[\oricycle^{(1)}]$, and an additional $C_2$-axion from $\oricycle^{(3)}$. A linear combination of $U(1)_1\times U(1)_2$ will eat the $C_2$ associated to $[\oricycle^{(1)}]$. The surviving $U(1)$ boson couples to the $[\oricycle^{(1)}]$ $C_4$-axion as well as the $C_2$-axion $c$ from $\oricycle^{(3)}$:
\begin{equation}
    \pazocal{S}_{EFT}\supset -\int \sqrt{-g}\;\;d^4x \bigg\{ \rho +\frac{\kappa_-(\wrap_1\magn_N + \nds^2\wrap_N\magn_1)}{\wrap_1+\nds^2\wrap_N} c \bigg\}\frac{2\wrap_1\alpha_1}{4} X^{(2)}_{\mu\nu}\widetilde{X}^{(2)\mu\nu}\fstop
\end{equation}
Here $X_{\mu\nu}^{(2)}$ is the field strength of the surviving $U(1)$ and its Chern-Simons couplings to the axions $\rho\, \&\, c$ are obtained by rotating the original gauge fields. In the above we have allowed for a non-zero magnetization $\magn_1$ of the $\oricycle^{(2)}$ D7-brane, and we have also used the relation $\wrap_N\alpha_N = \wrap_1\alpha_1$. We assume that the D7-brane stack on $\oricycle^{(1)}$ undergoes gaugino condensation. If we allow for magnetization $\magn_N$ on this stack, then both axions appear in the potential as in~\cref{eq:condaxpot} with $\{\wrap_{\pazocal{G}}, \magn_{\pazocal{G}, c(\pazocal{G})}\} = \{\wrap_N, \magn_N, \nds\}$. Defining decay constants $f_e = \frac{\nds}{2\pi\wrap_N}F_e$ and $f_o = \frac{\nds}{2\pi\wrap_N}\frac{F_o}{\kappa\magn_N}$ the scalar potential reads:
\begin{equation}
    \begin{aligned}
        V %&\supset -\Lambda_N^4 \cos\bigg[
    %\frac{2\pi\wrap_N}{\nds}(\rho + \kappa \magn_N c_2)\bigg]\\
        &=  -\Lambda^4_N \cos \bigg[ \frac{\chi_e}{f_e}  +\frac{\chi_o}{f_o}\bigg]\,.
    \end{aligned}
\label{eq:Vcont}
\end{equation}
The masses and eigenstates are those of~\cref{eq:condstates} with $\epsilon  =\kappa_- \magn_N F_e/F_o = f_e/f_o$. For the massive axion $\chi_2$ we define a re-scaled axion decay constant 
\begin{equation}
    f_2 = \frac{f_e}{\sqrt{1+\epsilon^2}}\,.
\end{equation}
Using $f_e$ and $f_o$ we can define Chern-Simons coupling parameters 
\begin{equation}
    \begin{aligned}
        \lambda_e &= \frac{\alpha_N\nds}{\pi}\\
        \lambda_o &= \frac{\alpha_N\nds}{\pi}\bigg\{\frac{(\wrap_1/\wrap_N)+\nds^2(\magn_1/\magn_N)}{\wrap_1+\nds^2\wrap_N}\bigg\}\,.
    \end{aligned}   
\end{equation}
The coupling of the massive axion $\chi_2$ to the massless gauge field $X^{(2)}$ in the form of~\cref{eq:MASAlag}, with decay constant $f_2$ will therefore read:
\begin{equation}
    \lambda_2 = \frac{\lambda_e + \epsilon^2\lambda_o}{1+\epsilon^2}\,.
\end{equation}
Since $\lambda_o$ is proportional to the magnetization $\magn_1$ of the brane wrapping $\oricycle^{(2)}$, we see that increasing $\magn_1$ will boost the Chern-Simons coupling of the axion $\chi_2$. However, there is a hurdle to overcome -- we expect $\epsilon <1$, and the magnetization must overcome this suppression in order to achieve the desired coupling strength. The $\chi_1$ axion couples to $X^{(2)}$, with strength
\begin{equation}
    \lambda_1 \simeq \lambda_o  - \lambda_e  \,,
\end{equation}
where we are approximating $1+\epsilon^2\simeq 1$ and $f_1\simeq f_o$. Thus one can boost the Chern-Simons coupling of $\chi_1$ without having to overcome suppression by $\epsilon$. Unfortunately, $\chi_1$ in the model above is massless and is not a viable spectator. 

This can be remedied by the inclusion of an ED1 instanton that gives rise to a potential of the form~\cref{eq:c2axionpot} for the $C_2$-axion $c$. In terms of $\chi_1$ and $\chi_2$, we have a contribution to the potential of the form
\begin{equation}
    V\supset -\Lambda_{ED1}^4 \cos\bigg[\frac{\nds}{\kappa_- \wrap_N\magn_N} \frac{\chi_1 + \epsilon \chi_2}{f_o\sqrt{1+\epsilon^2}}\bigg]\,.
\end{equation}
The ED1 introduces a mixing between $\chi_1$ and $\chi_2$, but to lowest order in $\epsilon$ this potential simply provides a mass for $\chi_1$. If we also take $\kappa_-\wrap_N\magn_N\simeq 1$, then indeed $f_1\simeq f_o$ defines the periodicity of $\chi_1$.

With the above ingredients, we see that we have two spectator axions coupled to the $U(1)$ gauge field $X^{(2)}$. The induced D3 tadpole depends on the string parameters as 
\begin{equation}
    \begin{aligned}
        Q_{D3,ind} &= \kappa_- \wrap_1 \magn_1^2 + \kappa_- \wrap_N \magn_N^2 \nds\fstop\\ 
    \end{aligned}
\end{equation}
From here, one can replace $m_1$, derive a DDE equation in analogy with~\cref{eq:QD3Drake-eq1}, and look for a valid region in parameter space. We can then imagine that $\chi_1$ is the spectator axion associated with an observable peak, while $\chi_2$ produced a lower signal. While there is a valid parameter space for $\lambda_1=22$, it lies dangerously close to saturating our tadpole condition and can therefore be realized only in the most extreme of compactifications.
\end{itemize}

We note here an observation relevant for future GW experiments: given that the GW peak amplitude is exponentially sensitive to $\lambda$, the CS coupling reacts only logarithmically to increasing the sensitivity of future GW detection experiments. Compared to the peak amplitude corresponding to the recently reported evidence for a nanoGRAV/PTA stochastic GW signal, even the most futuristic GW detection experiment currently envisioned -- DECIGO/BBO -- would increase sensitivity only such, that the required value of $\lambda$ for a matching axion-generated signal lowers only by a factor of two.

\begin{figure}[t!]  
    \centering
            \includegraphics[scale=0.6]{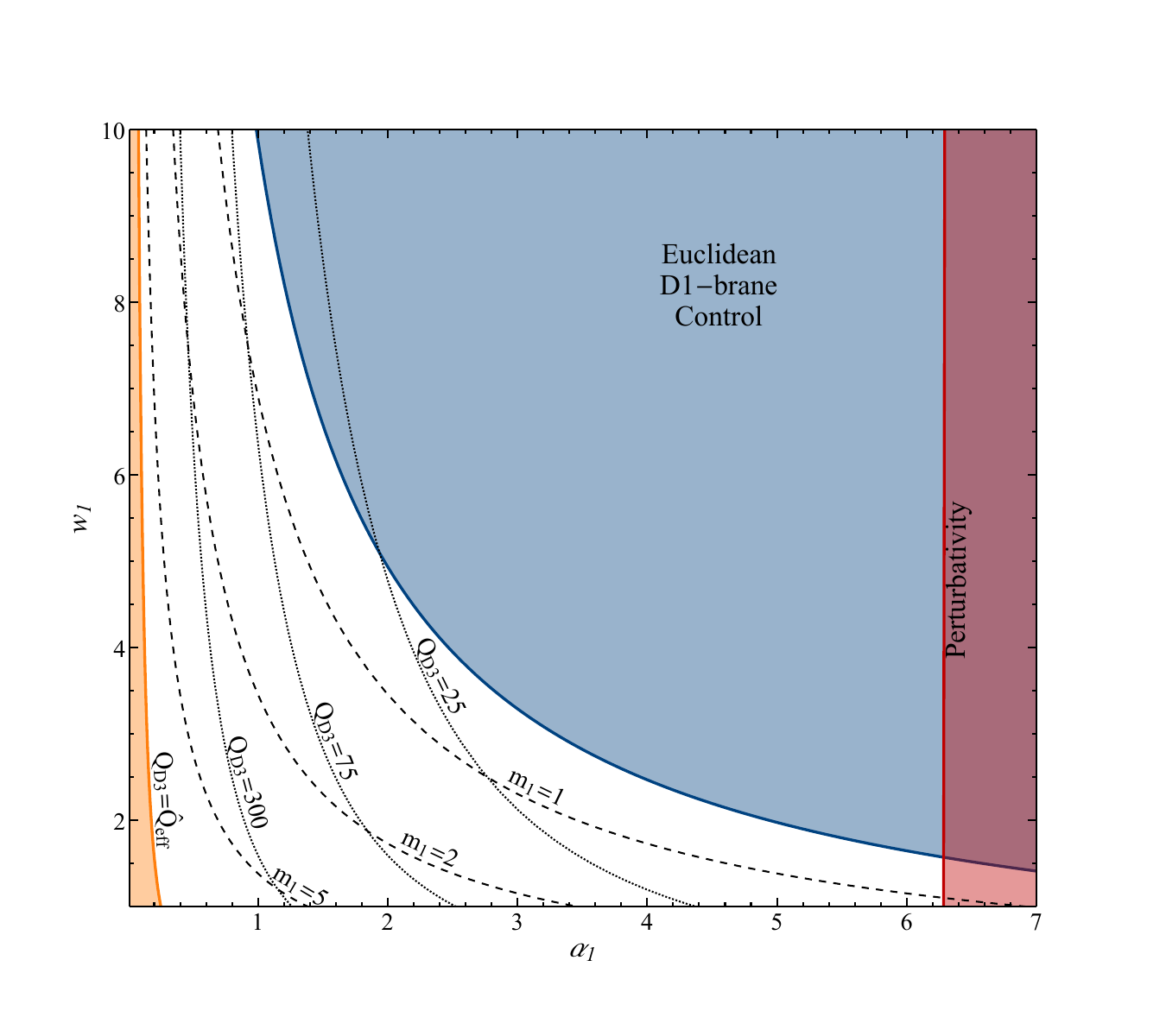}
       \caption{Example parameter space for Class I: Case 2 MASA models. The excluded shaded regions arise from constraints on control of the instanton expansion for ED1s (in blue), perturbativity of the $U(1)$ gauge theory (in red), and the tadpole constraint~\cref{eq:const:tadpole} (in orange). The dashed (dotted) contours correspond to parameter values with fixed magnetization $\magn_1$ (D3-tadpole $Q_{\rm D3}$). We chose parameters $\lambda_{GW} = 22$ and $\kappa_{+--}=10$.}
   \label{fig:stringparaspace}
\end{figure}

%==============================
\section{Conclusion}
\label{sec:conclusion}
%==============================
Inflationary models with spectator ALPs are an interesting framework for primordial cosmology, potentially allowing for observable signals from sectors that are highly sequestered from the Standard Model. In the context of string theory, a natural expectation is that the spectators may consist of multiple axions coupled to gauge fields. With this motivation, in the present paper we have extended the typical spectator models to account for multiple such fields, proposing the study of what we refer to as ``Multiple (non-)Abelian Spectator Axion" inflationary models.
Our primary objective has been to investigate the cosmological predictions from these models, exploring also their viability in string constructions and their connection to the axiverse. In this respect, models with Abelian gauge groups (MASA models) appear easier to realize in controlled string settings with respect to their non-Abelian (MnASA) counterpart.

Our analysis, focussed on the Abelian case, reveals distinct signatures in both the curvature and GWs power spectra, manifesting as multiple peaks at different scales.
We stress that our considerations and finding  do not necessarily apply to scenarios involving axions coupled to multiple gauge fields, or to cases where Abelian gauge fields interact via kinetic mixing. There is some indication that the former scenario may be rather rare in the type IIB landscape~\cite{Gendler:2023kjt}, while the latter may be entirely absent or quite suppressed ~\cite{Hebecker:2023qwl}.

The many possibilities granted by the choice of initial conditions, the number of spectators and their couplings chart an intriguing map of cosmological signatures, with multi-peaked spectra of varying width, amplitude, and position. Such map includes the possibility of a primordial GW signal at PTA scales, fully compatible \cite{Unal:2023srk} with the recently detected stochastic background in the nHz range and satisfying both perturbativity and backreaction constraints. Naturally, the presence of multiple spectator sectors has much more to offer: a gravitational wave forest which is ripe for the testing via existing as well as forthcoming GW probes. The sourcing mechanism for scalar and tensor fluctuations is analogous: the gauge fields non-linearly source both two-point functions. As a result, there is a scalar counterpart to the GW forest which, as we show, can be tested and constrained via CMB anisotropies, spectral distortions, and PBH bounds, depending on the scales involved. Scalar fluctuations also serve as a non-linear source of gravitational waves, something that provides a rather unique fingerprint that can potentially differentiate MASA models from other inflationary scenarios.

From the viewpoint of string theory, MASA models can provide a glimpse into the axionic content of a given compactification. However, this is not to say that observable MASA models are simple to realize. We have found that there are numerous restrictions that must be satisfied to have a viable spectator sector. One of the largest challenges is simply ensuring a suitable $U(1)$ factor exists in the gauge group of the EFT. The primary difficulty here is avoiding St\"{u}ckelberg couplings that would pair an axion and $U(1)$ gauge boson into a massive vector boson. Ostensibly, type IIB compactifications have two natural spectator axion candidates from dimensional reduction of the $10d$ p-forms $C_4$ and $C_2$. However, as argued above, it is quite difficult to generate sufficiently large CS couplings for the $C_4$-axions. This leaves $C_2$-axions as the sole viable spectator candidates.

On top of the difficulty in simply realizing the field content for spectator sectors, string constructions place further limitations on the parameters of the models. Boosting the CS coupling of $C_2$-axions requires non-zero magnetic flux in the worldvolume theory of D7-branes, which induces an effective D3-tadpole. This tadpole must be cancelled by other sources in the compactification manifold, such as orientifold planes, and is inherently limited. Furthermore, demanding perturbative and non-perturbative control of the construction places additional restrictions on the viable parameter space. 

Despite all the above constraints, there still appears to be viable scenarios. By far the least constrained scenarios are Class I: Case 2 models, where the St\"{u}ckelberg mechanism is sidestepped by appropriate features of the divisor wrapped by a D7-brane. An example of the parameter space of such models is presented in~\cref{fig:stringparaspace}. An interesting feature of this plot is that a viable spectator model can survive with relatively little magnetic flux - simply $\magn\simeq \mathcal{O}(1)$ can suffice so long as the fine-structure constant is sufficiently large. The root of this feature lies in the fact that Abelian spectators are not required to have extremely small gauge couplings to be viable. On the other hand, non-Abelian spectator models do feature small gauge couplings, which brings them in tension with tadpole cancellation as discussed above. 

So far, in the string context we have discussed only single-axion spectator models. To have multiple spectators, one needs several copies of the constructions that realize the Abelian gauge bosons and axions. Furthermore, one must heed the constraint of D3-tadpole cancellation. Since the parameter space in~\cref{fig:stringparaspace} allows for $\mathcal{O}(10)$ D3-tadpoles, it is not inconceivable that some compactifications may yield multiple visible spectator sectors. A distinct possibility is to stack a multitude of spectator sectors such that the collective signal is enhanced relative to that of a single constituent. This would require similar initial conditions for all axions involved, making it perhaps less generic. A precise distribution of initial conditions for the axions would be required to make this statement more precise.

One might interpret the above restrictions as an indication that visible spectator models are highly disfavored in string theory. An alternative viewpoint is that visible spectators serve as a powerful superselection rule in string vacua. If one is able to see a GW peak from a spectator axion, then one has learned valuable information regarding the topological structure of the manifold. For example, since $C_4$-axions are poor spectator candidates, a spectator GW peak is a direct probe of the \textit{odd} moduli content of a type IIB compactification. Previous studies~\cite{Cicoli:2012sz} considered the experimental signals of \textit{even} axions. If we take the optimistic viewpoint that we will be able to probe some features of the string compactification underlying our Universe, then direct coupling experiments will most likely probe even axions. Thus our MASA models provide a means to go beyond this and probe odd axions, which would reveal new information on the string embedding of our Universe. For example, a very optimistic scenario would be the future detection of several PTA-strength peaked GW signals at different frequencies. In this situation, by the argument above, we need to attribute all of them to odd sector string axions with GW signals increased sufficiently by choosing large enough magnetizing D7-brane fluxes. However, this potentially runs into the D3-brane charge tadpole bound, because for each odd sector string axion the induced D3-brane charge contribution on coupled D7-branes increases \emph{quadratically} in $m$. Hence, generating several strong GW peaks all by using odd sector string axions tends to use a significant fraction of the background D3-brane charge tadpole. This observational outcome would thus point us towards the F-theory compactifications on elliptic CY 4-folds with the largest Euler numbers, and may limit the number of strong GW wave peaks which we can explain using odd sector string axions. It is in this sense that a future ``wideband spectroscopy'' of GW signals may provide us with observational clues about the underlying topological structure and flux choices of the string compactification describing our universe.

Another interesting aspect of MASA models is that the GW signal is not directly dependent on the decay constants of the spectator axions, but is encoded in other parameters, such as $\delta$. On one hand, for fixed $\delta$ this removes the possibility of boosting the signal by lowering the decay constants. However, this feature is actually a boon from the string theory perspective. A persistent issue in probing string axions is that the axion decay constants are $\pazocal{O}(10^{16})$ GeV or above, making direct detection via terrestrial experiments extremely difficult, even if they couple to the Standard Model. Some haloscope experiments~\cite{Stern:2016bbw,Berlin:2020vrk,DMRadio:2022pkf,Ahyoune:2023gfw} will probe this interesting parameter space, but only for axions that i). couple directly to the Standard Model, ii). have sub-eV masses, and iii). constitute some non-trivial fraction of the observed dark matter. In contrast, MASA models more naturally probe heavier axions that need not couple to the Standard Model nor be present in dark matter. Thus one should consider searching for spectator GW signals as complementary to probing lighter elements of the type IIB string axiverse~\cite{Cicoli:2012sz}. \\

There are numerous directions that can be pursued from this work. First, an obvious direction is to attempt a bonafide embedding of Abelian MASA models into concrete string compactifications with specific orientifolds. Since the spectator mechanism hinges on the existence of an inflationary sector, a natural place to start is extensions of models with fibre inflation, K\"{a}hler inflation, or monodromy inflation. Along a similar vein, it would be interesting to perform scans of the type IIB orientifold landscape to determine how common are the structures required to realize the field content of Abelian spectator models. We have focused entirely on models with O7-planes, but it would be worthwhile to explore other scenarios, such as compactifications with both O3- and O7-planes\footnote{We thank Federico Carta for discussions on this point.}. Furthermore, we have only considered Chern-Simons couplings of axions to gauge fields, but one could also consider gravitational Chern-Simons couplings of the form $\propto \lambda_{GR} \chi R\tilde{R}$. 
We also note that we have entirely ignored any potential signals from axions with St\"{u}ckelberg couplings, but an interesting direction would be the study of spectator models with massive vector bosons.

%%%%%%%%%%%%%%%%%%%%%%%%%%%%%%%%%%%%%%%%%%%%%%%%%%%%%%%%%%%%%%%%%%%%%%%%%%%%%%%%%%%%%%%%%%%%%%%%%%%%

\section*{Acknowledgments}
We would like to thank Rafael \'Alvarez-Garc\'ia, Alessandro Mininno,  and Timo Weigand for many useful discussions. We also thank Timo Weigand for comments on the draft. ED and MF are indebted to Ogan {\"O}zsoy and Alex Papageorgiou for fruitful discussions.  
M. P. and J.M.L. would like to thank the University of Groningen for their hospitality
during the completion of this work. J.M.L. and M.P. are supported by the Deutsche Forschungsgemeinschaft under Germany's Excellence Strategy - EXC 2121 ``Quantum Universe'' - 390833306. MF acknowledges support from the
``Ramón y Cajal'' grant RYC2021-033786-I and the ``Consolidación Investigadora'' grant CNS2022-135590. MF’s work is partially supported by the Spanish Research Agency (Agencia Estatal de Investigación) through the Grant IFT Centro de Excelencia Severo Ochoa No CEX2020-001007-S, funded by MCIN/AEI/10.13039/501100011033. ED would like to acknowledge the Beate Naroska guest professorship programme of the \textsl{Quantum Universe } Cluster of Excellence at the University of Hamburg, and to thank the DESY Theory Group for warm hospitality whilst this work was in progress.

%%%%%%%%%%%%%%%%%%%%%%%%%%%%%%%%%%%%%%%%%%%%%%%%%%%%%%%%%%%%%%
% Appendix
%%%%%%%%%%%%%%%%%%%%%%%%%%%%%%%%%%%%%%%%%%%%%%%%%%%%%%%%%%%%%%
\appendix

%=========================================================
\section{Power Spectra Calculations}
\label{app:calculations}
%=========================================================
\subsection{Tensor Perturbations}

\noindent The mode functions in~\cref{eq:gaugemode1} satisfy the equation
\begin{equation}
    A_{ \pm}^{(l) \prime \prime }+\left(k^2 \mp k \frac{\lambda_l \chi_l^{\prime}}{f_l}\right) A_{ \pm}^{(l)}=0\,.
\end{equation}
with the helicity vectors obeying the following relations
\begin{equation}
    \vec{k} \cdot \vec{\epsilon}^{( \pm)}=0\,,\quad i \vec{k} \times \vec{\epsilon}^{( \pm)}= \pm k \vec{\epsilon}^{( \pm)}\,,\quad \vec{\epsilon}^{( \pm)} \cdot \vec{\epsilon}^{(\mp)}=1 \,, \quad\vec{\epsilon}^{( \pm)} \cdot \vec{\epsilon}^{(\pm)}=0\,.
\end{equation}
In a spatially flat, inflating Universe, the second term in parentheses reads $k\lambda_l\chi_l'/f_l =-2k\xi_l/\tau$ and~\cref{eq:gaugemodeEQ1} follows. \\
Plugging in the WKB solution in~\cref{eq:WKBsol} into~\cref{eq:gaugemode1} , we get the following expressions for the gauge fields 
\begin{equation}
\begin{aligned}
    \hat{A}_i^{(1)}(\tau, \vec{k})&=\int \frac{d^3 k}{(2 \pi)^{3 / 2}} \mathrm{e}^{i \vec{k} \cdot \vec{x}} \epsilon_i^{(+)}(\hat{k}) A_{+}^{(1)}(\tau, k)\left[\hat{a}_{+}(\vec{k})+\hat{a}_{+}^{\dagger}(-\vec{k})\right]\,,\\
    \hat{A}_i^{(2)}(\tau, \vec{k})&=\int \frac{d^3 k}{(2 \pi)^{3 / 2}} \mathrm{e}^{i \vec{k} \cdot \vec{x}} \epsilon_i^{(+)}(\hat{k}) A_{+}^{(2)}(\tau, k)\left[\hat{b}_{+}(\vec{k})+\hat{b}_{+}^{\dagger}(-\vec{k})\right]\,.
   \end{aligned}
\end{equation}
Next, to tie the above to the tensor perturbations, we write the metric as:
\begin{equation}
    ds^2=a^2(\tau)[-d\tau^2+\left(\delta_{ij}+\hat{h}_{ij}(\tau,\vec{x})\right)dx^idx^k]\,,
\end{equation}
where the $\hat{h}_{ij}$ have the mode expansion in~\cref{eq:hexpand}. Expanding the Einstein-Hilbert and gauge field action to second order in $\hat{h}_{ij}$, including the first order interaction term with the gauge field, one obtains 
\begin{equation}
    S_{\mathrm{GW}}=\int d^4 x\left[\frac{M_p^2 a^2}{8}\left(|\hat{h}_{i j}^{\prime}|^2-|\hat{h}_{i j, k}|^2\right)-\frac{a^4}{2} \hat{h}_{i j}\left[\left(\hat{E}_i \hat{E}_j+\hat{B}_i \hat{B}_j\right)_1+\left(\hat{E}_i \hat{E}_j+\hat{B}_i \hat{B}_j\right)_2\right]\right]\,.
\end{equation}
Here the electric and magnetic fields are defined in~\cref{eq:EMfields} and the subscripts ``1'' and ``2'' correspond to the spectators' labels.

\subsection{Curvature Perturbations}

\noindent Beginning with the action in~\cref{eq:curvmode}, one can derive the following equations of motion for the modes  $\hat{Q}_\phi$ and $\hat{Q}_\chi$
\begin{align}
\nonumber\left(\frac{\partial^2}{\partial \tau^2}+k^2+\tilde{M}_{\phi \phi}^2\right) &\hat{Q}_\phi+\tilde{M}_{\phi \chi_1}^2 \hat{Q}_{\chi_1}+\tilde{M}_{\phi \chi_2}^2 \hat{Q}_{\chi_2}=0\,, \\
\left(\frac{\partial^2}{\partial \tau^2}+k^2+\tilde{M}_{\chi_l \chi_l}^2\right) &\hat{Q}_{\chi_l}+\tilde{M}_{\chi_l \phi}^2 \hat{Q}_\phi+\tilde{M}_{\chi_l \chi_k}^2 \hat{Q}_{\chi_k}=\lambda_l \frac{a^3}{f_l} \int \frac{d^3 x}{(2 \pi)^{3 / 2}} \mathrm{e}^{-i \vec{k} \cdot \vec{x}} \vec{\hat{E}_l} \cdot \vec{\hat{B}_l}\,.
\end{align}
Taking the slow-roll expansion of the mass matrix in~\cref{eq:massmatrix}, one finds 
\begin{align}
    \tilde{M}_{ll }^2&= -\frac{2}{\tau^2}+\frac{3}{\tau^2}\eta_l-\frac{6}{\tau^2}\epsilon_l\,,\\
    \tilde{M}_{lk}^2&=-\frac{6}{\tau^2}\sqrt{\epsilon_l\epsilon_k}\,,
\end{align}
where slow-roll parameters have been introduced for each of the fields $\phi_l$, $(\phi_1,\phi_2,\phi_3)=(\phi,\chi_1,\chi_2)$. To leading order terms in slow roll, the mass matrix reads
\begin{equation}
    \tilde{M}_{lk}^2\simeq -\frac{1}{\tau^2}\begin{pmatrix}
2 & 6\sqrt{\epsilon_\phi \epsilon_{\chi_1}} & 6\sqrt{\epsilon_\phi \epsilon_{\chi_2}} \\
6\sqrt{\epsilon_{\chi_1} \epsilon_{\phi}} & 2& 6\sqrt{\epsilon_{\chi_1} \epsilon_{\chi_2}}\\
6\sqrt{ \epsilon_{\chi_2}\epsilon_\phi}&6\sqrt{ \epsilon_{\chi_2}\epsilon_{\chi_1}}&2
\end{pmatrix}\,.
    \end{equation}
To solve the equations of motion, one can introduce the retarded Green's function
\begin{equation}
    G_k\left(\tau, \tau^{\prime}\right)=\Theta\left(\tau-\tau^{\prime}\right) \frac{\pi}{2} \sqrt{\tau \tau^{\prime}}\left[J_{3 / 2}(-k \tau)Y_{3 / 2}\left(-k \tau^{\prime}\right)-Y_{3 / 2}(-k \tau) J_{3 / 2}\left(-k \tau^{\prime}\right)\right]\,,
\end{equation}
where J and Y denote the Bessel functions with real arguments. This yields the particular solution presented in~\cref{eqscalar1}, which can be rewritten more explicitly by defining the new vector $\vec{\tilde{p}}\equiv\frac{\vec{p}}{k}$ as:
\begin{align}
\nonumber\zeta^{(1)}(\tau, \vec{k})=\sum_{i=1,2}\frac{3 \pi^{3 / 2} H^2 \lambda_i \sqrt{\epsilon_{\chi_*^i}}}{8 M_p f_i} \int \frac{d^3 \tilde{p}}{(2 \pi)^{3 / 2}} \tilde{p}^{1 / 4}|\hat{k}-\vec{\tilde{p}}|^{1 / 4}&\left(\tilde{p}^{1 / 2}+|\hat{k}-\vec{\tilde{p}}|^{1 / 2}\right) 
N_i\left[\xi_*^i, \tilde{p}x_*^i, \delta_i\right] N_i\left[\xi_*^i,|\hat{k}-\vec{\tilde{p}}| x_*^i, \delta_i\right]\cdot\\
  &\cdot \hat{W}_i[\vec{p}, \vec{k}]\,
T_{\zeta}^i\left[\xi_*^i, x_*^i, \delta_i,
\sqrt{\tilde{p}}+\sqrt{|\hat{k}-\vec{\tilde{p}}}\right] \,,
\label{eq:curvepert}
\end{align}
where we introduced the quantities $x=-k\tau$ and $x_*=-k\tau_*$ and denoted with $\hat{k}$ the unit vector.
The coefficients $N_{i}$ appearing in Eq.~(\ref{eq:curvepert}) arise from the gauge field mode functions:
\begin{equation}
\tilde{A_i}(\tau,p)\tilde{A_i}(\tau,|\vec{k}-\vec{p}|)=N_i[\xi_*^i,-p\tau_*^i,\delta_i]N_i[\xi_*^i, -|\vec{k}-\vec{p}|\tau_*^i, \delta_i] \,\text{exp}\left[-\frac{4(\xi_*^{i})^{1/2}}{1+\delta_i}\left(\frac{\tau}{\tau_*^i}\right)^{\delta_i/2}\left(\sqrt{-p\tau}+\sqrt{-|\vec{k}-\vec{p}|\tau}\right)\right]\,.
\end{equation}
In Eq.~(\ref{eq:curvepert}), we also introduced the function 
\begin{equation}
    T_\zeta^i\left[\xi_*, x_*, d, Q\right]\equiv \int_0^{\infty} \frac{d x^{\prime}}{x^{\prime}} J_{3/2}\left(x^{\prime}\right) \sqrt{\frac{\epsilon_{\chi_i}\left(x^{\prime}\right)}{\epsilon_{\chi_{*i}}}} \int_{x^{\prime}}^{\infty} d x^{\prime \prime} x^{\prime \prime 3 / 2} \exp \left[-\frac{4 \xi_*^{1 / 2}}{1+\delta_i} \frac{x^{\prime \prime\left(1+\delta_i\right)/2}}{x_*^{\delta_i / 2}} Q\right] 
\left[J_{3 / 2}\left(x^{\prime}\right) Y_{3 / 2}\left(x^{\prime \prime}\right)-Y_{3 / 2}\left(x^{\prime}\right) J_{3 / 2}\left(x^{\prime \prime}\right)\right]\,.
\end{equation}
and defined the operators:
\begin{equation}
    \hat{W}_i[\vec{p}, \vec{k}]\equiv\epsilon_j^{(+)}(\vec{\tilde{p}}) \epsilon_j^{(+)}(\hat{k}-\vec{\tilde{p}})
   \left[\hat{a}_{+}(\vec{p})+\hat{a}_{+}^{\dagger}(-\vec{p})\right]
   \left[\hat{a}_{+}(\vec{k}-\vec{p})+\hat{a}_{+}^{\dagger}(-\vec{k}+\vec{p})\right] \,,
\end{equation}
for which the following relation holds (to leading order in slow-roll):
\begin{equation}
   \langle \hat{W}_1 \hat{W}_2\rangle = 0\,.
\end{equation}
As a result, the contributions from the different axions are decoupled from one another and one can compute the total power spectrum as the sum of two separate contributions.

%==============================
\section{Backreaction and perturbativity}
\label{app:backreaction}
%==============================

\noindent  Given a set of model parameters, such as those provided in Table~~\ref{table:cosmotable2}, consistency with the working assumptions in sections \ref{sectiontwo} and \ref{sectionthree} demands that one verifies that the backreaction of the spectator fields on the Friedmann equation and the backreaction on the evolution of the axion background remain negligible. In addition, a self-consistent analysis requires the implementation of perturbativity bounds \footnote{Perturbativity bounds are, ultimately, conditions imposed on loop corrections (induced by the Chern-Simons coupling) to the tree-level propagators of axion and gauge fields (see e.g. Fig.~3 of \cite{Peloso:2016gqs}).}.
The study of backreaction in the case of a single (Abelian) spectator sector was worked out in  \cite{Peloso:2016gqs}. Perturbativity constraints were also derived in \cite{Peloso:2016gqs}, following \cite{Ferreira:2015omg}. We will now generalise those bounds to our model.\\

Following \cite{Peloso:2016gqs}, we begin by requiring that the energy density of the axion fields gives a negligible contribution to the total energy density of the universe, in other words:
\begin{equation}
\sum_{i=1}^{N}\rho_{\chi_{i}}\equiv \sum_{i=1}^{N}\left(\frac{\dot{\chi_{i}}^{2}}{2}+V_{S_{i}}(\chi_{i})\right)\ll 3 H^{2} M_{P}^{2}\,,
\end{equation}
where $N$ is the number of spectator sectors. For each spectator, the maximum value of the kinetic energy is 
\begin{equation}\label{kinmax}
\frac{\dot{\chi_{i*}}^{2}}{2}=3 H^{2}M_{P}^{2}\,\frac{\epsilon_{\chi_{i*}}}{3}\,.
\end{equation}
The maximum of the potential, $V_{S_{i}}^{\text{max}}(\chi_{i})=\Lambda_{i}^4$, can also be written in terms of the slow-roll parameter $\epsilon_{\chi_{i*}}$ using the expression $\dot{\chi}_{i*}=f_{i} H \delta_{i}$ from Eq.~(\ref{chistar}), in combination with the relation $\delta_{i}=\Lambda_{i}^{4}/(6 H^{2} f_{i}^{2})$ (see Sec.~\ref{sectionthree}):
\begin{equation}\label{potmax}
V_{S_{i}}^{\text{max}}(\chi_{i})=3 H^{2}M_{P}^{2}\,\frac{4\epsilon_{\chi_{i*}}}{\delta_{i}}\,.
\end{equation}
With Eqs.~(\ref{kinmax}) and (\ref{potmax}), one obtains
\begin{equation}\label{ineq1}
\sum_{i=1}^{N}\rho_{\chi_{i}}^{\text{max}}=3 H^{2}M_{P}^{2}\sum_{i=1}^{N}\left(\frac{\epsilon_{\chi_{i*}}}{3}+\frac{4\epsilon_{\chi_{i*}}}{\delta_{i}}\right)\simeq 3 H^{2}M_{P}^{2}\sum_{i=1}^{N}\frac{4\epsilon_{\chi_{i*}}}{\delta_{i}}\ll 3 H^{2}M_{P}^{2}\quad\Longleftrightarrow \quad \sum_{i=1}^{N}\frac{\epsilon_{\chi_{i*}}}{\delta_{i}}\ll \frac{1}{4}\,.
\end{equation}
This condition can be rewritten as 
\begin{equation}\label{finalc1}
\sum_{i=1}^{N}\delta_{i}f_{i}^{2}\ll \frac{M_{P}^{2}}{2}\,.
\end{equation}
where the relation $\epsilon_{\chi_{i*}}=\delta_{i}^{2}f_{i}^{2}/(2 M_{P}^{2})$ was employed.
As an example, for values of the axion decay constant of order $10^{-3}M_{P}$ to $0.1\,M_{P}$, Eq.~(\ref{finalc1}) provides the corresponding upper bounds on the total number of axions, $N^{\text{max}}\in [10^{2},\,10^{6}]$ (assuming $\delta_{i}$ values of order $0.5$, as in the main text, and the same value of $f$ for all spectators).

The second backreaction constraint arises from requiring that the gauge fields amplification does not alter the motion of the axions. To this end, it was verified in \cite{Peloso:2016gqs} that it suffices to impose the condition $\rho_{A_{}}^{\text{max}}\ll\dot{\chi}_{*}^2/2$ ($\rho_{A_{}}$ being the gauge field energy density). 
Explicit expressions, and the corresponding inequalities, were derived for $\rho_{A_{}}^{\text{max}}$ in \cite{Peloso:2016gqs}. Those results straightforwardly apply to the case of multiple spectators, as the various sectors are only minimally coupled to one another. The same goes for the perturbativity constraints worked out in the same paper, these being inherent to each individual sector. These combined backreaction and perturbativity constraints lead to the following conditions \cite{Peloso:2016gqs}:
\begin{align}\label{delta02}
\delta=0.2&: \quad  2 \cdot 10^{-5} \mathrm{e}^{2.74 \xi_*} \sqrt{\epsilon_\phi} \lesssim \frac{f}{M_p} \lesssim 0.71 \,,\\\label{delta05}
\delta=0.5&:\quad \operatorname{Max}\left[1.4 \cdot 10^{-5} \mathrm{e}^{2.42 \xi_*} \sqrt{\epsilon_\phi}, \,5.1 \cdot 10^{-6} \mathrm{e}^{2.60 \xi_*} \sqrt{\epsilon_\phi}\right] \lesssim \frac{f}{M_p} \lesssim 0.28\,,
\end{align}
where $\delta=0.2,\,0.5$ are two of the sample values of $\delta$ considered also in the present manuscript. 
In each case, the second inequality is automatically satisfied given that $f \lesssim M_{p}$.\\
In the main text, for the computation of the power spectra, we used the value $\epsilon_{\phi}\sim 10^{-3}$. If we take as reference $f\sim 0.3 M_p$, the bounds in \cref{delta02} and \cref{delta05} can then be rewritten as: 
\begin{align}
\delta=0.2&: \quad  6.32 \cdot 10^{-7} \mathrm{e}^{2.74 \xi_*} \lesssim 0.3 \quad\Longrightarrow\quad \xi_*\lesssim 4.8 \,,\\
\delta=0.5&:\quad \operatorname{Max}\left[4.4 \cdot 10^{-7} \mathrm{e}^{2.42 \xi_*}, \, 1.6 \cdot 10^{-7} \mathrm{e}^{2.60 \xi_*} \right] \lesssim0.3\quad\Longrightarrow \quad \xi_*\lesssim 5.5\,,
\end{align}
These $\xi_{*}^{\text{max}}$ values are fairly close to the benchmark points in Table~\ref{table:cosmotable2}. A signal at the level of the stochastic background observed with PTA would therefore saturate both perturbativity and weak backreaction constraints. Similar conclusions can be drawn for the other cases we considered, $\delta=0.3$ and $\delta=0.6$.

%=====================================================

%==============================
\section{String Conventions}
\label{app:stringconv}
%==============================
\subsection{Type IIB Orientifolds}
In this subsection, we outline further details of $4d$ type IIB orientifold models. We will list only the essential details -- for more in-depth discussion, see~\cite{Jockers:2004yj,Jockers:2005pn,Jockers:2005zy,Cicoli:2012sz}.

We consider type IIB string theory on Calabi-Yau 3-fold $X_3$ with cohomology groups $H^{p,q}(X_3)$ and 
independent Hodge numbers $(h^{1,1},h^{2,1})$. The orientifold is defined by a projection operator composed of a holomorphic involution on $X_3$ and a worldsheet partity operator. Under the involution, the cohomology groups split into positive and negative eigenspaces as $H^{p,q} = H^{p,q}_+ \oplus H^{p,q}_-$ such that
\begin{equation}
    h^{1,1} = h^{1,1}_+ + h^{1,1}_- \qquad h^{2,1} = h^{2,1}_+ + h^{2,1}_- \,.
\end{equation}
As in the main text, we will use lower case Latin (Greek) indices to enumerate elements of the positive (negative) cohomologies as $\alpha,\beta,\gamma = 1,..,h^{1,1}_+$ and $a,b = 1,..,h^{1,1}_-$.

The particle content of the $4d$, $\pazocal{N}=1$ supergravity effective field theory includes $h^{2,1}_+$ vector multiplets and chiral supermultiplets for the axiodilaton, $h^{1,1}_+$ complexified K\"{a}hler moduli, $h^{1,1}_-$ odd moduli, and $h^{2,1}_-$ complex structure moduli. 
The vacuum expectation values of the $h^{1,1}_+$ scalars $\tau_\alpha$ control the size of 4-cycles in $X_3$ and are related to 2-cycle volumes $v^\alpha$ via the intersection numbers $k_{\alpha\beta\gamma}$ as $\tau_\alpha = \frac{1}{2}k_{\alpha\beta\gamma}v^\beta v^\gamma$. The axions of the model are $C_0$, $b^a$, $\rho_\alpha$, and $c^a$. The latter two sets have kinetic terms
\begin{equation}
    S_{axions}\supset -e^{\Phi} G_{ab}\; dc^a \wedge \star dc^b + \frac{1}{16\CYV^2}G^{\alpha\beta} \; d\rho_\alpha \wedge \star d\rho_\beta\fstop
\label{eq:appkin}
\end{equation}
Above we have introduced metrics on the space of harmonic 2-forms as
\begin{equation}
    \begin{aligned}
        G_{ab} &= -\frac{\pazocal{T}_{ab}}{4\CYV^2}\\
        G_{\alpha\beta} &= \frac{1}{4\CYV^2}\bigg(\frac{\tau_\alpha \tau_\beta}{\CYV^2} -\pazocal{T}_{\alpha\beta}\bigg)\,,
    \end{aligned}
\end{equation}
with $G^{\alpha\beta} = (G^{-1})^{\alpha\beta}$. Here $\CYV = \frac{1}{6}k_{\alpha\beta\gamma}v^\alpha v^\beta v^\gamma$ is the volume of $X_3$ and 
\begin{equation}
    \begin{aligned}
       % \pazocal{K}_\alpha &= k_{\alpha\beta\gamma}t^\beta t^\gamma\\ 
        \pazocal{T}_{\alpha\beta} &= k_{\alpha\beta\gamma}v^\gamma\\
        \pazocal{T}_{ab} &= k_{ab\gamma}v^\gamma\,.
    \end{aligned}
\end{equation}
These define the metrics used in~\cref{eq:axkinbas}. In the presence of D7-branes, the axions $c^a$ and $\rho_\alpha$ can become charged under gauge $U(1)$ interactions. In the Calabi-Yau $X_3$, we consider two divisors $\CYdiv$ and $\CYimdiv$ that are mapped to each other under the involution. One can then define $\pazocal{D}^+ = \CYdiv \cup \CYimdiv$ and $\pazocal{D}^- = \CYdiv \cup (-\CYimdiv)$, where the negative sign refers to orientation reversal. Note that in the main text, $\pazocal{D}^+$ is denoted as $\oridiv$. If we wrap one D7-brane on both $\CYdiv$ and $\CYimdiv$, the exterior derivatives in~\cref{eq:appkin} are promoted to the covariant derivatives 
\begin{align}
     \nabla c^a&= dc^a -q^a A\\
    \nabla \rho_\alpha&=d \rho_\alpha- i q_{\alpha}A \fstop
\label{eq:Stuck}
\end{align}
Where $A$ is the 1-form of the worldvolume gauge theory and the axion charges are 
\begin{align}
q^a&=\frac{1}{2\pi}N_{D7}\wrap^a\\
q_{\alpha}&= -\frac{\nds}{2\pi}(\kappa_{\alpha\beta\gamma}\tilde{\magn}^\beta \wrap^\gamma+\kappa_{\alpha bc}\magn^b\wrap^c)\,,
\end{align}%
where $\tilde{\magn}^\alpha$ ($\magn^c$) is worldvolume flux supported on even (odd) 2-cycles\footnote{Technically $\tilde{\magn}^\alpha$ is a combined flux of the worldvolume gauge theory and the pullback of $B_2$.}. The $\wrap^\alpha$ ($\wrap^a$) are wrapping numbers along the basis elements  $\tilde{\omega}^\alpha$ ($\tilde{\omega}^a$) of $H^{2,2}_{+}(X_3,\mathbb{Z})$ ($H^{2,2}_{-}(X_3,\mathbb{Z})$):
\begin{equation} 
\wrap^\alpha=\int_{\pazocal{D}^+}\tilde{\omega}^\alpha \,,\quad \quad \wrap^a=\int_{\pazocal{D}^-} \tilde{\omega}^a\,.
\label{eq:itsawrap}
\end{equation}
Notably, the charge $q^a$ is independent of flux and is determined by the odd wrapping number $\wrap^a$. 
This motivates the description of $C_2$-axion St\"{u}ckelberg couplings as being geometric in nature, as discussed in the main text. If $[\CYdiv] = [\CYimdiv]$, so that the divisor and image divisor are homologous, then $\wrap^a$ vanishes and the St\"{u}ckelberg mechanism can be avoided as discussed in~\cref{sec:stringy}.

\subsection{Worldvolume Theory of D7-branes \& Induced Tadpoles}
In this section, we briefly review the relation between $4d$ gauge theories and the worldvolume theory of D7-branes wrapped on 4-cycles. We will follow here the conventions used in~\cite{Polchinski:1998rr} as well as in~\cite{Jockers:2004yj}. In string frame, the bosonic portion of the low-energy $10d$ type IIB action in the democratic formulation is
\begin{equation}
    S_{IIB}= \frac{2\pi}{\ell_s^8}\left(\int e^{-2\phi}R \star_{10}1 - \frac{1}{2}\int e^{-2\phi}\bigg( 8d\phi\wedge \star_{10}d\phi - H_3 \wedge \star_{10} H_3 \bigg) +\frac{1}{4}\int \sum_{p=1,3,5,7,9}  G_{p} \wedge \star_{10}G_{p}\right) \fstop
\label{eq:IIBaction}
\end{equation}
Here $\phi$ is the dilaton, $H_3=dB_2$, and the various field strengths are defined is terms of the $p$-form gauge potentials $C_p$ by $G_1 = dC_0$ for $p=1$ and $G_p = dC_{p-1} - dB_2\wedge C_{p-2}$ otherwise. Here we define the string length as $\ell_s\equiv2\pi\sqrt{\alpha'}$. Dimensional reduction of~\cref{eq:IIBaction} on the $6d$ orientifold, and transition to the Einstein frame, yields the kinetic terms in~\cref{eq:Stuck} addition to the kinetic terms of moduli.

Contributions to the $10d$ bosonic effective action from the various supersymmetric (BPS) D$p$-branes take the form of the Dirac-Born-Infeld (DBI) action and the Chern-Simons action:
\begin{equation}
    S_{{\rm D}p}= -\frac{2\pi}{\ell_s^{p+1}}\int {\rm d}^{p+1}x\sqrt{-\det(\varphi^*[g_{10}+B_2]-\frac{\ell_s^2}{2\pi} F_2)} +\frac{2\pi}{\ell_s^{p+1}} \int e^{\frac{\ell_s^2}{2\pi} F_2-\varphi^*[B_2]}\wedge \sqrt{\frac{\hat{A}[\ell_s^2 R_T]}{\hat{A}[\ell_s^2R_N]}} \wedge \bigoplus_q \varphi^*[C_q] \fstop
\label{eq:Dp}
\end{equation}
Where $\varphi[\cdots]$ denotes the pullback to the D7-brane worldvolume, $\hat{A}[\cdots]$ is the A-roof genus, and $R_T$ ($R_N$) is the curvature 2-form of the tangent (normal) bundle of the brane worldvolume embedding. From here we set $\ell_s=1$ and restore proper mass dimensions in the usual manner.

We now consider the dimensional reduction of~\cref{eq:Dp} for the scenario of a D7-brane ($p=7$) filling 4D macroscopic space-time $M_4$ and wrapping a properly chosen 4-cycle $\oricycle$ of a suitable Calabi-Yau (CY) O7-orientifold compactification. Keeping only relevant terms, we look at the Kaluza-Klein (KK) zero mode of $C_4$ on $\Pi_4$ and set $C_4=c_4\omega_{\Pi_4}+\cdots$ and $F_2=\frac{1}{2}F_{\mu\nu}dx^\mu dx^\nu$. Using $F_2\wedge F_2=\frac{1}{4} \epsilon^{\mu\nu\rho\sigma}F_{\mu\nu}F_{\rho\sigma}d^4x$ and $F_2\wedge \star_4 F_2= \frac{1}{2} F_{\mu\nu}F^{\mu\nu} \sqrt{-g_4} d^4x$ as well as
\begin{equation}
\begin{aligned}
&\int\limits_{M_4\times \oricycle} {\rm d}^{8}x\sqrt{-\varphi^*[g_{10}]} = \underbrace{\int_{\oricycle}{\rm d}^4y\sqrt{g(\oricycle)}}_{=\tau}\cdot \int_{M_4}{\rm d}^4x\sqrt{-g_4}\\
&\int \sum_q e^{\frac{1}{2\pi} F_2-\varphi^*[B_2]}\wedge C_q\supset \underbrace{\int_{\Pi_4}C_4}_{=c_4}\cdot \int_{M_4}\frac12 \frac{1}{4\pi^2}F_2\wedge F_2\,,
\end{aligned}
\end{equation} 
the two-derivative part of the D7-brane action to be
\begin{eqnarray}
    S_{{\rm D}7-gauge}&=&- 2\pi \int \left[\frac{1}{4\pi^2} \tau \frac12 F_2\wedge\star_4 F_2+\frac{1}{4\pi^2} c_4 \frac12 F_2\wedge F_2\right]\nonumber\\
&=&\int {\rm d}^4x\sqrt{-g_4}\left[-\frac{1}{4} (\frac{1}{2\pi}\tau)  F_{\mu\nu}F^{\mu\nu}-\frac18 (\frac{1}{2\pi} c_4) \epsilon^{\mu\nu\rho\sigma}F_{\mu\nu}F_{\rho\sigma}\right]\,.
\label{eq:D7EM}
\end{eqnarray}
We can compare this to the standard form of a 4D ${\cal N}=1$ supersymmetric U(1) gauge theory with holomorphic gauge kinetic function $f(T)$ of a chiral superfield $T=\tau+ic_4$,  given by
\begin{equation}
    S_{U(1)}=\int_{M_4}{\rm d}^4x\sqrt{-g_4}\left[\frac14 \int {\rm d}^2\theta f(T) W_\beta W^\beta + h.c.\right]\,,
    \label{eq:SQED}
\end{equation}
where the $\theta^2$ component of the square of the super field strength $W_\alpha$ evaluates to ($\lambda$ denotes the gaugino superpartner of the gauge field)
\begin{equation}
\left. W_\beta W^\beta\right|_{\theta\theta}=-2i\lambda \sigma^\mu\partial_\mu\bar\lambda-\frac12 F_{\mu\nu}F^{\mu\nu}+D^2+\frac{i}{4} \epsilon^{\mu\nu\rho\sigma}F_{\mu\nu}F_{\rho\sigma}\,.
\end{equation}
Plugging this into eq.~\eqref{eq:D7EM} we get, for unbroken SUSY ($D=0$), for the bosonic sector
\begin{equation}
    S_{U(1),bos.}=\int_{M_4}{\rm d}^4x\sqrt{-g_4}\left[-\frac14 {\rm Re }\,f(T) F_{\mu\nu}F^{\mu\nu}-\frac18 {\rm Im}\,f(T)\epsilon^{\mu\nu\rho\sigma}F_{\mu\nu}F_{\rho\sigma}\right]\,.
   % \label{eq:D7bosons}
\end{equation}
Direct comparison with eq.~\eqref{eq:D7EM} reveals that we must choose
\begin{equation}
f_{\oricycle}^{(1)}(T) =\frac{1}{2\pi} T_{\oricycle}\,,
\end{equation}
to match the D7-brane effective action. Up to this point everything was done for a single D7-brane generating a single U(1)  supersymmetric gauge field theory. If we replace this with a stack of $N_{{\rm D}7}$ coincident D7-branes wrapping the same 4-cycle, this stack will generate a non-Abelian 4D super-Yang-Mills (SYM) gauge theory. What effectively changes in this case in the expressions above is that every occurrence of $F_2\wedge F_2$ and $F_2\wedge \star F_2$ gets replaced by $\tr F_2\wedge F_2$ and $\tr F_2\wedge \star F_2$, respectively, where the trace pertains to $F_2=F_2^a T^a$ (with $T^a$ the generators of the Lie group in a given representation) now being Lie algebra valued in the SYM gauge group. The additional trace will thus evaluate to $\tr T^aT^b$  producing an additional factor $1/2$ for the $T^a$ in the fundamental representation. \\
\indent As explained in~\cref{sec:stringy}, we can introduce an additional CS coupling of the gauge field to $C_2$-axions if we allow for the presence of magnetic flux in the D7-branes.  Inclusion of  these magnetic fluxes modifies the D7-brane gauge kinetic function to the expression in~\cref{eq:gaugkin2}.  However, there is one more effect of turning on gauge flux-the D7-branes contribute additional terms to the D3-brane tadpole . The relevant contribution can be determined from dimensional reduction of the CS term in~\cref{eq:Dp}:
\begin{equation}
\begin{aligned}
S_{{\rm D}7}^{CS}%&= 2\pi \int \sum_q e^{B_2+\frac{1}{2\pi} F_2}\wedge\sqrt{\frac{\hat{A}(T\Pi_4)}{\hat{A}(N\Pi_4)}}\wedge C_q\nonumber\\
&\supset 2\pi\wrap \frac12 \,\,\Bigg\{\int_{\Pi_4} \frac{1}{4\pi^2}F_2\wedge F_2 +\int_{\Pi_4} \frac{1}{4\pi^2 48}(p_1[R_N^{\oricycle}]-p_1[R_T^{\oricycle}])\Bigg\}\underbrace{\int_{M_4}C_4}_{=\text{D3-brane}\; \text{CS-term}}\\
&=\pi \Bigg\{\wrap (m_1)^2\int_{\Pi_4}\omega_{\Pi_2^G}\wedge \omega_{{\rm dual}(\Pi_2^G)}+\int_{\Pi_4}\wrap\frac{1}{24}c_2[R_T^{\oricycle}]\,\Bigg\}\cdot\int_{M_4}C_4\\
&=\pi\Bigg\{\underbrace{\wrap(\magn_1)^2 \kappa_{+--}+\wrap\frac{1}{24}\chi(\oricycle)}_{Q_{D3,ind}}\,\Bigg\}\cdot\int_{M_4}C_4 \quad.
\end{aligned}
\end{equation}

Where we have used the definition of the A-roof genus $\hat{A}[R]=1-\frac{1}{24}p_1[R]+...$ expressed in terms of Pontryagin classes $p_n$, while $c_2$ corresponds to the second Chern class.%~\cite{Junghans:2014zla}.  The subscripts $T,N$ denote the tangent and normal bundle of $\Pi_4$.  

The whole prefactor of $\int_{M_4}C_4$ thus constitutes a D3-brane charge induced on the D7-brane world volume.  
\begin{equation}
Q_{{\rm D}3,ind.}= \wrap \kappa_{+--} (m_1)^2 +\wrap\frac{1}{24}\chi(\Pi_4)\quad.
\end{equation}
The first term  comes from turning on quantized internal gauge flux, while the second corresponds to the intrinsic curvature-induced amount of D3-brane charge which any D7-brane or O7-plane wrapping a non-flat 4-cycle acquires. 
This induced D3-brane charge grows quadratically in the gauge flux quanta $m^G$ and linearly in the wrapping number. Now, D3-brane charge, like any localized charge sourcing a long-range gauge field strength, satisfies a Gauss' law constraint. Hence, in the compact 6 dimensions of the CY the field lines emanating from $D_{{\rm D}3,ind.}$ must end on equal in magnitude and sign-opposite D3-charge. In a consistent type IIB string theory compactification on CY orientifolds this balancing D3-charge is generated by higher-curvature couplings in the CS terms of single D7-branes wrapping all the 4-cycles of the CY. Any such consistent type IIB CY orientifold compactification has a lift to F-theory, where the orientifolded CY 3-fold of type IIB string theory gets lifted into an elliptically fibred CY 4-fold $X_4$. The total D3-brane charge from curvature couplings on D7-branes wrapping 4-cycles in type IIB becomes in F-theory equal to $Q_{{\rm D}3, tot.}(X_4)=1/24 \chi(X_4)$ and is thus completely fixed by the topology of $X_4$.

\bibliographystyle{JHEP}
\bibliography{MASA_Refs}

\end{document}